\documentclass[twocolumn]{aastex631}
\newcommand\aastex{AAS\TeX}

\newcommand{\cha}{\textit{Chandra }}
\def\xmm{{XMM-{\it Newton }}}
\def\fermi{{{\it Fermi}--LAT }}

\def\aap{A\&A}

\def\apjl{ApJL}
\def\apjs{ApJS}

\shorttitle{A Systematic Search for MeV--GeV PWNe}
\shortauthors{The \fermi Collaboration.}

\PassOptionsToPackage{colorlinks,linkcolor={blue},citecolor={blue},urlcolor={red}}{hyperref}
\usepackage{hyperref}
\usepackage{natbib}
\usepackage{float}
\usepackage{color}
\usepackage{graphicx}
\usepackage[utf8]{inputenc}
\usepackage{gensymb}
\usepackage{amsmath}
\usepackage{enumitem}
\usepackage{multirow,helvet}
\usepackage{todonotes}
\usepackage{lipsum,wrapfig} 
\usepackage{makecell}
\usepackage{xr}

\begin{document}
\title{A Systematic Search for MeV--GeV Pulsar Wind Nebulae without Gamma-ray Detected Pulsars}

\collaboration{108}{The Fermi-LAT collaboration}

\newcommand\blfootnote[1]{%
  \begingroup
  \renewcommand\thefootnote{}\footnote{#1}%
  \addtocounter{footnote}{-1}%
  \endgroup
}
\blfootnote{*Corresponding authors. Contact:}


\author[0000-0002-2028-9230]{A.~Acharyya}
\affiliation{Center for Cosmology and Particle Physics Phenomenology, University of Southern Denmark, Campusvej 55, DK-5230 Odense M, Denmark}
\author{A.~Adelfio}
\affiliation{Istituto Nazionale di Fisica Nucleare, Sezione di Perugia, I-06123 Perugia, Italy}
\author[0000-0002-6584-1703]{M.~Ajello}
\affiliation{Department of Physics and Astronomy, Clemson University, Kinard Lab of Physics, Clemson, SC 29634-0978, USA}
\author[0000-0002-9785-7726]{L.~Baldini}
\affiliation{Universit\`a di Pisa and Istituto Nazionale di Fisica Nucleare, Sezione di Pisa I-56127 Pisa, Italy}
\author[0000-0002-8784-2977]{J.~Ballet*}
\affiliation{Universit\'e Paris-Saclay, Universit\'e Paris Cit\'e, CEA, CNRS, AIM, F-91191 Gif-sur-Yvette Cedex, France}
\author[0000-0001-7233-9546]{C.~Bartolini}
\affiliation{Istituto Nazionale di Fisica Nucleare, Sezione di Bari, I-70126 Bari, Italy}
\affiliation{Universit\`a degli studi di Trento, via Calepina 14, 38122 Trento, Italy}
\author[0000-0002-6729-9022]{J.~Becerra~Gonzalez}
\affiliation{Instituto de Astrof\'isica de Canarias and Universidad de La Laguna, Dpto. Astrof\'isica, 38200 La Laguna, Tenerife, Spain}
\author[0000-0002-2469-7063]{R.~Bellazzini}
\affiliation{Istituto Nazionale di Fisica Nucleare, Sezione di Pisa, I-56127 Pisa, Italy}
\author[0000-0001-9935-8106]{E.~Bissaldi}
\affiliation{Dipartimento di Fisica ``M. Merlin" dell'Universit\`a e del Politecnico di Bari, via Amendola 173, I-70126 Bari, Italy}
\affiliation{Istituto Nazionale di Fisica Nucleare, Sezione di Bari, I-70126 Bari, Italy}
\author[0000-0002-4264-1215]{R.~Bonino}
\affiliation{Istituto Nazionale di Fisica Nucleare, Sezione di Torino, I-10125 Torino, Italy}
\affiliation{Dipartimento di Fisica, Universit\`a degli Studi di Torino, I-10125 Torino, Italy}
% ...

\begin{abstract}
An increasing number of pulsar wind nebulae (PWNe) are being identified in the TeV band by ground-based Imaging Air Cherenkov Telescopes such that they constitute the dominant source class of Galactic TeV emitters. 
However, MeV--GeV PWN counterparts are still largely lacking. 
To date, only a dozen PWNe are identified by the {\it Fermi}--Large Area Telescope (LAT) in the MeV--GeV band. Most PWNe are located along the Galactic plane embedded within the prominent, diffuse Galactic $\gamma$-ray emission, which makes these sources difficult to disentangle from the bright diffuse background. 
We present a systematic search for $\gamma$-ray counterparts to known PWNe in the 300\,MeV -- 2\,TeV energy band using the {\it Fermi}--LAT. 
We target the locations of previously identified PWNe
that lack detected \fermi pulsars to minimize associated pulsar contamination. The sample includes 6 previously identified {\it Fermi} PWNe and 8 \fermi sources associated with PWNe. 
We report the analysis of 58 regions of interest and classify \fermi detected sources as either a likely PWN or a candidate PWN counterpart based on their morphological and spectral characteristics across the broadband spectrum. 
There are 9 unidentified \fermi sources that we consider as likely PWN counterparts, which, if confirmed to be PWNe, would greatly increase the PWN population detected by the {\it Fermi}--LAT from 12 to 21. The remaining \fermi detected sources are considered weaker PWN candidates. A second approach in the systematic search for $\gamma$-ray emitting PWNe will involve studying the off-pulse phases of \fermi detected pulsars for the presence of an obscured PWN and will be reported in a subsequent paper.

\end{abstract}

\section{Introduction}

A pulsar wind nebula (PWN) is composed of a highly magnetized, relativistic plasma powered by an energetic, rapidly rotating neutron star descending from a core collapse supernova (CC SN). The neutron star loses angular momentum as rotational energy is converted to a relativistic particle wind made up of mostly electrons and positrons. Particles are injected by the central pulsar and accelerated at the termination shock, where the ram pressure of the cold pulsar wind is balanced by the pressure of the relativistic plasma. Altogether, the nebula is confined by the surrounding supernova (SN) ejecta \citep[e.g.,][]{rees1974,kennel1984}. Synchrotron emission from the relativistic leptons is observed from the majority of PWNe, from radio wavelengths to hard X-rays, while the same leptons scatter off local photon fields, resulting in Inverse Compton (IC) emission at $\gamma$-ray energies. 

A PWN evolves together with its host supernova remnant (SNR) and is influenced by the properties of the central pulsar, the SNR, and the structure of the surrounding interstellar medium \citep[ISM,][]{slane2006}. A variety of both magnetohydrodynamical and semi-analytic radiative evolutionary models have been developed to explore the PWN evolution inside a nonradiative SNR, see, e.g., \cite{blondin2001, swaluw2004, gelfand2009,Martin2012,Bandiera2023}. Constraining the physical properties of the particles is difficult, requiring a combination of simulation tools and observable properties over the entire electromagnetic spectrum. MeV--GeV measurements are therefore critical for characterizing accurately the particle spectrum at injection and its consequent evolution. 
Further, evolutionary studies of PWNe indicate that the $\gamma$-ray luminosity increases with time, suggesting that many evolved PWNe may be emitting brightly in the {\it Fermi}--Large Area Telescope (LAT) band. Indeed, so far we have seen that several identified \fermi PWNe are old \citep[$\tau \gtrsim 5\,$kyr, see e.g.,][]{hess2012,devin2018,principe2020}.

The termination shock of the PWN is a potential site for efficient particle acceleration. As such, PWNe provide unique laboratories to study relativistic particle acceleration processes and the required environmental conditions, especially during the PWN interaction with its surroundings. Understanding the PWN population and the interactions that take place are crucial for identifying how the relativistic particles rejoin the ISM, how they contribute to replenishing the electron Galactic cosmic ray (CR) population \citep{renaud2009,karg2013}, and whether they are responsible for local enhancements in the e$^-$e$^+$ flux \citep{malyshev_2009}.

Historically, PWNe like the Crab were discovered in droves in the radio ($\gtrsim 30$) and X-ray ($\gtrsim 60$) wavelengths and now $\gamma$-rays are becoming key to finding and characterizing PWNe \citep{karg2013,green2019}. The H.E.S.S. Galactic Plane Survey \citep{hessgps2018} has been particularly impactful towards TeV PWN discovery, as well as earlier MAGIC and VERITAS observations, all of which are included in the TeVCat\footnote{\url{tevcat.uchicago.edu}} \citep[$\sim 36$ listed as PWNe,][]{tevcat2008}. Most recently, the LHAASO catalog reported 90 $E > 1\,$TeV sources, 43 of them also detected as ultra-high-energy (UHE, $E > 100\,$TeV), and many of them classified as PWNe or candidate PWNe \citep{lhaaso2023}. 
%
%
%
%
The observed TeV $\gamma$-ray spectra from many of these sources indicate that the energy peak occurs in the MeV--GeV band, where the \fermi provides the best sensitivity and sky coverage. 
%
%
MeV--GeV observations are also effective in detecting PWNe that may peak in the $\gamma$-ray band, but appear faint in other wavelengths \citep[e.g.,][]{ackermann2011}. 

\citet{ackermann2011} analyzed the off-pulse emission of 54 \fermi detected pulsars for $E>100\,$MeV and detected 2 PWNe: the Crab and Vela-X. The 2FGL catalog \citep{2fgl} detected two more \fermi PWNe and the second pulsar catalog \citep{2pc} reported the detection of PWN 3C~58 which is later characterized in detail by \citet{li2018}. 
Several PWNe with no associated $\gamma$-ray bright pulsar have been identified by the \fermi through spatial coincidence using observations in other wavelengths. An analysis was conducted in 58 regions around TeV PWNe and unidentified TeV sources within 5\,$^\circ$ of the Galactic plane using 45 months of \fermi data \citep{acero2013} that resulted in the detection of 30 \fermi sources, 3 of which were clearly identified as PWNe and 11 as PWN candidates. The latest comprehensive \fermi source catalog, 4FGL \citep[data release 4 (DR4),][]{4fgl-dr4} currently lists a total of 12 firm PWNe and 9 PWN associations. 

We expand the \fermi PWN search effort by analyzing 138 months of \fermi data in the direction of PWNe identified in radio, X-ray, GeV, and TeV observations. The systematic search reported here targets the locations of 58 PWNe and PWN candidates that lack an associated $\gamma$-ray detected pulsar \citep[Section~\ref{sec:sample}, see also][]{3pc}. 
The sample construction comprising the 58 regions of interest (ROIs) is described in the following section (Section \ref{sec:sample}). In Section \ref{sec:fermi_analysis} we describe the \fermi data selection, reduction, and analysis. In Section \ref{sec:fermi_results} we report the results of the 58 analyzed ROIs with detailed discussions. 
We provide an overview of the results and implications in Section~\ref{sec:discuss} and conclude in Section \ref{sec:conclude}.

\section{Source Selection}\label{sec:sample}

\begin{figure*}[htbp]
\begin{minipage}[b]{1.0\textwidth}
\includegraphics[width=1.0\linewidth]{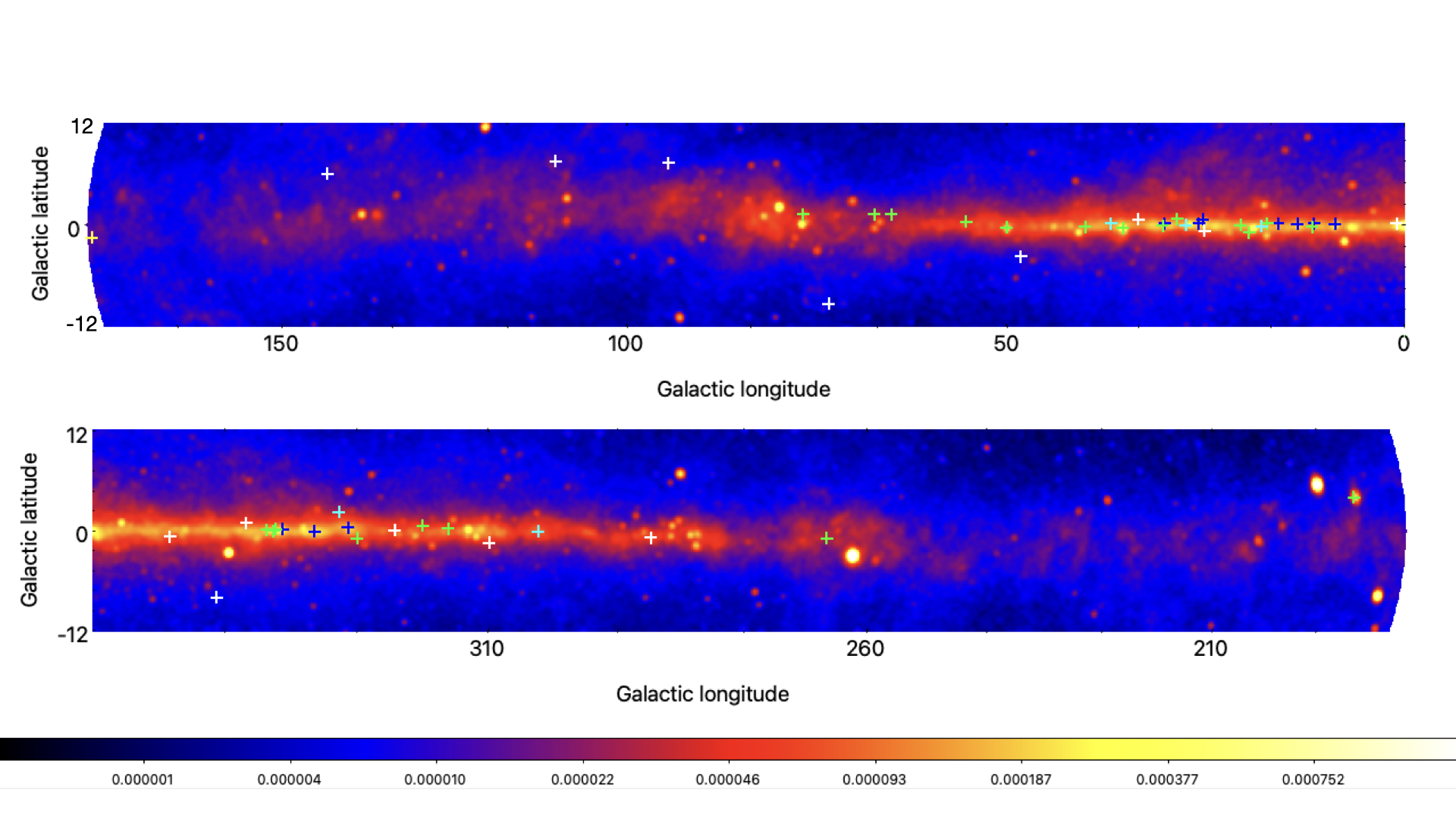}
\end{minipage}
\vspace{-.40cm}
\caption{The \fermi intensity map of the Galactic plane for $|b| < 12\,\degree$ using 12\,years of observational data with energies $E > 1$\,GeV, based on \texttt{P8R3\_SOURCE} class and \texttt{PSF3} event type. 56 sources are indicated as crosses and their color indicates whether the sources are detected (point-like sources in green and extended in blue) and nondetections in white. Two sources are located in the Large Magellanic Cloud and are not shown (see instead Figure~\ref{fig:lmc}). The units of the color scale are ph cm$^{-2}$ s$^{-1}$ sr$^{-1}$. 
}\label{fig:12yr_allsky_map}
\end{figure*}

There are $\sim$ 125 PWNe and PWN candidates\footnote{The estimate is from a compilation of plerionic SNRs and PWNe reported in \citet{mallory,manitoba2012,tevcat2008}.} that have been discovered from radio to TeV $\gamma$-rays, the majority of which were first identified in radio or X-ray surveys\footnote{\url{http://snrcat.physics.umanitoba.ca/index.php?}} \citep{manitoba2012} with an increasing number of discoveries in the TeV band.
Indeed, the majority of the TeV Galactic source population is found to originate from PWNe as observed by Imaging Air Cherenkov Telescopes \citep[IACTs,][]{hessgps2018}. However, \fermi PWN counterparts are still lacking even after 15 years of observing the entire sky every 3 hours, with only 21 PWNe currently noted as associated with sources in the comprehensive 4FGL-DR4 catalog. \citep[][]{4fgl-dr4}. 12 are extended sources considered firm PWN associations, 3 are extended sources coincident with TeV PWN candidates and 6 are point-like sources spatially coincident with PWNe. Most of these objects are located along the Galactic plane embedded within the prominent Galactic diffuse $\gamma$-ray emission (e.g., Figure~\ref{fig:12yr_allsky_map}), which makes these sources difficult to find. Additionally, nearly 150 rotation-powered pulsars that are capable of generating a PWN also emit brightly in the \fermi energy range, potentially outshining and obscuring their fainter PWNe\footnote{A publicly available list that is continuously updated can be found here: \url{https://confluence.slac.stanford.edu/display/GLAMCOG/Public+List+of+LAT-Detected+Gamma-Ray+Pulsars}} \citep{3pc}. 

Of the $\sim$ 125 PWNe and PWN candidates currently known, 62 of these have no detected $\gamma$-ray pulsar. To minimize associated pulsar contamination and the need to consider pulsar timing solutions, we remove the 63 PWNe that have detected \fermi $\gamma$-ray pulsars from the source selection. We note that removing systems with associated $\gamma$-ray detected pulsars does not necessarily avoid contamination of magnetospheric emission (i.e., pulsar). In the case of no detected $\gamma$-ray pulsar, intrinsic pulsed emission can be present but too faint to detect in a pulsation search, whether using a timing solution or not. 
For more details on this impact, see Section~\ref{sec:psrs}. A second approach in the systematic search for $\gamma$-ray emitting PWNe will involve studying the off-pulse phases of \fermi detected pulsars for the presence of a PWN and will be reported in a subsequent paper.

\begin{figure}[htbp]
\hspace{-.25cm}
\includegraphics[width=1.05\linewidth]{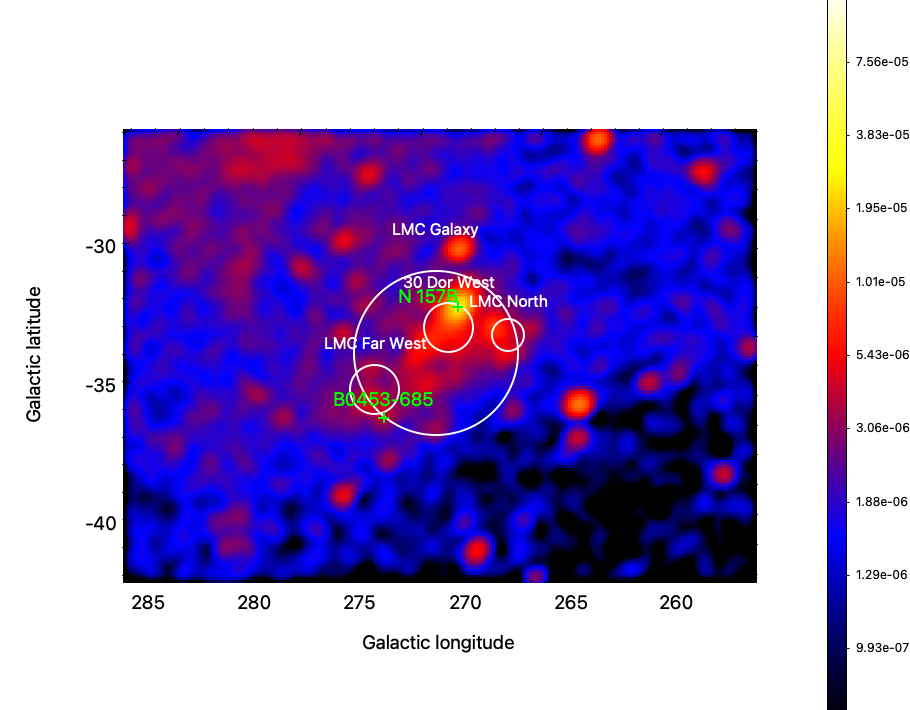}
\caption{The \fermi intensity map for $E > 1\,$GeV of the LMC centered on ($l$,$b$) = (278.76$\,\degree$, --33.46$\,\degree$). The two sources analyzed in the LMC are indicated as green crosses. The four extended source components representing the diffuse LMC emission  \citep[following the {\it{emissivity model}} developed in][]{lmc2016} are indicated in white. Based on \texttt{P8R3\_SOURCE} class and \texttt{PSF3} event type. The units of the color scale are ph cm$^{-2}$ s$^{-1}$ sr$^{-1}$.
}\label{fig:lmc}
\end{figure}

Four additional PWNe are omitted due to their proximity to the Galactic center (within $1\,\degree$ or less). The remaining 58 PWNe analyzed in this search are listed with relevant parameters in Table~\ref{tab:all_rois}.
%
Each ROI is indicated on the 12-year \fermi all-sky map for $E > 1$\,GeV in Figure~\ref{fig:12yr_allsky_map}. We note that two additional ROIs located in the Large Magellanic Cloud (LMC), see Figure~\ref{fig:lmc}, are included in the analysis, each with an identified PWN.

\begingroup
\setlength\LTcapwidth{1.0\textwidth}
\setlength{\tabcolsep}{0.04cm}
\begin{longtable*}{c | llcccc}
ROI & Galactic  & 4FGL Name & R.A. & Dec. & Extent & $\lambda$  \\
 & PWN Name  & (CLASS1)   \\
\hline
\hline
			1 & G0.87+0.08  & -- & 266.84 & --28.15 & 0.04 & Radio \\
			2 & G8.40+0.15 & J1804.7--2144e (spp) & 271.18 & --21.74 & 0.25 & TeV \\
			3 & G11.03--0.05c & J1810.3--1925e (spp) & 272.52 & --19.41 & 0.05 & Radio \\
			4 & G11.09+0.08c & J1810.3--1925e (spp) & 272.46 & --19.21 & 0.05 & Radio \\
			5 & G11.18--0.35 & J1811.5--1925 (psr) & 272.87 & --19.43 & 0.04 & Radio \\
			6 & G12.82--0.02 & J1813.1--1737e (spp) & 273.40 & --17.83 & 0.05 & TeV \\
			7 & G15.40+0.10c & J1818.6--1533 (spp) & 274.50 & --15.47 & $<$ 0.04 & TeV \\
			8 & G16.73+0.08 & J1821.1--1422 (spp) & 275.24 & --14.33 & 0.015 & Radio \\
			9 & G18.00--0.69 & J1824.5--1351e (PWN) & 276.26 & --13.97 & 0.46 & TeV \\
			10 & G18.90--1.10c & J1829.4--1256 (spp) & 277.36 & --12.97 & 0.15 & Radio \\
			11 & G20.20--0.20c & J1828.0--1133 (spp) & 277.03 & --11.59 & 0.05 & Radio \\
			12 & G23.50+0.10c & -- & 278.41 & --8.454 & 0.017 & X-ray \\
			13 & G24.70+0.60c & J1834.1--0706e (SNR) & 278.55 & --7.04 & 0.12 & Radio \\
			14 & G25.10+0.02c & J1838.9--0704e (pwn) & 279.51 & --6.93 & 0.02 & X-ray \\
			15 & G25.24--0.19 & J1836.5--0651e (pwn) & 279.34 & --6.87 & 0.02 & X-ray \\
			16 & G26.60--0.10 & J1840.9--0532e (PWN) & 280.14 & --5.72 & 0.4 & TeV \\
			17 & G27.80+0.60c & J1840.0--0411 (spp) & 279.98 & --4.29 & 0.11 & Radio \\
			18 & G29.40+0.10c & J1844.4--0306 (unk) & 281.14 & --3.12 & 0.06 & Radio \\
			19 & G29.70--0.30 & J1846.4--0258 (pwn, DR4) & 281.60 & --2.97 & $<$ 0.03 & TeV \\
			20 & G32.64+0.53c & -- & 282.24 & --0.04 & 0.09 & TeV \\
			21 & G34.56--0.50 & -- & 284.04 & 1.22 & 0.05 & X-ray \\
			22 & G36.01+0.10 & J1857.7+0246e (PWN) & 284.34 & 2.76 & 0.26 & TeV \\
			23 & G39.22--0.32 & J1903.8+0531 (spp) & 286.02 & 5.45 & 0.01 & X-ray \\
			24 & G47.38--3.88  & -- & 293.03 & 10.92 & 0.08 & X-ray \\
			25 & G49.20--0.30c & J1922.7+1428c (unk/blank) & 290.70 & 14.27 & 0.02 & X-ray \\
			26 & G49.20--0.70c & -- & 290.83 & 14.06 & 0.02 & X-ray \\
			27 & G54.10+0.27 & J1930.5+1853 (DR3) (pwn) & 292.61 & 18.84 & $<$ 0.05 & TeV \\
			28 & G63.70+1.10 & J1947.7+2744 (pwn) & 296.99 & 27.74 & 0.05 & Radio \\
			29 & G65.73+1.18 & J1952.8+2924 (spp) & 298.21 & 29.48 & 0.1 & TeV \\
			30 & G74.00--8.50c  & -- & 312.33 & 29.02 & 0.05 & X-ray \\
			31 & G74.94+1.11 & J2016.2+3712 (snr) & 304.04 & 37.19 & 0.13 & Radio \\
			32 & G93.30+6.90c  & -- & 313.06 & 55.29 & 0.01 & X-ray \\
			33 & G108.60+6.80  & -- & 336.42 & 65.60 & 0.05 & X-ray \\
			34 & G141.20+5.00  & -- & 54.30 & 61.89 & 0.03 & Radio \\
			35 & G179.72--1.69  & -- & 84.60 & 28.28 & 0.05 & X-ray \\
			36 & G189.10+3.00 & -- & 94.28 & 22.37 & 0.02 & X-ray \\
			37 & G266.97--1.00 & -- & 133.90 & --46.74 & 0.04 & Radio \\
			38 & G279.60--31.70 & J0537.8--6909 (pwn) & 84.45 & --69.17 & 0.01 & TeV \\
			39 & G279.80--35.80 & -- & 73.41 & --68.49 & 0.01 & Radio \\
			40 & G290.00--0.93 & -- & 165.44 & --61.02 & 0.08 & Radio \\
			41 & G304.10--0.24 & J1303.0--6312e (PWN) & 195.76 & --63.19 & 0.18 & TeV \\
			42 & G310.60--1.60  & -- & 210.19 & --63.43 & 0.02 & Radio \\
			43 & G315.78--0.23 & J1435.8--6018 (spp) & 219.34 & --60.02 & 0.07 & Radio \\
			44 & G318.90+0.40c & J1459.0--5819 (unk) & 224.71 & --58.41 & 0.05 & Radio \\
			45 & G322.50--0.10c  & -- & 230.80 & --57.10 & 0.08 & Radio \\
			46 & G326.12--1.81 & J1552.4--5612e (PWN) & 238.11 & --56.21 & 0.08 & Radio \\
			47 & G327.15--1.04 & J1554.4--5506 (DR3) (pwn) & 238.67 & --55.08 & 0.05 & Radio \\
			48 & G328.40+0.20 & J1553.8--5325e (blank) & 238.89 & --53.28 & 0.04 & Radio \\
			49 & G332.50--0.30 & J1616.2--5054e (PWN) & 244.40 & --50.94 & 0.02 & X-ray \\
			50 & G332.50--0.28c & J1616.2--5054e (PWN) & 244.41 & --51.04 & 0.02 & X-ray \\
			51 & G336.40+0.10 & J1631.6-4756e (pwn) & 247.98 & --47.77 & 0.18 & TeV \\
			52 & G337.20+0.10c & -- & 248.98 & --47.32 & 0.03 & X-ray \\
			53 & G337.50--0.10c & J1638.4--4715c (DR3) (blank) & 249.51 & --47.23 & 0.01 & X-ray \\
			54 & G338.20--0.00 & J1640.6--4632 (DR1), J1640.7--4631e (DR3) (spp) & 250.19 & --46.52 & 0.04 & X-ray \\
			55 & G341.20+0.90  & -- & 251.75 & --43.77 & 0.06 & Radio \\
			56 & G350.20--0.80c  & -- & 260.91 & --37.58 & 0.01 & X-ray \\
			57 & G358.29+0.24c  & -- & 265.32 & --30.38 & $<$0.07 & TeV \\
			58 & G358.60--17.20c & -- & 284.15 & --37.91 & $<$ 0.01 & X-ray \\
\hline
\hline
\caption{All PWNe and PWN candidate ROIs (marked with ``c'' at the end of their Galactic name) analyzed in this paper. Coincident 4FGL sources and their classifications are listed in the third column, considering the 4FGL--DR3 and DR4 catalogs \citep{4fgl-dr3,4fgl-dr4}\tablenotemark{a}. ``PWN'' = firm PWN counterpart, ``pwn'' = association with known PWN, ``psr'' = association with known pulsar, ``SNR'' = firm SNR counterpart, ``snr'' = association with known SNR, ``spp'' = association with known SNR or PWN, and ``unk'' or ``blank'' = unknown/unassociated. The observed right ascension (R.A.) and declination (Dec.) in J2000 equatorial degrees are listed in the fourth and fifth columns as well as the observed extent (radius) in degrees in the sixth column. The last column specifies the wavelength of the observed positions and extents. TeV data are taken from \url{http://tevcat.uchicago.edu/} \citep{tevcat2008} and radio or X-ray data from \url{http://snrcat.physics.umanitoba.ca/SNRtable.php} \citep{manitoba2012}. \tablenotemark{a}\footnotesize{The 4FGL--DR3 and DR4 include new or updated sources from DR1 or DR2 and are indicated using the appropriate data release in parantheses.}}
\label{tab:all_rois}
\end{longtable*}
\endgroup

\section{\fermi Data Analysis}\label{sec:fermi_analysis}

\subsection{Data Selection}
We use 11.5\, years (from 2008 August to 2020 January) of Pass~8 \texttt{SOURCE} class data \citep{atwood2013,pass82018} between 300\,MeV and 2\,TeV. Photons detected at zenith angles larger than 100\,$\degree$ were excluded to limit the contamination from $\gamma$-rays generated by CR interactions in the upper layers of Earth's atmosphere. We perform binned likelihood analyses with the Fermitools package\footnote{\url{https://fermi.gsfc.nasa.gov/ssc/data/analysis/software/}} (v.2.0.8) and FermiPy Python~3 package \citep[v.1.0.1,][]{fermipy2017}, 
utilizing the \texttt{P8R3\_SOURCE\_V3} instrument response function (IRF) and account for energy dispersion, to perform data reduction and analysis. We organize the events by point spread function (PSF) type, defined by the quality of their angular reconstruction\footnote{\url{https://www.slac.stanford.edu/exp/glast/groups/canda/lat_Performance.htm}}, using \texttt{evtype=4,8,16,32} to represent \texttt{PSF0, PSF1, PSF2}, and \texttt{PSF3} components. We consider separately all event types and then exclusively consider only \texttt{PSF3} events (\texttt{evtype=32}) for each ROI. The binned likelihood analyses are performed on each event type and, for the all event analyses, are then combined into a global likelihood function for the ROI to represent all events\footnote{See FermiPy documentation for details: \url{https://fermipy.readthedocs.io/en/0.6.8/config.html}}. Performing a separate binned likelihood analysis for each event type increases the sensitivity of the joint likelihood fit and optimizes the spatial resolution of the LAT, a useful tool in resolving faint and/or extended structures in crowded regions. The $\gamma$-ray data are binned using a pixel bin size $0.1\,\degree$ and 10 bins per decade in energy (38 total bins). 

The maximum likelihood technique and the resulting test statistic (TS) are used to analyze the $\gamma$-ray data compared to the best-fit source model (see Section~\ref{sec:source_models} for details on source model construction). The TS value is defined to be twice the natural logarithm of the ratio between the likelihood of one hypothesis $\mathcal{L}_1$ (i.e., presence of one additional source) and the likelihood for the null hypothesis $\mathcal{L}_0$ (i.e., absence of source):
\begin{equation}
    \text{TS} = 2 \times \log\big({\frac{{\mathcal{L}_{1}}}{{\mathcal{L}_{0}}}}\big)
\end{equation}
The TS quantifies the significance of a source detection for a given set of location and spectral parameters and the significance of such a detection can be estimated by taking the square root of the TS \citep{mattox1996}. 
For the analysis of coincident residual emission that is point-like and faint (TS $\lesssim 25$ in \texttt{PSF3}), we must consider all events. For the analysis of coincident residual emission that is significantly detected and/or exhibits extension, we consider only events of \texttt{PSF3} type. This choice is motivated first and foremost by the source confusion and complexity of the majority of regions as observed by {\it Fermi}-LAT, with the added benefit of preserving computing time. The largest drawback of removing 75\% of the photon events are the uncertainties in the spectral measurements for each source, which is estimated to improve by a factor of two when including all event types. Of the 58 ROIs, four have faint (TS $\lesssim 25$ in \texttt{PSF3}), point-like detections that instead make use of all events: G54.10+0.27, B0453--685, G327.15--1.04, and G318.90+0.40 (see Section~\ref{sec:fermi_results}). The remaining ROIs utilize \texttt{PSF3} events only. 

This analysis is complex and time intensive, during which \fermi continues to accumulate data. We have checked explicitly that additional integration time does not yield a change of conclusions for 5 of the ROIs using more of the available \fermi data (14 years) and considering the 4FGL--DR4 catalog \citep{4fgl-dr4}. Event weighting is also explored for both 14 and 11.5\,year datasets, showing that the results reported in this work are in good agreement (in the Appendix, Section~\ref{sec:system_checks}). 

\subsection{Building the Source Models}\label{sec:source_models}
We model each ROI considering the comprehensive \fermi source catalog based on 10\,years of data, 
4FGL-DR2 \citep{4fgl-dr2}, for point and extended sources\footnote{\url{https://fermi.gsfc.nasa.gov/ssc/data/access/lat/10yr_catalog/}.} and the latest Galactic diffuse template to model the Galactic interstellar emission (\texttt{gll\_iem\_v07.fits}). For the four ROIs that require consideration of all events, we employ the isotropic diffuse template \texttt{iso\_P8R3\_SOURCE\_V3\_v1.txt}. For the remainder of ROIs that only consider \texttt{PSF3} events, the isotropic diffuse template employed is \texttt{iso\_P8R3\_SOURCE\_V3\_PSF3\_v1.txt}\footnote{LAT background models and appropriate IRFs: \url{https://fermi.gsfc.nasa.gov/ssc/data/access/lat/BackgroundModels.html}.}.

For most sources, a $10 \degree \times 10 \degree$ ROI is used to construct the source model considering 4FGL sources and backgrounds within $15 \degree$ of the ROI center, which is always chosen to be the PWN position. 
The only sources that require larger ROIs ($15-20 \degree$ while considering sources within $20-25 \degree$ of ROI center) are the PWNe that overlap with SNR~S~147 (G179.72--1.69) and the Cygnus Loop (SNR~G74.00--8.50). We instead consider events within $20 \degree$ for G179.72--1.69 and within $15 \degree$ for G74.00--8.50 due to the uniquely large SNR radii and the extended 4FGL counterparts. The shell of SNR~S147 is nearly 5 degrees in diameter and is an identified $\gamma$-ray emitter, 4FGL~J0540.3+2756e, characterized best using the H-$\alpha$ SNR shell morphology that has radius $r \sim 2.5\,\degree$. SNR~G74.00--8.50 is nearly 8 degrees in diameter and has a large extended source, 4FGL~J2051.0+3049e \citep[outer annulus radius $\sim 1.6 \degree$,][]{katagiri2011,tutone2021}, overlapping in location that may be associated with the SNR shell. 

Two of the 58 ROIs are located within the Large Magellanic Cloud (LMC, see Figure~\ref{fig:lmc}): N~157B (G279.60--31.70, 4FGL~J0537.8--6909) and newly detected B0453--685 (G279.80--35.80). Their location requires accounting for a third diffuse emission component from the LMC. The source models include four extended source components to reconstruct the {\it{emissivity model}} developed in \citet{lmc2016} to represent the diffuse LMC emission. The four additional sources are 4FGL~J0500.9--6945e (LMC Far West), 4FGL~J0519.9--6845e (LMC Galaxy), 4FGL~J0530.0-6900e (30 Dor West), and 4FGL~J0531.8--6639e (LMC North).

\subsection{Detection Method}\label{sec:detection}

With the source models described above, we allow the isotropic and Galactic diffuse background components and sources with TS $\geq 25$ and a distance from the ROI center $\leq3.0$\,$\degree$ to vary in normalization and in spectral index (except for the isotropic background which has no spectral index parameter). Sources that have TS $< 25$ are removed from the source model. We compute a series of diagnostic TS and count maps in order to search for and understand any residual $\gamma$-ray emission. We generate the counts and TS maps in the following energy ranges: 300\,MeV--2\,TeV, 1--10\,GeV, 10--100\,GeV, and 100\,GeV--2\,TeV. The motivation for increasing energy cuts stems from the improving PSF of the \fermi instrument with increasing energies\footnote{See \url{https://www.slac.stanford.edu/exp/glast/groups/canda/lat_Performance.htm} for a review on the dependence of the PSF with energy for Pass 8 data.}. We inspect the TS maps for any significant residuals (TS $\geq 25$) in the ROI and point sources are added to the source model to account for these. 

In most cases, unassociated, unknown, or plausibly PWN/SNR-related (i.e., ``spp'' class) 4FGL sources are already found to be coincident and likely counterparts to sources in the selected sample (Table~\ref{tab:all_rois}). We consider any 4FGL sources as coincident if their centroid, extent, or uncertainty in position (95\% confidence level, C.L.) overlap with the position of the PWN. For the majority of ROIs, the total number of coincident 4FGL sources is one. Other crowded regions may have an overlapping unidentified extended 4FGL source (e.g., G8.40+0.15, G11.18--0.35, G25.10+0.02, G25.24--0.19, G336.40+0.10, and G337.20+0.10). In either case, the sources falling on the PWN position are re-characterized when finding the best source model. The ROIs where the latter becomes relevant are explicitly described in Section~\ref{sec:results_by_roi}. In the final source model of each ROI, there is only one source (whether point-like or extended) that is associated with the PWN.

\begingroup
\begin{table*}
\centering
\begin{tabular}{cccc}
\hline
\hline
Galactic PWN Name & 4FGL Name & TS & Refs.$^a$ \\
\hline
G29.70--0.30 (Kes~75) & J1846.4--0258 (DR4) & 20.7 & \citet{straal2022}\\  
G54.10+0.27 	& J1930.5+1853 (DR3)  & 31.1 & \citet{abeysekara2018} \\ 
\multirow{2}{*}{G279.60--31.70 (N~157B)} & \multirow{2}{*}{J0537.8--6909} & \multirow{2}{*}{168.3} & \multirow{2}{*}{\citet{lmc2016}} \\
\ & & & \citet{saito2017} \\
G279.80--35.80 (B0453--685) &  --  & 18.4 & \citet{eagle2023} \\ 
G315.78--0.23 (Frying Pan) & J1435.8--6018 & 37.4 & \\
G327.15--1.04 & J1554.4–5506 (DR3) & 34.1 & \citet{eagle2022} \\
\hline
G11.18--0.35 & J1811.5--1925 & 53.9 & \\
G16.73+0.08 & J1821.1--1422 & 142.5 & \\
G18.90--1.10 & J1829.4--1256 & 101.8 & \\
G20.20--0.20 & J1828.0--1133 & 153.9 & \\
G27.80+0.60 & J1840.0--0411 & 131 & \\
G39.22--0.32 & J1903.8+0531 & 129.7 & \\
G49.20--0.30 (in W~51C) & J1922.7+1428c & 157.2 & \\
G49.20--0.70 (in W~51C) & -- & 28.0 & \\
G63.70+1.10 & J1947.7+2744 & 93.9 & \\
G65.73+1.18 & J1952.8+2924 & 123.3 & \\
G74.94+1.11 & J2016.2+3712 & 66.3 & \\
G318.90+0.40 & J1459.0--5819 & 55.6 & \\
G337.20+0.10 & -- & 18.4 & \\
G337.50--0.10 & J1638.4--4715c (DR3) & 42.7 & \\
G338.20--0.00 & J1640.6--4632 (DR1), J1640.7--4631e (DR3) & 174.8 & \\
\hline
\hline
\end{tabular}
\caption{{\bf (Top) Likely point-like PWNe:} Results of the maximum likelihood fits for newly identified point-like LAT PWNe along with the Galactic PWN name, 4FGL counterpart, 
and the TS for a fixed position at the PWN location (2 DOF). The final column lists the prior reports of broadband investigations that support a PWN origin. {\bf (Bottom) Point-like PWN candidates:} Results of the maximum likelihood fits for point-like $\gamma$-ray source detections that coincide with known PWNe.}
\label{tab:points}
\end{table*}
\endgroup

\subsection{Localization and Extension}\label{sec:ext}

To model residual $\gamma$-ray emission that meets the \fermi significance threshold ($\text{TS}=25$ for 4 DOF), we first add a fixed point source characterized by a power-law spectrum and index $\Gamma = 2$ to the 300\,MeV--2\,TeV global source model, after removing any unidentified 4FGL sources coincident with the source position. We localize the point source executing \texttt{GTAnalysis.localize} to find the best-fit position. If no better position is found, it remains fixed to the PWN location. Table~\ref{tab:points} reports the results of detected point sources fixed to the PWN locations and Table~\ref{tab:points_r95} provides the 95\% positional uncertainty when the position is freed. Those without any positional uncertainty are fixed to the PWN location. 

For all detected PWNe, we run extension tests in FermiPy utilizing \texttt{GTAnalysis.extension} and the two spatial templates supported in the FermiPy framework, radial disk and radial Gaussian. Both of these extended templates assume a symmetric 2D shape with width parameters radius $r$ and sigma $\sigma$, respectively. We allow the position to remain free when finding the best-fit spatial extension of both templates. We measure the significance for extension to be $\text{TS}_{\text{ext}} = 2 \times \log\big(\frac{\mathcal{L}_{ext}}{\mathcal{L}_{ps}}\big)$ where $\mathcal{L}_{ext}$ is the hypothesis for an extended source and $\mathcal{L}_{ps}$ is the point source hypothesis, following \citet{lande2012,fges2017}. We consider the source as extended when TS$_{\text{ext}} > 16$. The results of extended sources are provided in Table~\ref{tab:extent}. 
Of the 15 extended sources, 13 were already extended in the 4FGL catalogs. The extensions found for these 13 sources are comparable to the sizes reported in the 4FGL, where the only differences are in the best-fit template used (see also Figure~\ref{fig:extension_plots} in Appendix~\ref{sec:pop}). For all extended sources in this paper except for MSH~15--56, we find a radial Gaussian can better fit or fit just as well the $\gamma$-ray emission as the radial disk template. MSH 15--56 is the only extended source that is best fit with a custom spatial template, a template already adopted by the 4FGL catalogs and based on the PWN shape in radio \citep{devin2018}. 

The new best fit for 4FGL~J1836.5--0651e can model well the extended emission coincident with the TeV PWN~HESS~J1837--069 \citep[$r = 0.36\,\degree$,][]{hessgps2018} such that the TS of the nearby extended source 4FGL~J1838.9--0704e, a second possible PWN associated with HESS~J1837--069, subsequently declines to TS\,$ = 0$. We therefore suggest that there is only one extended source associated with HESS~J1837--069, and it is most likely to be 4FGL~J1836.5--0651e, though two X-ray PWNe are coincident in location with this extended source, see Section~\ref{sec:new}. 
There are two new extended sources in Table~\ref{tab:extent} that are not considered extended in the 4FGL catalogs: 
4FGL~J1818.6–-1533 (PWN candidate G15.40+0.10) and 4FGL~J1844.4--0306 (PWN candidate G29.40+0.10). Furthermore, we find that the detection for PWN G338.20--0.00 is point-like whereas the 4FGL--DR3 characterizes the emission with extension $r \sim 0.1\,\degree$, based on the extension of its probable TeV counterpart HESS~J1640--465 \citep{mares2021}. Our results for G338.20--0.00 are in line with what is reported in the 4FGL--DR3, finding a 2D Gaussian yields the best fit with $r = 0.08\,\degree$ and TS$_{\text{ext}} = $ 7.6. 

The smallest extended source is MSH 15--56 with a radius $r \sim 0.08\, \degree$ and the largest extended source is 4FGL~J1824.5--1351e (HESS~J1825--137) with  $r = 0.83\,\degree$. The average extension assuming a radial Gaussian for the sample of 15 extended sources is $r = 0.35\,\degree$. Of the 36 total detections, 15 are found to be extended, which make up $\sim$40\% of the entire sample analyzed here, see Appendix~\ref{sec:pop} for more discussion on extension. We explore the systematics on the extension in Section~\ref{sec:sys}.  

\begingroup
\begin{table}
\centering
\begin{tabular}{cccc}
\hline
\hline
Galactic PWN Name & R.A. & Dec. & R95 \\
\hline
G29.70--0.30 (Kes~75) & 281.60 & --2.96 & 0.09\\  
G54.10+0.27  & 292.65 & +18.90 & 0.10\\ 
G279.60--31.70 (N~157B) & 84.40 & --69.18  & 0.04 \\
G279.80--35.80 (B0453--685)  & 73.46 & --68.47 & 0.15 \\
G315.78--0.23 (Frying Pan) &	219.36 & --60.11 & 0.14\\
G327.15--1.04 & 238.59 & --55.11 & 0.08\\
\hline
G11.18--0.35 & 272.88 & --19.44 & 0.05\\
G16.73+0.08 & 275.29 & --14.35 & 0.05\\
G18.90--1.10 & 277.34 & --12.88 & 0.06\\
G20.20--0.20 & 277.02 & --11.57 & 0.06\\
G27.80+0.60 & 279.97 & --4.27 & 0.04 \\
G39.22--0.32 & 285.97 & +5.50 & 0.05\\
G49.20--0.30 (in W~51C) & 290.70  & +14.27 & -- \\
G49.20--0.70 (in W~51C) & 290.74 & +14.09 & 0.03\\
G63.70+1.10 & 296.96 & +27.73 & 0.07\\
G65.73+1.18 & 298.13 & +29.46 & 0.05\\
G74.94+1.11 & 304.04 & +37.20 & 0.04\\
G318.90+0.40 & 224.69 &	--58.38 & 0.10\\
G337.20+0.10 & 248.88 & --47.19 & 0.06\\
G337.50--0.10 & 249.67 & --47.28 & 0.09\\
G338.20--0.00 & 250.19 & --46.57 & 0.19\\
\hline
\hline
\end{tabular}
\caption{Same as Table~\ref{tab:points}, listing R.A. and Dec. in J2000 equatorial degrees and the 95\% uncertainty radius in degrees of the best-fit position. G49.20--0.30 remains fixed to the PWN location (see text for details).}
\label{tab:points_r95}
\end{table}
\endgroup

\begingroup
\begin{table*}
\hspace{-2.7cm}
\scalebox{0.8}{
\begin{tabular}{ccccccccc}
\hline
\hline
Galactic PWN Name & 4FGL Name & R.A. & Dec. & TS & TS$_{\text{ext}}$ & $r$ ($\degree$) & 95\% U.L. $r^a$ ($\degree$) & Refs. \\
\hline
G18.00--0.69 (HESS~J1825--137) & J1824.5--1351e & 275.93 & 	--13.88 & 151.5 & 88.29 & $0.83 \pm 0.07 \pm 0.10$ & 1.01 & \citet{grondin2011} \\
G26.60--0.10 & J1840.9--0532e & 280.18 & --5.52 & 288.6 & 74.32 & $0.36 \pm 0.03 \pm 0.04$ & 0.41 & \citet{ahag262008} \\
G36.01+0.10 & J1857.7+0246e & 284.22 & +2.65 & 227.8 & 93.71 & $0.45 \pm 0.05 \pm 0.16$ & 0.55 & \citet{ahag262008} \\
G304.10--0.24 (HESS~J1303--631) & J1303.0--6312e & 195.81 & --63.16 & 142.6 & 79.45 & $0.35 \pm 0.03 \pm 0.12$ & 0.40  & \citet{hess2012} \\
G326.12--1.81 (MSH 15--56) & J1552.4--5612e & 238.11 & --56.21 & 53.3 & -- & -- & --   & \citet{devin2018} \\
\hline
G8.40+0.15 & J1804.7--2144e & 271.11 & --21.74 & 104.8 & 59.65 & $0.29 \pm 0.02 \pm 0.11$ & 0.34 & \citet{liu2019} \\
G25.10+0.02 (HESS~J1837--069) & \multirow{2}{*}{J1836.5--0651e} & \multirow{2}{*}{279.24} & \multirow{2}{*}{--6.91} & \multirow{2}{*}{1174} & \multirow{2}{*}{616.2} & \multirow{2}{*}{$0.53 \pm 0.02 \pm 0.06$} & \multirow{2}{*}{0.56} &   \\
G25.24--0.19  (HESS~J1837--069) & & & & & & & &  \citet{lande2012} \\
G332.50--0.28 & \multirow{2}{*}{J1616.2--5054e}$\dag$ & \multirow{2}{*}{244.20} & \multirow{2}{*}{--50.99} & \multirow{2}{*}{485.1} & \multirow{2}{*}{187.3} & \multirow{2}{*}{$0.31 \pm 0.02 \pm 0.03$} & \multirow{2}{*}{0.35} &   \\
G332.50--0.30 (RCW~103) & & & & & & & & \citet{fges2017}  \\
G336.40+0.10 & J1631.6--4756e &  248.14 &  --47.91 & 94.5 & 24.56 & $0.19 \pm 0.03 \pm 0.83$ & 0.23  & \citet{fges2017}  \\
\hline
G11.03--0.05 & \multirow{2}{*}{J1810.3--1925e} & \multirow{2}{*}{272.39} & \multirow{2}{*}{--19.42} & \multirow{2}{*}{84.0} & \multirow{2}{*}{25.88}& \multirow{2}{*}{$0.41 \pm 0.05 \pm 0.05$} &  \multirow{2}{*}{0.49} & \\
G11.09+0.08 &  &  &  &  &  &  &   &  \\
G12.82--0.02 & J1813.1–1737e & 273.47 & --17.65 & 854.6 & 195.6 & $0.41 \pm 0.02 \pm 0.03$ & 0.45  &  \\
G15.40+0.10 & J1818.6--1533 & 274.60 & --15.56 & 394.5 & 20.60 & $0.19 \pm 0.03 \pm 0.10$ & 0.24 &   \\
G24.70+0.60 & J1834.1--0706e & 278.53 & --7.12 & 290.8 & 72.57 & $0.19 \pm 0.02 \pm 0.02$ & 0.22 &  \citet{fges2017} \\
G29.40+0.10 & J1844.4--0306 & 281.15 & --3.12 & 195.7 & 18.77 & $0.27 \pm 0.04 \pm 0.17$ & 0.34  &  \\
G328.40+0.20 (MSH~15--57) & J1553.8--5325e & 238.61 & --53.38 & 1070 & 436.5 & $0.43 \pm 0.02 \pm 0.03$ & 0.46  &  \\ 
\hline
\hline
\end{tabular}}
\caption{{\bf (Top) Extended PWNe:} Results of the maximum likelihood fits for previously identified extended LAT PWNe along with the Galactic PWN name, 4FGL counterpart, R.A. and Dec. in J2000 equatorial degrees, the TS of the best fit and the TS$_{\text{ext}}$ using the radial Gaussian spatial template (see Section~\ref{sec:ext} for details). The seventh and eighth columns quote the best-fit extension using the radial Gaussian template and the 95\% upper limit. The first quoted error on the extension $r$ corresponds to the symmetric $1 \sigma$ statistical error and the latter corresponds to the systematic error. The final column lists the prior reports of broadband investigations that support a PWN origin. G326.12--1.81 uses the custom spatial template from the 4FGL \citep[see also][]{devin2018}. {\bf (Middle) Likely Extended PWNe:} Results of the maximum likelihood fits for identifications of likely extended PWNe. {\bf (Bottom) Extended PWN candidates:} Results of the maximum likelihood fits for source detections that coincide with extended \fermi sources. \footnotesize{$^a$ In all cases, $r = r_{68}$ such that $\sigma = \frac{r}{1.51}$ \citep{lande2012}.} $\dag$ RCW~103 is a known {\it Fermi} PWN, but may have contamination from a nearby PWN, G332.50--0.28, overlapping in location, see Section~\ref{sec:new}.}
\label{tab:extent}
\end{table*}
\endgroup

\subsection{Spectral Shapes, Fluxes, and Upper Limits}\label{sec:spectral_tests}
We assume the source spectrum can be characterized as a power-law for the full energy range\footnote{For a review of Fermi source spectral models see \url{https://fermi.gsfc.nasa.gov/ssc/data/analysis/scitools/source_models.html}.}:
\begin{equation}
    \frac{dN}{dE} = N_{0} \left(\frac{E}{E_0}\right)^{-\Gamma}
\end{equation}\label{eq:1} 
allowing the index and normalization to vary. The scale $E_0$ is fixed to 1\,GeV for simplicity for all sources except N~157B and G304.10--0.24 where we use their catalog values $E_0$ of 5.3 and 11\,GeV, respectively.  We additionally perform curvature tests for all sources testing a Log Parabola spectrum, 
\begin{equation}
    \frac{dN}{dE} = N_{0} \left(\frac{E}{E_b}\right)^{-(\alpha+\beta\log{E/E_b})}
\end{equation}\label{eq:2}
fixing $E_b = 1\,$GeV except for G318.90+0.40 and G326.12--1.81 which use their catalog values $E_b$ of 2.2 and 5.5\,GeV, respectively. 
In 4FGL-DR3 and later, TS$_\text{LogP} = 2 \log\big({\frac{{\mathcal{L}_{LogP}}}{{\mathcal{L}_{PL}}}}\big) > 4$ ($> 2\sigma$) are considered preferentially curved spectra. To follow this method, we model the sources with TS$_\text{LogP} > 4$ using the best-fit Log Parabola spectrum that we measure. The remainder of sources are characterized by a power-law spectrum. The spectral and spatial properties are summarized in Table~\ref{tab:tev_table} alongside the properties of TeV counterparts for all detected sources. We provide the curvature metric in the second to last column. The MeV--GeV spectral values are in good agreement with those in the 4FGL catalogs.


For detected sources, we measure the energy flux from 300\,MeV--1\,TeV in seven energy bins using \texttt{GTAnalysis.sed}
, where only the background components and the source of interest are left free to vary. Other sources are fixed to their values from the best-fit source model. We measure the systematic errors on the flux for all 36 source detections and describe the method and results in Section~\ref{sec:sys}. The systematic errors for the spectral fluxes $E^2\frac{dN}{dE}$ per energy bin are listed in the Appendix, Table~\ref{tab:spectral_fluxes}. 

For non-detections, we place a point source characterized by a power-law spectrum at the location of the PWN, a spatial choice that corresponds to the size of the systems from multiwavelength observations (see Table~\ref{tab:all_rois}) and perform a global fit with only the spectral parameters of the backgrounds and the point source free to vary.  If the point source has TS\,$< 25$ for 4 DOF, we measure the 95\% C.L. upper limit spectral fluxes for the 300\,MeV--2\,TeV energy range and list them in the final column of Table~\ref{tab:nondetect}, which lists the sources not detected in this analysis. We also calculate the 95\% C.L. upper limit spectral flux values for 300\,MeV--2\,TeV across nine energy bins. 
All upper limit flux values for the 19 undetected sources are listed in the Appendix, Table~\ref{tab:spectral_ul_fluxes}.

\subsection{Sources of Systematic Errors}\label{sec:sys}

We account for systematic uncertainties introduced by the choice of the interstellar emission model (IEM) and the IRFs, which mainly affect the spectrum of the measured $\gamma$-ray emission. We have followed the prescription developed by \cite{depalma2013} and \cite{acero2016}, who generated eight alternative IEMs using a different approach than the standard IEM. For this analysis, we employ the eight alternative IEMs (aIEMs) that were generated for use on Pass 8 data in the \fermi Galactic Extended Source Catalog \citep[FGES,][]{fges2017}. 
We re-generate the best-fit model and perform independent fits for each ROI using the 8 aIEMs for a total of 9 fits per ROI including the standard model fit. We then compare the flux values to the standard model following equation (5) in \citet{acero2016}.

We estimate the systematic uncertainties introduced by the uncertainty in the effective area\footnote{\url{https://fermi.gsfc.nasa.gov/ssc/data/analysis/LAT_caveats.html}} while enabling energy dispersion as follows: $\pm 3\%$ for $E < 100$\,GeV, $\pm 4.5 \%$ for $E = 175\,$GeV, and $\pm 8\%$ for $E = 556\,$GeV. Since the IEM and IRF systematic errors are taken to be independent, we can evaluate both and perform the quadratic sum for the total systematic error. This method has been applied in two prior catalogs and both reported systematic errors on the same order as the $1 \sigma$ statistical errors in flux \citep[see Fig.~29 in][]{acero2016}, which is consistent with our results (see Table~\ref{tab:spectral_fluxes} in the Appendix). In general, it is found that the systematics introduced by the uncertainty in the IEMs and IRFs are most important for sources that lie along the Galactic plane ($|b| < 1.0\,\degree$) and for energies $E < 5\,$GeV, largely dominated by the uncertainty in the IEMs.

For the two sources located in the LMC (Figure~\ref{fig:lmc}), N~157B (4FGL~J0537.8--6909) and newly detected B0453--685, we find that the systematic errors are negligible for all energy bins, which is not surprising given the location of the LMC with respect to the bright, diffuse $\gamma$-ray emission along the Galactic plane. However, we must also account for the systematic error that is introduced by having an additional diffuse background component for the two LMC ROIs. 
We can probe these effects by employing the method described in \citet{lmc2016}. This requires replacing the four extended sources that represent the diffuse LMC in this analysis \citep[the {\it emissivity model},][]{lmc2016} with four different extended sources to represent an alternative template for the diffuse LMC \citep[the {\it analytic model},][]{lmc2016}. N~157B and B0453--685 are both refit with the alternative diffuse LMC template using the same approach as for the aIEMs described above. 
The uncertainty in the choice of the diffuse LMC template is similar in impact to the standard IEM, where the systematic errors are on the same order as the $1 \sigma$ statistical errors in flux. Similarly, we find the errors are largest in the energy bins below 5\,GeV, but are negligible for higher-energy bins.

Finally, we consider the use of weights particularly for the lowest-energies $E < 1$\,GeV where the Galactic ridge is most prominent \citep{4fgldr1}. The effect from Galactic ridge emission decreases with increasing energy, becoming small above $1$\,GeV. The choice to use PSF3 events minimizes this effect, but it can still be noticeable in the softest sources as a systematic effect on the spectral index if the lowest-energy bin in the spectral fit has the largest influence. For PSF3 events, the weights increase the lowest-energy bin statistical errors at 316\,MeV by a factor of $\sim$ 3. For all events, the effect is larger ($\sim 8$). To further explore this, we look at the spectral index 
of all sources, comparing their values to those of the 4FGL counterparts for both 300\,MeV--2\,TeV and $E > 1$\,GeV energy bands. No discrepancy is found in the spectral index. 
However, we still account for the potential bias effect without event weighting. 
This is accounted for in the systematic uncertainty of the flux in Table~\ref{tab:spectral_fluxes} in the lowest-energy bin for all sources. 
The LMC ROIs are not located in the Galactic ridge and thus do not suffer from this bias, having weights $\sim 0.5$ below 1\,GeV and $\sim 0.9$ above 1\,GeV. Nevertheless, the systematic uncertainty includes the conservative estimate ($\sim 0.12$).

To explore the uncertainties on the extension influenced by the IEM and effective area, we follow the same method outlined for the spectral flux above: we re-generate the best-fit point-source model for all extended sources and perform extension tests. Each PWN is tested for extension considering the 8 aIEMs for a total of 9 extension tests per ROI including the standard model fit. We then compare the best-fit extension and symmetric error values to the standard model following equation (5) in \citet{acero2016}. This method has been similarly applied in prior catalogs \citep{acero2016,fges2017} and reported similar trends on the systematic error on extension as was found on the flux. The systematic errors for extension are generally also on the same order as the $1 \sigma$ statistical errors \citep[e.g., Section~2.4.2 in][]{acero2016}, which is generally consistent with our results that are displayed in Table~\ref{tab:extent}. For crowded regions, a larger error on both the flux and the extension is expected \citep[see Tables~\ref{tab:extent}, \ref{tab:spectral_fluxes}, and][]{fges2017}. 

We note that the systematic uncertainties on extension for 4FGL~J1631.6--4756e and 4FGL~J1824.5--1351e are incompatible to prior systematic studies. For 4FGL~J1631.6--4756e, we find this source is best fit as a radial Gaussian with $r = 0.19 \pm 0.027 \pm 0.83\,\degree$, where the first quoted error represents the $1 \sigma$ statistical error and the latter is the systematic error. For comparison, the FGES catalog \citep{fges2017} 
found an extension using a radial disk template $r = 0.26 \pm 0.02 \pm 0.08\,\degree$ for 4FGL~J1631.6--4756e. However, this source lies in a crowded region among several other extended sources, including a larger, unknown extended 4FGL source (J1633.0--4746e) coincident in location, and therefore the large systematic errors are not unexpected. 4FGL~J1824.5--1351e, which we find is better fit as a radial Gaussian with $r = 0.83 \pm 0.074 \pm 0.10\,\degree$, has a smaller systematic error than what is reported in the FGES catalog by 40\%, but this is likely due to the larger radius found for the extension of 4FGL~J1824.5--1351e, a 
radial disk with $r = 1.05\,\degree$ in the FGES catalog. For the remainder of the extended sources, we find compatible systematic uncertainties on the extension as those reported in the FGES catalog, see Table~\ref{tab:extent}.

\section{Comprehensive Results}\label{sec:fermi_results}

Six of the 12 \fermi sources classified as PWNe, each detected as extended, are included in this sample (Table~\ref{tab:all_rois}):
\begin{itemize}
    \item G18.00--0.69 (4FGL~J1824.5--1351e)
    \item G26.60--0.10 (4FGL~J1840.9--0532e)
    \item G36.01+0.10 (4FGL~J1857.7+0246e)
    \item G304.10--0.24 (4FGL~J1303.0--6312e)
    \item G326.12--1.81 (4FGL~J1552.4--5612e)
    \item HESS~J1616--508 (4FGL~J1616.2--5054e)
\end{itemize}
and eight \fermi PWN associations are also analyzed here:
\begin{itemize}
    \item N~157B (4FGL~J0537--6909)
    \item G29.70--0.3 (Kes~75, 4FGL~J1846.4--0258, DR4)
    \item G54.10+0.27 (4FGL~J1930.5+1853, DR3)
    \item G63.70+1.10 (4FGL~J1947.7+2744)
    \item G327.15--1.04 (4FGL~J1554.4–5506, DR3) 
    \item G336.40+0.10 (4FGL~J1631.6--4756e)
    \item HESS~J1837--069 (4FGL~J1836.5–-0651e and 4FGL~J1838.9-0704e)
\end{itemize}
The ninth known PWN association, G0.13--0.11 (4FGL~J1746.4–2852), is not considered here due to its proximity to the Galactic center. The remainder of known \fermi PWNe: 4FGL~J0205.6+6449e \citep[3C~58 powered by PSR~J0205+6449,][]{li2018}, 4FGL~J0534.5+2201e (Crab), 4FGL~J0833.1--4511e (Vela), 4FGL~J1355.2--6420e, 4FGL~J1420.3--6046e, and 4FGL~J1514.2--5909e are all powered by \fermi detected pulsars and are not included in this search. 


\subsection{Source Classification}

We consider three different properties to analyze the likelihood of our associations, with the notable caveat that the criteria do not provide an unambiguous PWN classification (see below):
\begin{itemize}
    \item Positional overlap with a PWN identified in radio, X-ray or TeV surveys,
    \item Source extent as observed by the \fermi and the observed extent in radio, X-ray, or TeV surveys,
    \item The energetics of the central pulsar, PWN, and host SNR.
\end{itemize}


The first criterion is positional coincidence, though alone it does not achieve a confident classification. For many Galactic sources detected by the {\it Fermi}--LAT, positional coincidence may include multiple sources. We must also consider the evolution and hence age of PWNe, since the broadband spectrum and morphology can change significantly depending on the evolutionary phase of the PWN. The physical consequences can become dramatic for PWNe believed to be in the reverberation phase, where the SNR reverse shock is crushing the PWN. The influences and conditions of this phase are not well-known, though they are a current focus of ongoing studies \citep[e.g.,][]{reynolds1984,torres2017,Bandiera2023a,Bandiera2023}. The findings of current studies suggest that radiative losses dominate once the PWN is crushed by the reverse shock, which can result in the generation of a compact nebula of high-energy electrons separate from a diffuse nebula of low-energy electrons. The components can have different spectral and spatial characteristics, a reflection of the different particle energy losses. Therefore, the observed morphology is not necessarily a rigorous indicator for the dominant $\gamma$-ray origin. In general, evolved PWNe should have comparable extent in radio and GeV bands, often being larger than the X-ray and TeV extents. 
%
%

Confidently classifying \fermi sources as PWNe is one of the biggest challenges since it requires a combination of available multiwavelength observations and constraints on the energetics of the system. It is often the case that the central pulsar and(or) the SNR shell are not identified, which prevents reliable source classification. 
Given the challenge around understanding PWNe and their environments together with the energy-dependent angular resolution of {\it Fermi}-LAT, no robust classification method currently exists that is free of reasonable doubt, especially when considering the possibility that the central pulsar and(or) host SNR may contribute to the observed $\gamma$-rays or may even dominate the $\gamma$-ray signal. In fact, pulsars make up the largest Galactic source population in the 4FGL catalogs \citep{3pc,4fgl-dr4} and SNRs outnumber the PWN population by a factor two, with 24 detected by the \fermi and another 19 listed as SNR candidates in the 4FGL \citep{4fgl-dr4}. Because source classification for high-energy PWNe is so difficult, we can only consider a small number of $\gamma$-ray sources here as likely PWNe, while the remainder require in-depth multiwavelength analyses to determine a likely origin. These are considered weaker PWN candidates. 

We discuss the implications of pulsar contributions for the detected sources, particularly for energies $E < 10\,$GeV where the spectral energy distributions (SEDs) for several of the sources indicate the possible presence of a lower spectral component, such as a pulsar, in the following section.

\subsection{Pulsar Contributions}\label{sec:psrs}

The presence of pulsar contributions particularly at lower energies ($E < 10\,$GeV) is indicated by the 300\,MeV--2\,TeV SED for a number of likely PWNe and candidate PWNe, see Table~\ref{tab:tev_table}, the final column. The possible pulsar contributions are listed as either a Y (yes) or N (no). 
The pulsar contribution possibility is assessed by the presence of any one of three features: 1) if the best-fit flux data points suggest more than one spectral component when compared to the best-fit spectral model, 2) if the source emission is primarily detected at low energies, similar to pulsars, and 3) if an energetic pulsar is known to the system. See Appendix~\ref{sec:seds} for a sample of source spectra regarding features 1) and 2). 

Of course, the key features we consider are not fully representative of the possibility for pulsar contributions. We are limited in evaluating many of the $\gamma$-ray sources (statistics will not yield reliable pulsation search results) and their associations (not all central pulsars have been identified), and the problem is also compounded by the complexity of the nature of $\gamma$-ray pulsars \citep[unpulsed emission may be present but would not be detected in a pulsation search e.g.,][]{dormody2011}. As a final note, the presence of two spectral components does not confirm the presence of a pulsar contribution. Many PWNe, such as Vela--X \citep[e.g.,][]{tibaldo2018}, have multiple emission components in the \fermi band. 

\tablewidth{0pt}
\tabletypesize{\footnotesize}
\setlength{\tabcolsep}{1.25pt}
\renewcommand*{\arraystretch}{1.0}
\setlength{\LTcapwidth}{\linewidth}
\begingroup
\begin{longrotatetable}
\begin{deluxetable*}{c|c|c|c|c|c|c|c|c|c|c|c|c}
\tablecaption{\label{tab:tev_table} The spectral and spatial properties measured from this work compared to the TeV counterparts (if present) for both likely (top) and weak candidates (bottom). 
}
\tablecolumns{13}
\tablehead{
\colhead{PWN Name} &
\colhead{GeV Template} &
\colhead{GeV Spectrum} &
\colhead{GeV Flux} &
\colhead{GeV Index or $\alpha$} &
\colhead{$\beta$} &
\colhead{TeV ID} &
\colhead{TeV Class} &
\colhead{TeV Template} &
\colhead{TeV Extension} &
\colhead{TeV Index} &
\colhead{TS$_\text{LogP}$} &
\colhead{PSR?} 
}
\movetabledown=1.0cm
\startdata
                {\footnotesize Likely} & & & & & & & & & & & \\
                & & & eV cm$^{-2}$ s$^{-1}$ & (at 1\,GeV) & & & & & & & & \\
                \hline
			G8.40+0.15 & Gaussian & PowerLaw & 79.9$\pm$ 10.3 & 1.96$\pm$ 0.04 & -- & HESS~J1804--216 & UNID & 2-Gaussian & 0.24$\pm$ 0.03 & 2.69$\pm$ 0.04 & 2.4 & N \\
			G18.00-0.69 & Gaussian & LogParabola & 139.0$\pm$ 23.2 & 1.58$\pm$ 0.06 & 0.04$\pm$ 0.01 & HESS~J1825--137 & PWN/TeV halo & 3-Gaussian & 0.46$\pm$ 0.03 & 2.15$\pm$ 0.06 & 32.0 & N \\
			\multirow{2}{*}{G25.24$-$0.19} & \multirow{2}{*}{Gaussian} & \multirow{2}{*}{PowerLaw} & \multirow{2}{*}{285.0$\pm$ 16.6} & \multirow{2}{*}{1.97$\pm$ 0.02} & \multirow{2}{*}{--} & \multirow{2}{*}{HESS~J1837--069} & \multirow{2}{*}{PWN} & \multirow{2}{*}{3-Gaussian} & \multirow{2}{*}{0.36$\pm$ 0.03} & \multirow{2}{*}{2.54$\pm$ 0.04} &  \multirow{2}{*}{--0.5} & \multirow{2}{*}{N} \\
            		G25.10+0.02 & & & & & & & & & & & & \\
			G26.60-0.10 & Gaussian & PowerLaw & 72.3$\pm$ 6.98 & 2.15$\pm$ 0.01 & -- & HESS~J1841--055 & PWN & 2-Gaussian & 0.41$\pm$ 0.03 & 2.21$\pm$ 0.07 & 0.32 & N \\
			G29.70--0.30 & PointSource & PowerLaw & 7.29$\pm$ 1.65 & 2.41$\pm$ 0.15 & -- & HESS~J1846--029 & PWN & PointSource & 0.01 & 2.41$\pm$ 0.09 & --0.03 & Y \\
			G36.01+0.10 & Gaussian & PowerLaw & 65.1$\pm$ 6.94 & 2.15$\pm$ 0.05 & -- & HESS~J1857+026 & UNID & 2-Gaussian & 0.26$\pm$ 0.06 & 2.57$\pm$ 0.06 & 0.23 & N \\
			G54.10+0.27 & PointSource & PowerLaw & 3.82$\pm$ 0.92 & 2.09$\pm$ 0.12 & -- & HESS~J1930+188 & PWN & PointSource & 0.03 & 2.59$\pm$ 0.26 & 0.70 & N \\
			N157B & PointSource & PowerLaw & 10.1$\pm$ 1.61 & 2.11$\pm$ 0.07 & -- & LHA~120--N157B & PWN & PointSource & 0.01 & 2.80$\pm$ 0.10 & 0.5 & Y \\
			B0453--685 & PointSource & PowerLaw & 0.75$\pm$ 0.22 & 2.27$\pm$ 0.18 & -- & -- & -- & -- & -- & -- & 3.7 & Y \\
			G304.10--0.24 & Gaussian & PowerLaw & 32.8$\pm$ 6.03 & 1.88$\pm$ 0.06 & -- & HESS~J1303--631 & PWN & 2-Gaussian & 0.18$\pm$ 0.01 & 2.04$\pm$ 0.06 & --0.2 & N \\
			G315.78--0.23 & PointSource & PowerLaw & 5.66$\pm$ 0.99 & 2.76$\pm$ 0.16 & -- & -- & -- & -- & -- & -- &  2.1 & Y \\
			G326.12-1.81 & SpatialTemplate & LogParabola &  9.70$\pm$ 2.03 & 1.23$\pm$ 0.14 & 0.16$\pm$ 0.04 & -- & -- & -- & -- & -- & 6.7 & N \\
			G327.15--1.04 & PointSource & LogParabola & 2.48$\pm$ 0.49 & 1.73$\pm$ 0.28 & 0.41$\pm$ 0.15 & HESS~J1554--550 & PWN & PointSource & 0.02 & 2.19$\pm$ 0.17 & 5.0 & Y \\
			\multirow{2}{*}{RCW 103} & \multirow{2}{*}{Gaussian} & \multirow{2}{*}{PowerLaw} & \multirow{2}{*}{117.5$\pm$ 9.61} & \multirow{2}{*}{1.98$\pm$ 0.03} & \multirow{2}{*}{--} & \multirow{2}{*}{HESS~J1616--508} & \multirow{2}{*}{PWN} & \multirow{2}{*}{2-Gaussian} & \multirow{2}{*}{0.23$\pm$ 0.03} & \multirow{2}{*}{2.32$\pm$ 0.06} &  \multirow{2}{*}{--0.4} & \multirow{2}{*}{N} \\
            		G332.50--0.28 & & & & & & & & & & & & \\
			G336.40+0.10 & Gaussian & PowerLaw & 44.7$\pm$ 6.04 & 2.05$\pm$ 0.04 & -- & HESS~J1632--478 & PWN & Gaussian & 0.18$\pm$ 0.02 & 2.52$\pm$ 0.06 & 1.2 & Y \\
			\hline
                {\footnotesize Weaker} & & & & & & & & & & & & \\
                \hline
			\multirow{2}{*}{G11.03--0.05} & \multirow{2}{*}{Gaussian} & \multirow{2}{*}{LogParabola} & \multirow{2}{*}{20.9$\pm$ 3.24} & \multirow{2}{*}{1.88$\pm$ 0.23} & \multirow{2}{*}{0.40$\pm$ 0.19} & \multirow{2}{*}{HESS~J1809--193} & \multirow{2}{*}{UNID} & \multirow{2}{*}{3-Gaussian} & \multirow{2}{*}{0.40$\pm$ 0.05} & \multirow{2}{*}{2.38$\pm$ 0.07} & \multirow{2}{*}{11.9} & \multirow{2}{*}{N} \\
			G11.09+0.08 & & & & & & & & & & & & \\
			G11.18--0.35 & PointSource & PowerLaw & 13.3$\pm$ 2.90 & 2.03$\pm$ 0.01 & -- & -- & -- & -- & -- & -- & 2.9 & Y \\
			G12.82--0.02 & Gaussian & LogParabola & 74.1$\pm$ 3.30 &2.24$\pm$ 0.04 & 0.10$\pm$ 0.03 & HESS~J1813--178 & PWN & 2-Gaussian & 0.05$\pm 0.004$ & 2.07$\pm$ 0.05 & 10.9 & Y \\
			G15.40+0.10 & Gaussian & LogParabola & 29.7$\pm$ 1.69 & 2.65$\pm$ 0.07 & 0.18$\pm$ 0.08 & 2HWC~J1819--150* & UNID & PointSource & 0.09 & 2.21$\pm$ 0.15 & 8.8 & Y \\
			G16.73+0.08 & PointSource & PowerLaw & 16.5$\pm$ 1.55 & 2.72$\pm$ 0.09 & -- & -- & -- & -- & -- & -- & 2.5 & Y \\
			G18.90--1.10 & PointSource & LogParabola & 8.26$\pm$ 0.97 & 1.46$\pm$ 0.24 & 0.66$\pm$ 0.16 & -- & -- & -- & -- & -- & 25.7 & Y \\
			G20.20--0.20 & PointSource & LogParabola & 15.4$\pm$ 1.35 & 2.27$\pm$ 0.13 & 0.50$\pm$ 0.16 & -- & -- & -- & -- & -- & 21.2 & N \\
   			G24.70+0.60 & Gaussian & PowerLaw & 56.5$\pm$ 6.37 & 2.02$\pm$ 0.04 & -- & -- & -- & -- & -- & -- & 2.9 & N \\
			G27.80+0.60 & PointSource & LogParabola & 9.65$\pm$ 1.13 & 1.62$\pm$ 0.16 & 0.34$\pm$ 0.07 & -- & -- & -- & -- & -- & 12.3 & Y \\
			G29.40+0.10 & Gaussian & PowerLaw & 35.9$\pm$ 2.71 & 2.55$\pm$ 0.06 & -- & HESS~J1844--030 & UNID & PointSource & 0.02 & 2.48$\pm$ 0.12 & 3.8  & Y \\
			G39.22--0.32 & PointSource & PowerLaw & 15.5$\pm$ 1.63 & 2.53$\pm$ 0.09 & -- & -- & -- & -- & -- & -- & 1.5 & N \\
			G49.20--0.30 & PointSource & LogParabola &  37.2$\pm$ 1.86 & 2.35$\pm$ 0.06 & 0.26$\pm$ 0.07 & -- & -- & -- & -- & -- & 8.8 & N \\
			G49.20--0.70 & PointSource & PowerLaw & 18.5$\pm$ 2.41 & 2.38$\pm$ 0.09 & -- & -- & -- & -- & -- & -- & 2.4 & N \\
			G63.70+1.10 & PointSource & LogParabola & 4.38$\pm$ 0.60 & 1.84$\pm$ 0.22 & 0.37$\pm$ 0.14 & -- & -- & -- & -- & -- & 11.0 & Y \\
   			G65.73+1.18 & PointSource & LogParabola & 5.63$\pm$ 0.63 & 2.13$\pm$ 0.19 & 0.58$\pm$ 0.20 & 2HWC~J1953+294 & PWN & PointSource & 0.24 & 2.78$\pm$ 0.15 & 9.6 & Y \\
			G74.94+1.11 & PointSource & PowerLaw & 13.0$\pm$ 1.75 & 2.24$\pm$ 0.09 & -- & VER~J2016+371 & UNID & PointSource & 0.08 & 2.30$\pm$ 0.30 & 0.3 & Y \\
			G318.90+0.40 & PointSource & LogParabola & 2.47$\pm$ 0.41 & 0.24$\pm$ 0.83 & 1.16$\pm$ 0.44 & -- & -- & -- & -- & -- & 21.9 & N \\
			G328.40+0.20 & Gaussian & LogParabola & 103.0$\pm$ 6.79 & 2.00$\pm$ 0.06 & 0.05$\pm$ 0.01 & -- & -- & -- & -- & -- & 15.5 & N \\
			G338.20--0.00 & PointSource & PowerLaw & 37.6$\pm$ 6.73 & 1.82$\pm$ 0.06 & -- & HESS~J1640--465 & Composite SNR & 2-Gaussian & 0.11 $\pm$ 0.03 & 2.57 $\pm$ 0.04 & --0.02 & Y \\
			G337.20+0.10 & PointSource & PowerLaw & 9.03$\pm$ 2.04 & 2.26$\pm$ 0.12 & -- & HESS~J1634--472 & UNID & Gaussian & 0.11$\pm$ 0.03 & 2.31$\pm$ 0.05 & 0.2 & N \\
			G337.50--0.10 & PointSource & PowerLaw & 15.4$\pm$ 1.67 & 2.69$\pm$ 0.10 & -- & -- & -- & -- & -- & -- & --1.23  & N \\
\enddata
\tablecomments{The quoted errors are the 68\% C.L. values. TeV extension is in $\degree$. For point-like TeV sources we quote the 68\% containment radius. 
All TeV values are taken from the TeVCat catalog \citep{tevcat2008}. Those listed as X-Gaussian refer to multiple Gaussian components \citep[see][for details]{hessgps2018}. The next to last column provides the curvature metric, see Section~\ref{sec:spectral_tests} for details. The final column provides the sources that have a possible additional low-energy spectral component such as the central pulsar (yes or no, Y/N). See Section~\ref{sec:psrs} for details.}
\end{deluxetable*}
\end{longrotatetable}
\endgroup

\begingroup
\begin{table*}
\centering
\begin{tabular}{ccccccc}
\hline
\hline
Galactic PWN Name & 4FGL Name & TeV Name & R.A. &  Dec. & TS & Flux Upper Limit (MeV cm$^{-2}$ s$^{-1}$) \\
\hline
G0.87+0.08 & -- & HESS~J1747--281 & 266.84 & --28.15  & 0.05 & $1.62 \times 10^{-7}$ \\
G23.50+0.10 & -- & -- & 278.42 & --8.46 & 10.4 & $9.30 \times 10^{-7}$ \\
G25.10+0.02 & J1838.0--0704e & HESS~J1837--069 & 279.37 & --6.96 & 2.48$^\dag$ & $3.83 \times 10^{-7}$ \\
G32.64+0.53 & --  & IGR~J18490--0000 & 282.25 & --0.02  & 11.16 & $5.13 \times 10^{-7}$ \\
G34.56--0.50 (in W44) & -- & -- & 284.04 & +1.22 & 4.58 & $1.12 \times 10^{-6}$ \\ 
G47.38--3.88 & -- & -- & 293.03 & +10.92  & 7.85 & $1.85 \times 10^{-7}$ \\
G74.00--8.50 (in the Cygnus Loop) & -- & -- & 312.33 & +29.02  & 0.05 & $8.09 \times 10^{-8}$ \\
G93.30+6.90 & -- & -- & 313.06 & +55.29  & 7.43 & $1.77 \times 10^{-7}$ \\
G108.60+6.80 & -- & -- & 336.42 & +65.60  & 3.74 & $4.48 \times 10^{-8}$ \\
G141.20+5.00 & -- & -- & 54.29 & +61.89  & 0.00 & $4.34 \times 10^{-8}$ \\
G179.72--1.69 (in S~147) & -- & -- & 84.60 & +28.28  & 10.3 & $9.91 \times 10^{-8}$ \\
G189.10+3.00 (in IC~443) & -- & -- & 94.28 & +22.37 & 25.9$^\ddag$ & $1.26 \times 10^{-6}$ \\
G266.97--1.00 & -- & -- & 133.90 & --46.74 & 16.7 & $2.58 \times 10^{-7}$ \\
G290.00--0.93 (IGR~J11014–6103) & -- & -- & 165.44 & --61.02  & 0.00 & $9.91 \times 10^{-8}$ \\
G310.60--1.60 & -- & -- & 210.19 & --63.42  & 0.01 & $2.02 \times 10^{-8}$ \\
G322.50--0.10 & -- & -- & 230.86 & --57.10  & 8.62 & $2.40 \times 10^{-7}$ \\
G341.20+0.90 & -- & -- & 251.87 & --43.75  & 0.00 & $4.24 \times 10^{-8}$ \\
G350.20--0.80 & -- & -- & 260.85 & --37.61  & 14.9 & $4.96 \times 10^{-7}$ \\
G358.29+0.24 & -- & HESS~J1741--302 & 265.32 & --30.38  & 0.00 & $1.85 \times 10^{-7}$ \\
G358.60--17.20 & -- & -- & 284.15 & --37.91  & 0.00 & $2.18 \times 10^{-8}$ \\
\hline
\hline
\end{tabular}
\caption{{\bf Sources not detected by the LAT:} Results of the maximum likelihood fits for PWNe and PWN candidates not detected by the LAT along with the Galactic PWN Name, TeV name if applicable, R.A. and declination in J2000 equatorial degrees, and the detection significance (TS) of a point source at the specified location. The last column provides the 95\% C.L. flux upper limit for the 300\,MeV--2\,TeV energy range. \footnotesize{Notes: $^\dag$ This source is classified as a potential PWN in the 4FGL--DR2 catalog associated with TeV PWN HESS~J1837--069, but a detailed analysis of this region shows only one extended source (4FGL~J1836.5--0651e) is required to model residual emission here (see Section~\ref{sec:new} and Section~\ref{sec:nondetect} for details). $^\ddag$ Interpreted as a non-detection. See Section~\ref{sec:nondetect} for details.}}
\label{tab:nondetect}
\end{table*}
\endgroup

G29.70--0.3, B0453--685, and G327.15--1.04 have been independently analyzed in separate reports \citep[see][]{straal2022,eagle2022,eagle2023}, where a pulsar contribution is explored or questioned. In these cases, the PWN is expected to dominate above a few GeV. Many of the sources reported here are significantly detected above 10\,GeV and/or have significant extension. Despite the possible presence of a low-energy spectral contribution, a dominant PWN contribution at higher energies is indicated. 

The likely PWNe with possible pulsar components include: G29.70--0.30 (Kes~75), G315.78--0.23, G327.15--1.04, G336.40+0.10, N~157B, and B0453--685, which is the majority of likely PWNe (6/9). Similarly for the other \fermi PWN candidates: G11.18--0.35, G12.82--0.02, G15.40+0.10, G16.73+0.08, G18.90--1.10, G27.80+0.60, G29.40+0.10, G63.70+1.10, G65.73+1.18, G74.94+1.11, and G338.20--0.00 have possible additional spectral components arising in the lower energy band ($E < 10\,$GeV), which correspond to 11/21 candidate PWNe. This brings the total number of sources with a possible pulsar component in this sample to 17. We attempt to characterize any additional spectral component, described below.

The 6 likely \fermi PWNe with a potential second component are tested with two different additional spectral shapes to the source spectrum: 1) a simple power-law and 2) a power-law with an exponential cut-off (PLEC4 from the 4FGL-DR3 catalog\footnote{\url{https://fermi.gsfc.nasa.gov/ssc/data/analysis/scitools/source_models.html}}).  Likely due to low statistics in most of the PWNe, however, no secondary contributions could be detected, except for G29.70--0.30, which has a likely MeV pulsar \citep{kuiper2018,straal2022}, and yields a $3\,\sigma$ detection for a second source at the pulsar location (TS $\sim 13$). The pulsar contribution tests for G327.15--1.04 and B0453--685 imply that a single PLEC4 source can marginally improve the fit over a source assuming a simple power-law, which may indicate PWN entanglement with any present pulsar contribution. Both PWNe lack the statistics and the detection of a central pulsar to explore this possibility further. 

A second test for a possible low-energy spectral component is to split the energy range into two: 300\,MeV--3\,GeV and 3\,GeV--2\,TeV. 
If a pulsar or low-energy contribution exists, one would expect the spectral index $E<3\,$GeV to be softer than above 3\,GeV. No significant differences in the spectral indices are found for the sources that are fit using a power law. The sources that are better fit using a Log Parabola spectrum tend to indicate slightly softer spectral indices for 3\,GeV--2\,TeV, than for 300\,MeV--3\,GeV, as expected for curved spectra.

As a final check, we also compare our best-fit spectral index for each source to the 4FGL values for those that have associated 4FGL counterparts. We do this for two different datasets of each ROI: \texttt{ALL} event types and then again for just the \texttt{PSF3} event type. We verify that no source deviates from the 4FGL spectral index value unless dramatic changes were adopted in this work with respect to the 4FGL catalogs aside from the different energy range (changes to location, spatial templates, size, etc.). We also inspect the spatial count profiles in both Galactic latitude and longitude around each source. These checks reasonably rule out the presence of systematic contamination such as the Galactic background. All sources reported here have a distinct spatial separation from the diffuse background. While we cannot rule out or confirm the presence of a lower-energy component such as a contributing pulsar, we can rule out systematic contamination.

\begin{figure*}[!ht]
\begin{minipage}[b]{.5\linewidth}
\centering
\includegraphics[width=1.0\linewidth]{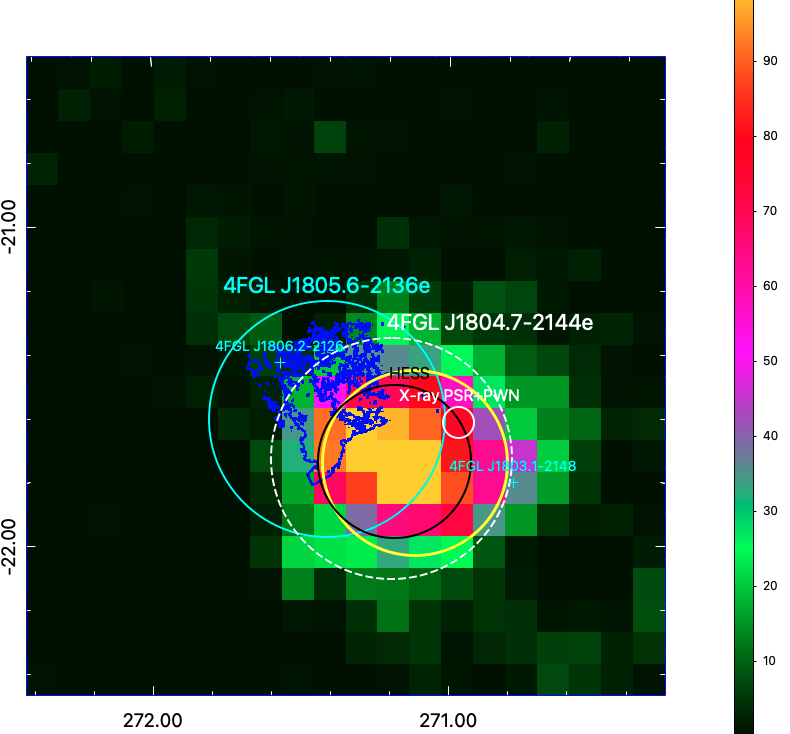}
\end{minipage}
\begin{minipage}[b]{0.5\linewidth}
\centering
\includegraphics[width=1.0\linewidth]{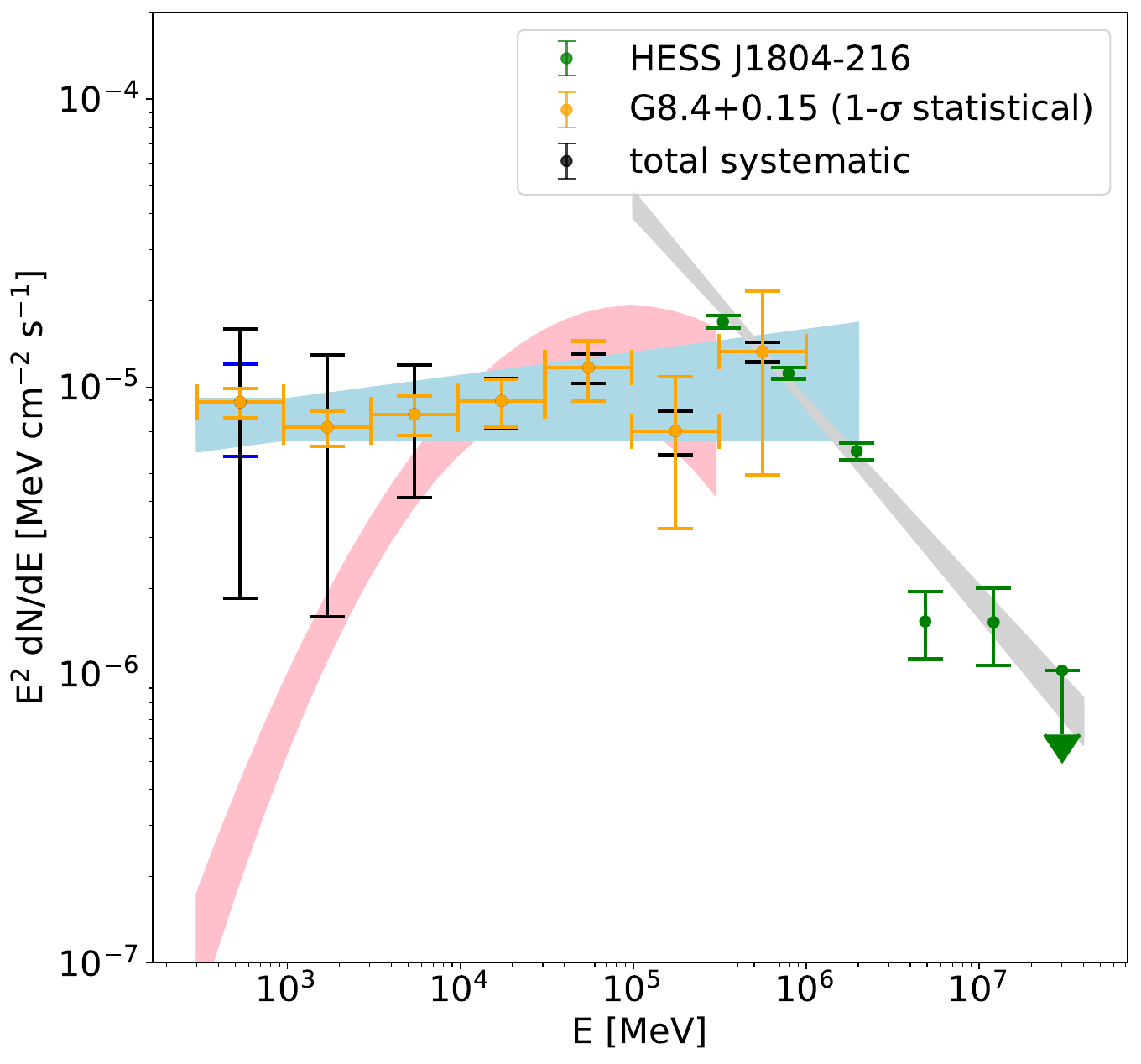}
\end{minipage}
\caption{{\it Left:} A $2\,\degree \times 2\,\degree$ 300\,MeV--2\,TeV TS map of \texttt{PSF3} events for PWN~G8.40+0.15. The color scale reflects the TS value. The TeV PWN HESS~J1804--216 is displayed as a black circle, $r = 0.24\,\degree$. The position of PSR~J1803--2137 and the size of the extended nebula observed in X-ray are marked and labeled as the white circle. The blue contours represent the SNR~G8.7--1.4 in radio. The 4FGL~J1804.7--2144e position and extent is indicated by the white dashed circle but is not included in the source model. The best-fit position and extent of the radial Gaussian template is indicated by the yellow circle. The maximum TS is $\sim$ 122. 4FGL sources in the field of view are labeled in cyan. The coordinates are in J2000 equatorial degrees. {\it Right:} The best-fit \fermi spectral model (blue band) and data (yellow points) for PWN~G8.40+0.15 are plotted beside the best-fit spectral model of its TeV counterpart HESS~J1804--216 in grey from \citet{hessgps2018}. The blue flux error for $E< 1$\,GeV is the additional systematic error as discussed in Section~\ref{sec:sys}. The Log Parabola model from \citet{liu2019} is plotted in pink.}\label{fig:hess_fermi_liu_g8.4_sed}
\end{figure*} 

\subsection{Results by ROI}\label{sec:results_by_roi}

We discuss the results of a majority of the new source detections and classifications along with their morphological and spectral characteristics in this section. In Section~\ref{sec:new}, we describe each of the likely PWN classifications. In Section~\ref{sec:pwnc}, we discuss the majority (15/21) of the weaker PWN candidates, chosen by either the multiwavelength evidence supporting follow-up investigations to constrain the origin or the unclear nature of the GeV emission due to being in complex regions of the $\gamma$-ray sky. The latter motivates follow-up analysis in the high-energy regime. For the sources where there is significant source emission between 1--10\,GeV or E $>10$\,GeV, we display TS images in these energy ranges. This choice avoids the declining angular resolution below these energies. For sources where this is not possible, TS maps from the full energy range 300\,MeV--2\,TeV are displayed.

\subsubsection{Likely PWNe}\label{sec:new}

\begin{figure}[hbt]
\centering
\includegraphics[width=0.95\linewidth]{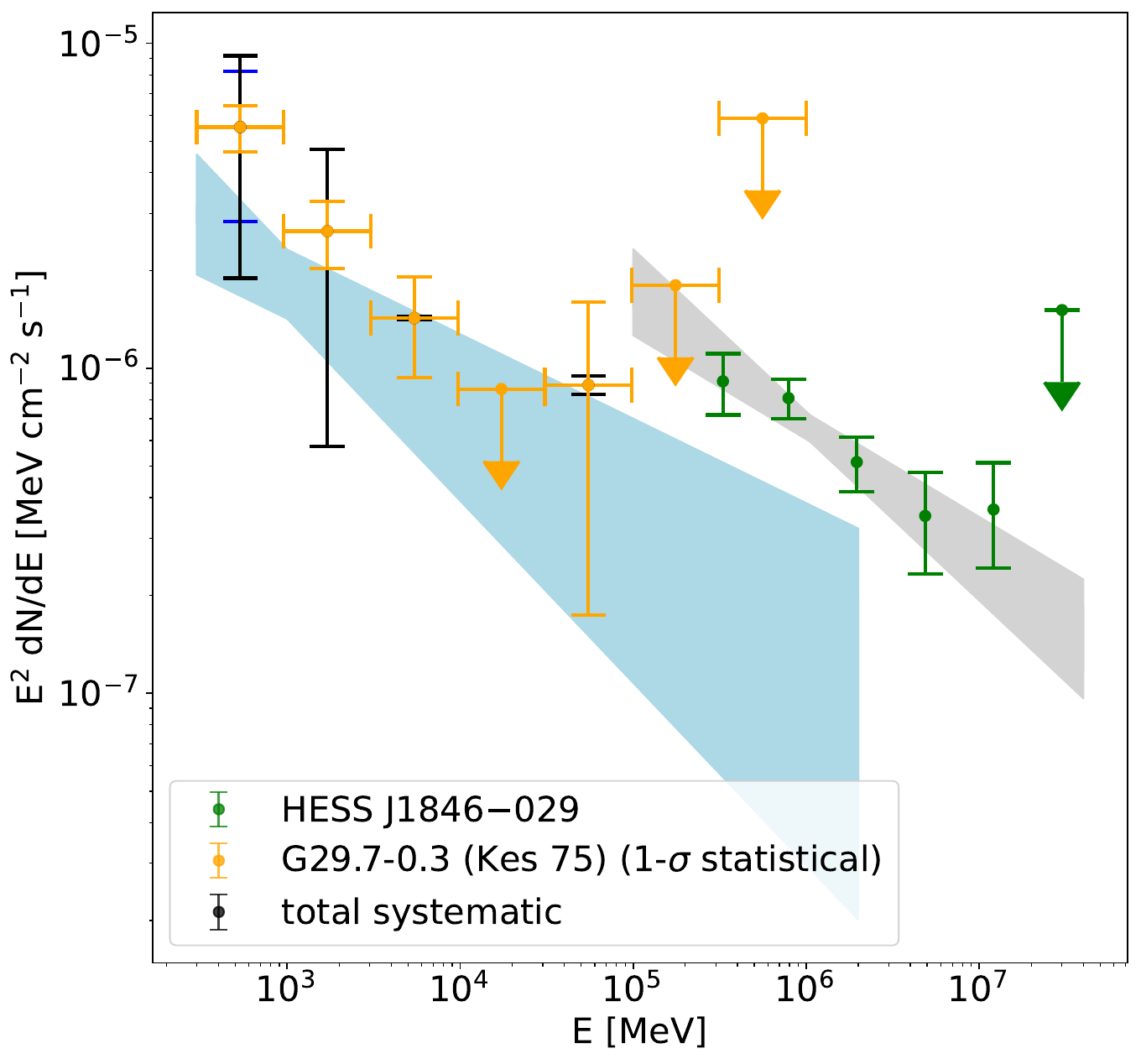}
\caption{
The best-fit \fermi spectral model (blue band) and data in yellow for PWN~G29.70--0.30 (Kes 75) are plotted beside the best-fit spectral model (grey band) and data in green of its TeV counterpart HESS~J1846--029 from \citet{hessgps2018}. The blue flux error for $E< 1$\,GeV is the additional systematic error as discussed in Section~\ref{sec:sys}.} \label{fig:g29.7}
\end{figure}

\begin{figure*}
\begin{minipage}[b]{.5\linewidth}
\centering
\includegraphics[width=0.95\linewidth]{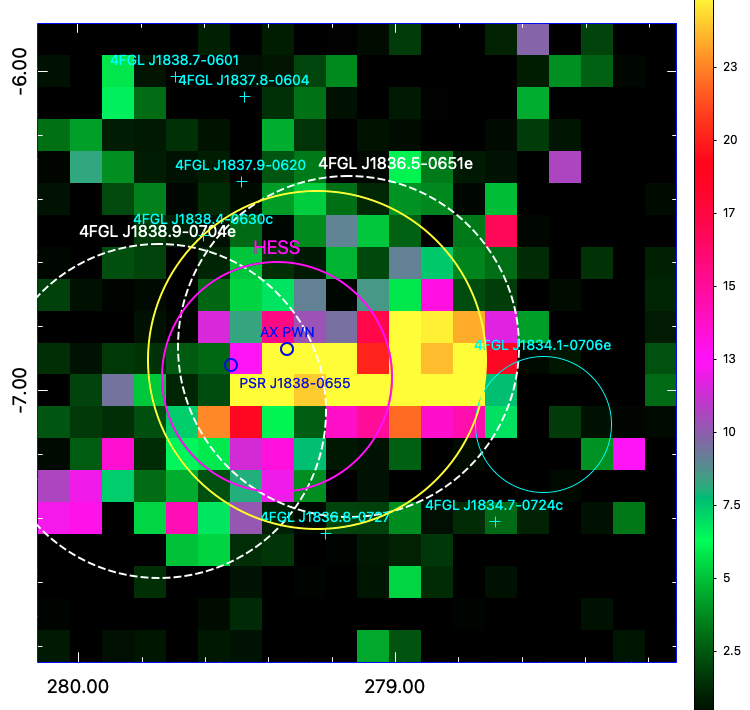}
\end{minipage}
\begin{minipage}[b]{.5\linewidth}
\centering
\includegraphics[width=1.0\linewidth]{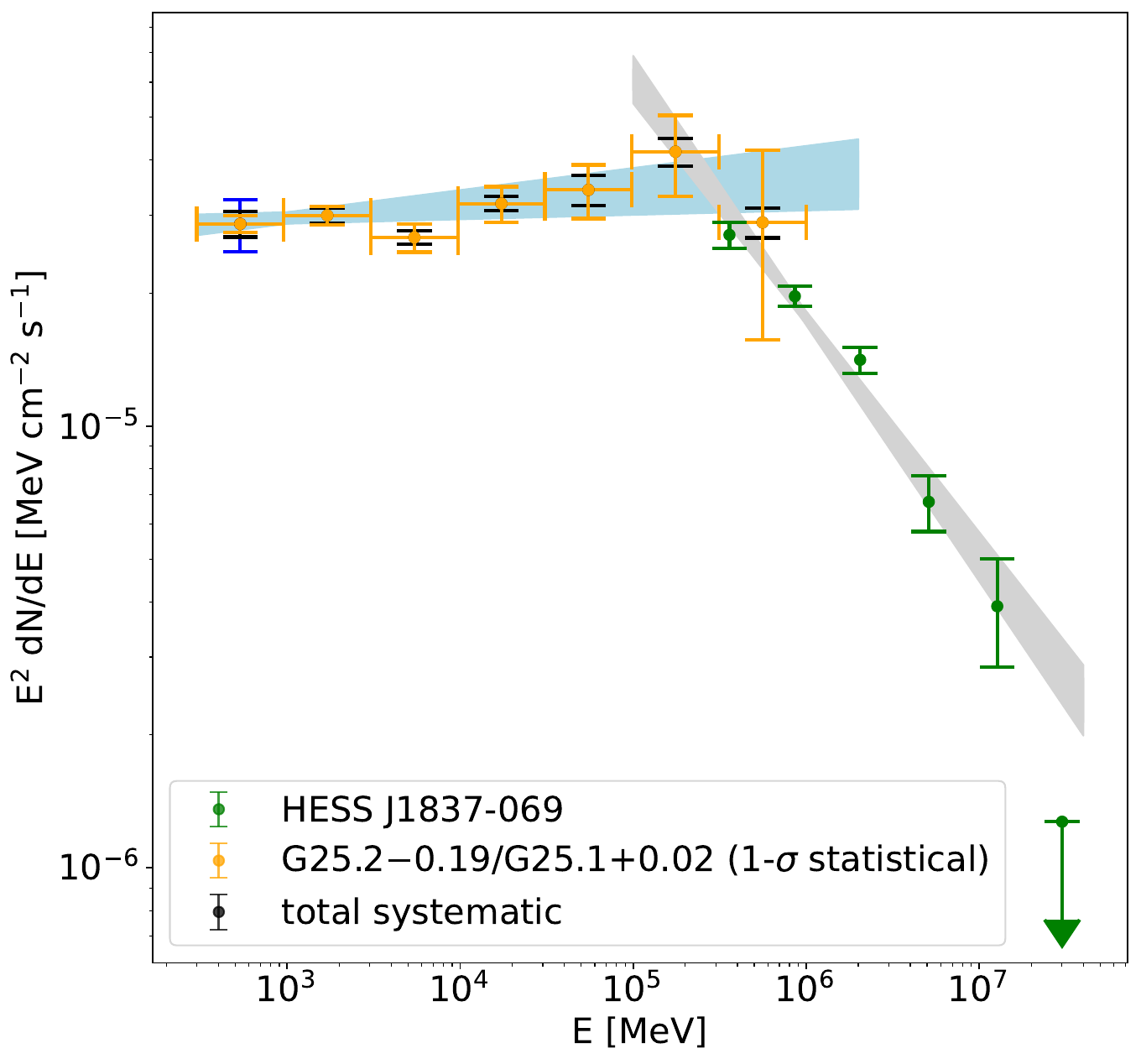}
\end{minipage}
\caption{{\it Left:} A $2\,\degree \times 2\,\degree$ TS map for $E > 10\,$GeV of the TeV PWN HESS~J1837--069 (magenta circle). The 4FGL~J1836.5--0651e and 4FGL~J1838.9--0704e positions and extents are indicated by the white dashed circles but are not included in the source model. The best-fit position and extent of the radial Gaussian template is indicated by the yellow circle.  The blue circles correspond to the location and size of the X-ray nebulae G25.24--0.19 (``AX PWN'') and G25.10+0.02 powered by PSR~J1838--0655. The maximum TS at the PWN positions is $\sim$ 46. Unrelated 4FGL sources are indicated in cyan. {\it Right:} The best-fit \fermi spectral model (blue band) and data in yellow for the extended source coincident with both PWN~G25.10+0.02 and G25.24--0.19 (J1836.5--0651e) as well as the best-fit spectral model (grey band) and data in green of its TeV counterpart HESS~J1837--069 from \citet{hessgps2018}. The blue flux error for $E< 1$\,GeV is the additional systematic error as discussed in Section~\ref{sec:sys}.}\label{fig:hess_j1837_fermi_sed}
\end{figure*}

{\bf G8.40+0.15:} First detected in the TeV band \citep{hess2005}, HESS~J1804--216 is an extended unidentified source with semi-major and semi-minor axes 0.24\,$\degree$ and 0.16\,$\degree$, respectively. In \citet{liu2019}, the corresponding extended GeV emission is found to most likely originate from two unrelated sources: PWN~G8.40+0.15 and SNR~G8.7--1.4. 4FGL~J1805.6-2136e is plausibly the SNR~G8.7--1.4 and 4FGL~J1804.7-2144e, coincident with TeV source HESS~J1804--213, is argued to be the PWN~G8.40+0.15. \citet{liu2019} and the 4FGL catalogs characterize the GeV PWN counterpart as a radial disk with $r = 0.38\,\degree$ \citep{fges2017}. A careful analysis of the \fermi data in this region confirms that the two extended sources reported in \citet{liu2019} provide the best-fit spatial model, with a small adjustment on the likely PWN 4FGL~J1804.7-2144e. 
We find that a smaller extension using the radial Gaussian template $r = 0.29 \,\degree$ characterizes the emission just as well, see Table~\ref{tab:extent}. The best-fit spectrum is a power law and the spectral index is softer than the spectrum measured in \citet{liu2019}, but is consistent with what is reported in the 4FGL catalogs, $\Gamma = 1.96 \pm 0.04$ \citep{fges2017,4fgl-dr4}. We plot the best-fit Log Parabola spectral model from \citet{liu2019} alongside the \fermi and HESS flux data points in Figure~\ref{fig:hess_fermi_liu_g8.4_sed}, right panel, showing that the spectral measurements, overall, agree well. The TeV emission is characterized as a power law with $\Gamma = 2.69 \pm 0.04$ \citep{hessgps2018}, indicating that the peak of $\gamma$-ray emission is likely occurring in the \fermi band. Given the broadband results of \citet{liu2019}, the agreement between the H.E.S.S. and \fermi extensions, spectral properties, and the positional coincidence with an identified PWN that extends 12$^{\prime\prime}$ from the pulsar PSR~J1803--2137 as observed by \cha \citep{karg2007}, we classify 4FGL~J1804.7--2144e as a likely PWN, see Figure~\ref{fig:hess_fermi_liu_g8.4_sed}, left panel. 
This does not preclude, however, that contamination from other sources in the complex (SNR, PSR) contribute to the $\gamma$-ray data.


\smallskip
{\bf G29.70--0.30 (Kes~75):} G29.70--0.30 is the youngest known PWN in the Milky Way Galaxy with an estimated age $\tau \sim 500\,$yr \citep{kes752018}. 
It is powered by a very energetic pulsar that may be a magnetar based on its high surface magnetic field strength and having emitted several magnetar-like short X-ray bursts \citep{kes752018}. The magnetar-like pulsar is detected in 30--100\,MeV \fermi data at the 4.2\,$\sigma$ significance level \citep{kuiper2018} with a peak energy at $E \sim 3 \,$MeV. The $E < 100\,$MeV pulsed emission is characterized as a super-exponential cut-off power law with a hard photon index before the break $\Gamma \sim 0.9$ and a cut-off energy at $E_c \sim 0.009\,$MeV. Pulsations disappear before $E = 100$\,MeV \citep{kuiper2018}. Combined with a faint source detection, this prevents a pulsation search from being feasible above this energy. \citet{straal2022} characterize a pulsar component $\Gamma \sim 1.5 \pm 0.4$ and $E_c \sim 1\,$GeV for 100\,MeV $< E <$ 5\,GeV. The PWN is visible in radio ($\sim 0.6^\prime$ in size) and X-ray as a bright, compact nebula encompassed by an incomplete radio SNR shell ($\sim 4^\prime$ in size). 

The PWN is a known TeV emitter HESS~J1846--029. 
There is a nearby unknown source 4FGL~J1846.9--0247c that is plausibly associated with Kes~75, however, the diffuse residuals are improved when this source remains in the model; thus it is not the likely counterpart. A significant detection for an additional point source at the position of Kes~75 is found. In 4FGL-DR4 \citep{4fgl-dr4}, this corresponds to the new source 4FGL~J1846.4--0258. 
A detailed broadband investigation was performed in \citet{straal2022} who argue 4FGL~J1846.9--0247c is indeed the PWN from Kes~75. Despite slight differences in the source model they use, we measure consistent spectral results for the PWN as in \citet{straal2022}, who find a photon index for the PWN above $E > 5\,$GeV of $\Gamma_\gamma = 2.49 \pm 0.38$.  The best-fit photon index measured between 300\,MeV--2\,TeV is $\Gamma = 2.41 \pm 0.15$, and is also in agreement with the H.E.S.S. spectral index $\Gamma = 2.41 \pm 0.09$, see Figure~\ref{fig:g29.7}. We classify Kes~75 as a likely \fermi PWN, although it is probable that both the pulsar and the PWN contribute to the observed \fermi signal.

\smallskip
{\bf HESS~J1837--069:} First detected as an unidentified TeV source \citep{hess2005}, it was investigated as a PWN candidate after the discovery of two compact ($\lesssim 1^\prime$) X-ray PWNe inside the extended H.E.S.S. source \citep[$r \sim 0.36\,\degree$,][]{gotthelf2008,hessgps2018}. The subsequent detection of an extended \fermi counterpart \citep{acero2013} further supported HESS~J1837--069 as a strong TeV PWN candidate. The investigation into the local environment solidified the PWN origin as the most probable source class \citep{fujita2014}. 
Displayed in Figure~\ref{fig:hess_j1837_fermi_sed}, left panel, is the HESS~J1837--069 TS map and all potential counterparts indicated. 
Two X-ray PWNe are coincident in location with both the extended TeV and GeV sources: PWN AX~J1837.3--0652 (``AX PWN'' or G25.24--0.19) 
and the closeby compact PWN~G25.10+0.02 powered by PSR~J1838--0655 \citep{karg2012}. PSR~J1838--0655 has a spin-down power $\dot{E} \sim 5 \times 10^{36}$\,erg s$^{-1}$ and characteristic age $\sim 23\,$kyr. AX~J1837.3--0652, while much fainter in X-ray than PSR~J1838--0655, may have similar characteristics \citep{gotthelf2008}, making both plausible contributors to the TeV emission. There are two extended sources associated with HESS~J1837--069 in the 4FGL: 4FGL~J1838.9--0704e and 4FGL~J1836.5--0651e. 
4FGL~J1836.5--0651e is modeled with $r = 0.53\,\degree$ and 4FGL~J1838.9--0704e is similarly fit as $r = 0.52\,\degree$, both as a radial disk in the 4FGL \citep{fges2017,4fgl-dr4}. We find that there is only one extended source required for the observed emission and corresponds to 4FGL~J1836.5--0651e, but is significantly better fit as a radial Gaussian with extension $r = 0.53\,\degree$. The GeV spectral index $\Gamma = 1.97 \pm 0.02$ is harder than the TeV PWN spectral index $\Gamma = 2.54 \pm 0.04$, see Figure~\ref{fig:hess_j1837_fermi_sed}, right panel, connecting the GeV and TeV spectra.

Given the PWN classification of the TeV source HESS~J1837--069 and the positional coincidence of the TeV and GeV emission with two compact X-ray PWNe, it seems likely the GeV counterpart is a PWN. 
It is possible one or both X-ray nebulae contribute. Therefore we classify the extended emission modeled by 4FGL~J1836.5--0651e as a likely PWN, but multiwavelength investigations are needed to determine the X-ray counterpart(s). 

\begin{figure*}[!ht]
\begin{minipage}[b]{.5\linewidth}
\centering
\includegraphics[width=1.0\linewidth]{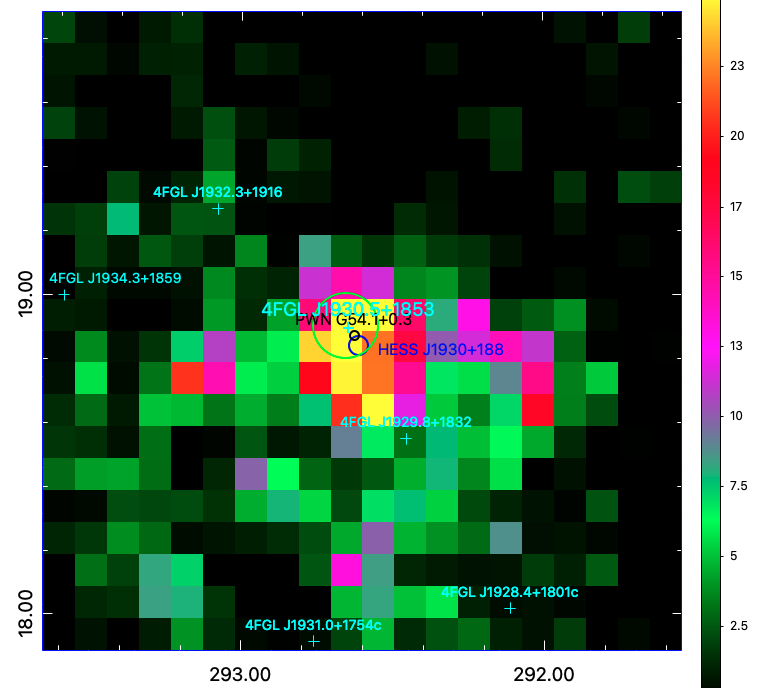}
\end{minipage}
\begin{minipage}[b]{.5\linewidth}
\centering
\includegraphics[width=1.0\linewidth]{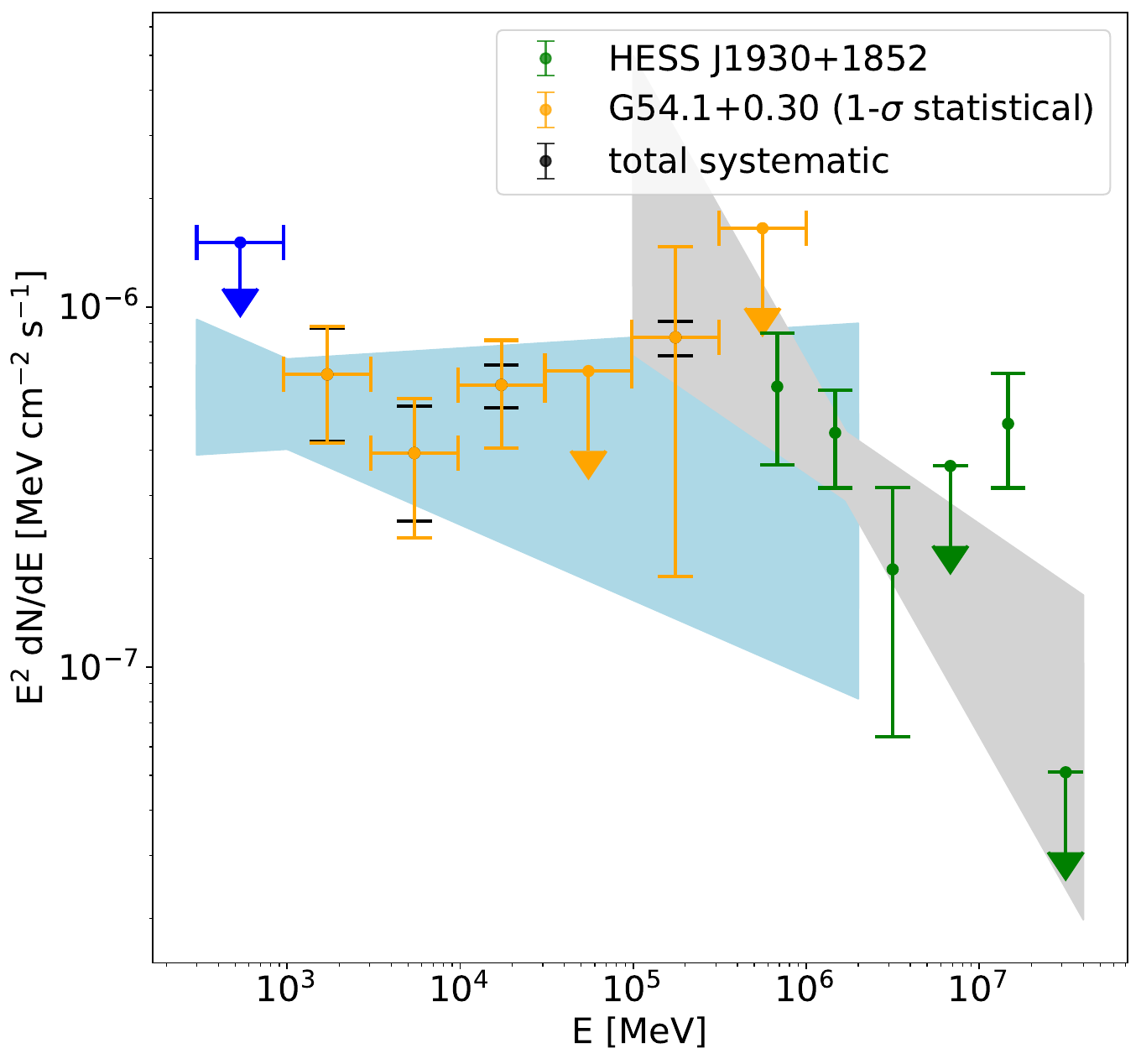}
\end{minipage}
\caption{{\it Left:} A $2\,\degree \times 2\,\degree$ 300\,MeV--2\,TeV TS map of \texttt{ALL} events for PWN~G54.10+0.27. The green circle represents the 95\% positional uncertainty of a point source at the PWN position and corresponds to the DR3 source 4FGL~J1930.5+1853. The black circle represents the size and location of the X-ray PWN. The 68\% uncertainty region for the TeV PWN HESS~J1930+188 is indicated in blue. The maximum TS at the PWN position is $\sim$ 32. Unrelated 4FGL sources are indicated in cyan. {\it Right:} The best-fit spectral model (blue band) and data in yellow of J1930.5+1853 as well as the best-fit spectral model (grey band) and data in green of its TeV counterpart HESS~J1930+188 from \citet{hessgps2018}. The blue flux upper limit for $E< 1$\,GeV is from the additional systematic error as discussed in Section~\ref{sec:sys}.}\label{fig:hess_fermi_g54.1+0.30_sed}

\begin{minipage}[b]{.5\linewidth}
\centering
\includegraphics[width=1.04\linewidth]{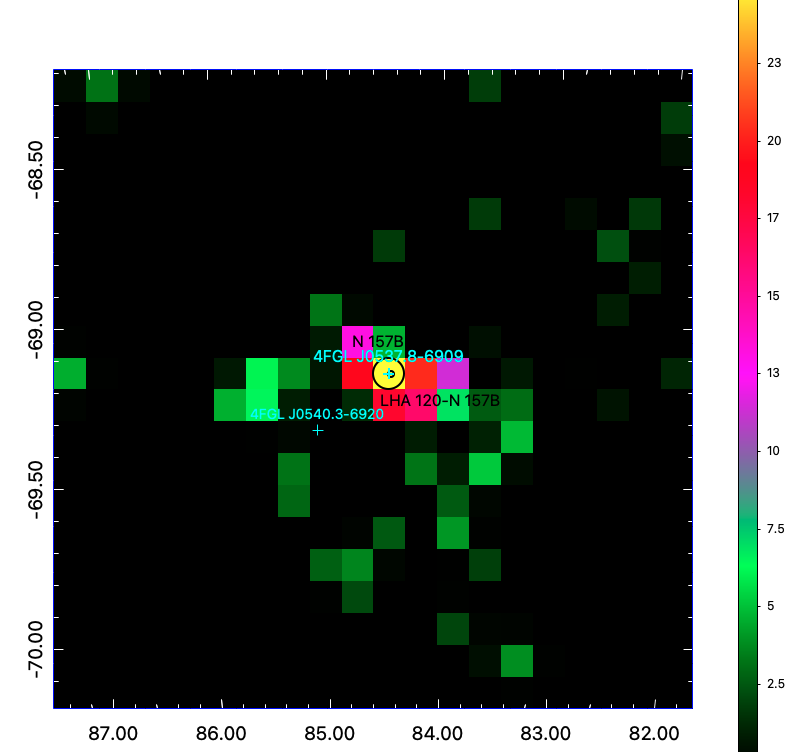}
\end{minipage}
\begin{minipage}[b]{.5\linewidth}
\centering
\includegraphics[width=0.98\linewidth]{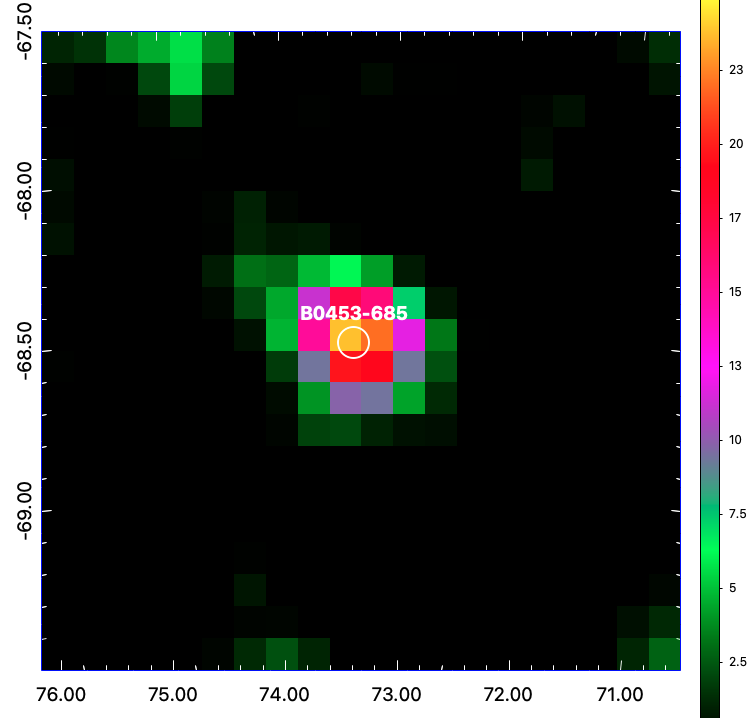}
\end{minipage}
\caption{{\it Left:} A $2\,\degree \times 2\,\degree$ 10\,GeV--2\,TeV TS map of \texttt{PSF3} events for PWN~N~157B. The 4FGL counterpart is 4FGL~J0537.8--6909, indicated in cyan with the 95\% positional uncertainty (outer black circle). The TeV PWN ``LHA 120--N 157B'' (inner black circle) is indicated. Unrelated nearby 4FGL sources are labeled in cyan. The maximum TS at the PWN/SNR position is $\sim$ 35 for $E > 10$\,GeV.  {\it Right:} A $2\,\degree \times 2\,\degree$ TS map of \texttt{ALL} events centered on B0453--685 (white circle) between 1--10\,GeV. 
The maximum TS at the PWN/SNR position is $\sim$ 24 in the 1--10\,GeV energy range.}\label{fig:n157b_b0453}
\end{figure*} 

\smallskip
{\bf G54.10+0.27:} G54.10+0.27 is a young, Crab-like SNR (i.e., presence of pulsar and diffuse PWN emission, but lacks a SNR shell) powered by pulsar J1930+1852 and is roughly $\sim 2^\prime$ in size as observed by \cha \citep{temim2010}. The central pulsar and PWN have been studied in detail in both radio and X-ray \citep[e.g.,][]{camilo2002,gelfand2015, temim2010}. Point-like TeV emission HESS~J1930+188 is identified as the TeV counterpart to the PWN \citep{hessgps2018}. 
4FGL~J1930.5+1853 is probably the PWN~G54.10+0.27 considering the positional coincidence. The agreement in position and spectral index between the GeV and TeV emission \citep[$\Gamma = 2.09 \pm 0.12$ in GeV and $\Gamma = 2.59 \pm 0.26 $ in TeV, see also both panels of Figure~\ref{fig:hess_fermi_g54.1+0.30_sed} and][]{hessgps2018} classifies the source as a likely PWN. Confirmation of the true source class will require an in-depth multiwavelength investigation to rule out source contamination, whether related (pulsar, PWN, or SNR) or unrelated (nearby source).

\smallskip
{\bf N~157B: } N~157B is a young PWN powered by PSR~J0537--6910 (characteristic age $\tau_c \sim 4.9\,$kyr) located within the LMC and is the first MeV--GeV  \citep{lmc2016,saito2017} and TeV  \citep{n157b2012} detection for $\gamma$-ray emission associated with a PWN outside of the Milky Way Galaxy. The pulsar has a spin-down luminosity even larger than that of the Crab pulsar ($L \sim 5 \times 10^{38}\,$erg s$^{-1}$), making it the most rapidly spinning and most powerful young pulsar known, with a spin period $\sim 16\,$ms \citep{marshall2004}. The $\gamma$-ray detection of N~157B is possible given the central pulsar's enormous power output, in addition to potentially rich local photon fields that facilitate a large IC flux. Point-like GeV emission coincident with the system could plausibly be from the energetic pulsar, the PWN, or the SNR \citep{lmc2016}. The morphology of the PWN/SNR is not clear: the lack of thermal X-rays and the missing limb-brightened outer SNR shell imply a Crab-like morphology, but the diffuse X-rays beyond the cometary nebula have a weak thermal component, implying possible re-heated SN ejecta from the passage of the reverse shock \citep{chen2006}.  Even though the observed \fermi emission cannot firmly rule out a SNR scenario when considered alone, it seems unlikely given the lack of observational evidence for an energetic SNR shell in other wavebands. Pulsations from the central pulsar were searched for in the observed $\gamma$-ray emission, but none were evident above a 1-$\sigma$ C.L. \citep{lmc2016}. The identified TeV counterpart of the PWN combined with no detectable pulses in the \fermi signal and a photon index similar to other \fermi PWNe ($\Gamma = 2.11 \pm 0.07$) guide us to classify this source as a likely PWN, see also Figure~\ref{fig:n157b_b0453}, left panel.

\begin{figure*}[hbtp]
\begin{minipage}[b]{.49\linewidth}
\centering
\includegraphics[width=0.93\linewidth]{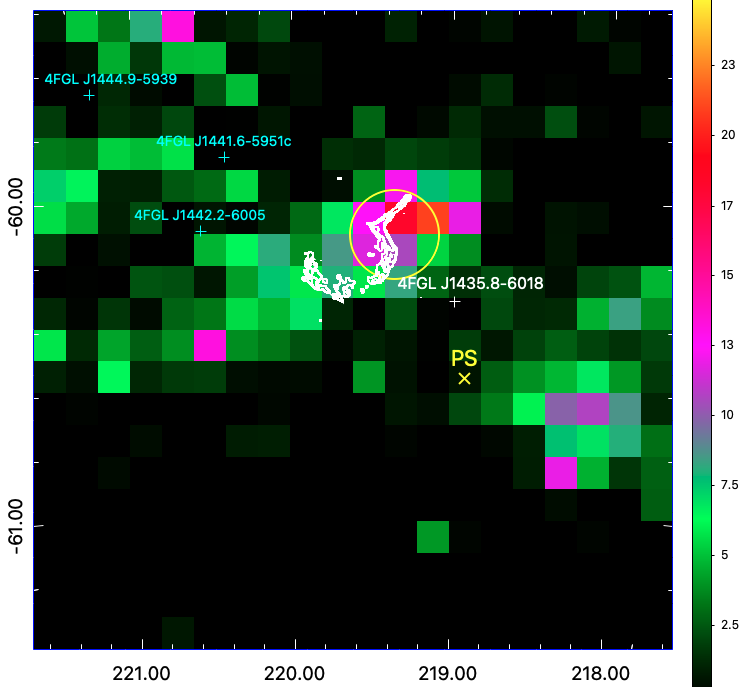}
\end{minipage}
\begin{minipage}[b]{.49\linewidth}
\centering
\includegraphics[width=1.0\linewidth]{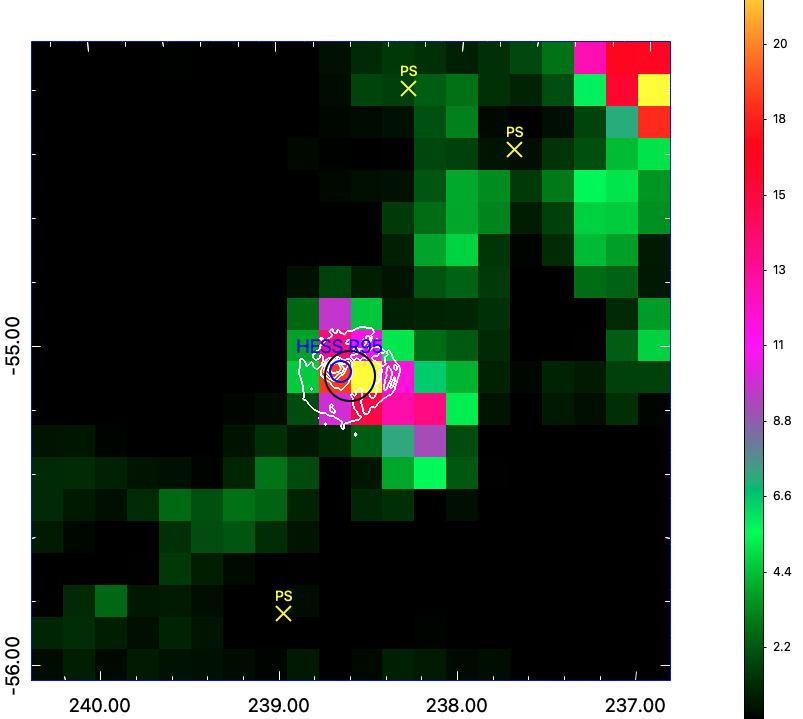}
\end{minipage}
\caption{{\it Left:} A $2\,\degree \times 2\,\degree$ 300\,MeV--2\,TeV TS map of \texttt{PSF3} events for PWN~G315.78--0.23. There is one possibly associated \fermi source, 4FGL~J1435.8--6018. The 95\% uncertainty region of the new best-fit position of 4FGL~J1435.8--6018 is the yellow circle, which coincides with the ``handle'' of the Frying Pan radio morphology. The ``handle'' consists of the supersonic pulsar and the trailing PWN. The yellow cross labeled ``PS'' represents the addition of a point source to model residual emission unrelated to the SNR. The second unrelated point source (see text) is just out of view to the West. Unrelated nearby 4FGL sources are labeled in cyan.
The maximum TS at the PWN/SNR position is $\sim$ 21. {\it Right:} A $2\,\degree \times 2\,\degree$ 1--10\,GeV TS map of \texttt{ALL} events centered on SNR~G327.15--1.04 (denoted by 843\,MHz radio contours in white) accompanied by the 95\% confidence regions for the TeV PWN HESS~J1554--550 (blue) and for a point source modeling the residual MeV--GeV emission (black). The maximum TS at the PWN/SNR position is $\sim$23 in the 1--10\,GeV energy range.}\label{fig:g315_g327}
\end{figure*} 

\smallskip
{\bf B0453--685:} SNR~B0453--685 is a middle-aged ($\tau \sim 14\,$kyr) composite SNR located in the LMC, on the opposite (Western) side from PWN N~157B \citep[e.g., Figure~\ref{fig:lmc},][]{gaensler2003}. \fermi $\gamma$-ray emission coincident with SNR~B0453--685 is detected at a significance level $>$4\,$\sigma$, see Figure \ref{fig:n157b_b0453}, right panel. The $\gamma$-ray emission displays no evidence for extension and the best-fit spectral index $\Gamma = 2.27 \pm 0.18$ characterizes a power-law spectrum. There is no known TeV counterpart for this system. While the central pulsar has not been detected, a detailed multiwavelength investigation described in \citet{eagle2023} finds the most likely origin to be the PWN with a possible pulsar contribution below $5\,$GeV. The host SNR displays energetics that are inconsistent if the SNR is the $\gamma$-ray origin such as no detectable non-thermal emission in X-ray nor any indication of an interaction with ambient media, but this does not rule out possible SNR contamination to the $\gamma$-ray data. The position, extent, and energetics favor a PWN dominant origin.

\smallskip
{\bf G315.78--0.23:} Also known as the Frying Pan, this ancient system features a radio SNR shell and an exiting supersonic pulsar that powers a trailing PWN in its wake \citep{ng2012}. An unidentified point source 4FGL~J1435.8--6018 is in close proximity to the radio position of the bow-shock PWN and pulsar. We investigate the possible association with PWN~G315.78--0.23 by replacing 4FGL~J1435.8--6018 with a point source at the PWN position, in addition to two point sources to the Southwest of the SNR shell in order to model unrelated, persisting diffuse residual $\gamma$-ray emission. 
Localizing 4FGL~J1435.8--6018 to the ``handle'' of the Frying Pan SNR morphology, which is where the supersonic pulsar and its bow-shock nebula are located, see Figure~\ref{fig:g315_g327}, left panel, can provide a better global fit to the data.  The TS for a point source at the PWN/PSR position is TS $= 39$ with no evidence for extension. The placement of the $\gamma$-ray source with the bow-shock nebula, its spectral index $\Gamma = 2.76 \pm 0.16 $ being similar to other \fermi PWNe, as well as considering the energetic nature of the bow-shock nebula \citep{ng2012}, all support a PWN classification, despite the lack of an identified TeV counterpart. A SNR component seems least likely considering the age of the system and the lack of molecular material observed in the region \citep{ng2012}. We consider this a likely PWN, but we cannot rule out a pulsar contribution. 

\smallskip
{\bf G327.15--1.04:} The first MeV--GeV detection of this source was reported in \citet{xiang2021}. The $\gamma$-ray emission is detected at $>4$\,$\sigma$ significance with the \fermi between 300\,MeV--2\,TeV and is best characterized with photon index $\Gamma = 2.45 \pm 0.13$ \citep{eagle2022}. There is $>2\,\sigma$ significance for curvature, consistent to both \citet{eagle2022} and the 4FGL--DR3. 
A detailed multiwavelength investigation of the associated emission was carried out \citep{eagle2022}, where the point-like MeV--GeV $\gamma$-ray emission is consistent with a PWN origin from G327.15--1.04, supported by its TeV counterpart HESS~J1554--550.  The position, extent, and energetics of the system favor a PWN origin. A 1--10\,GeV TS map demonstrating the source detection is displayed in Figure~\ref{fig:g315_g327}, right panel. 
While the current IC model predictions \citep{eagle2022} can characterize well the MeV--TeV emission for PWN~G327.15--1.04, there remains uncertainty in the properties of ambient photon fields in addition to physical properties of the system (i.e., true age, maximum particle energy, energy break in particle spectrum, etc.) that influence the IC shape. Moreover, the central pulsar remains undetected at any wavelength, such that a MeV--GeV pulsar contribution cannot be reliably ruled out, but the TeV PWN counterpart supports a high-energy PWN contribution to the \fermi data. 

\begin{figure*}[hbtp]
\begin{minipage}[b]{.49\linewidth}
\centering
\includegraphics[width=1.0\linewidth]{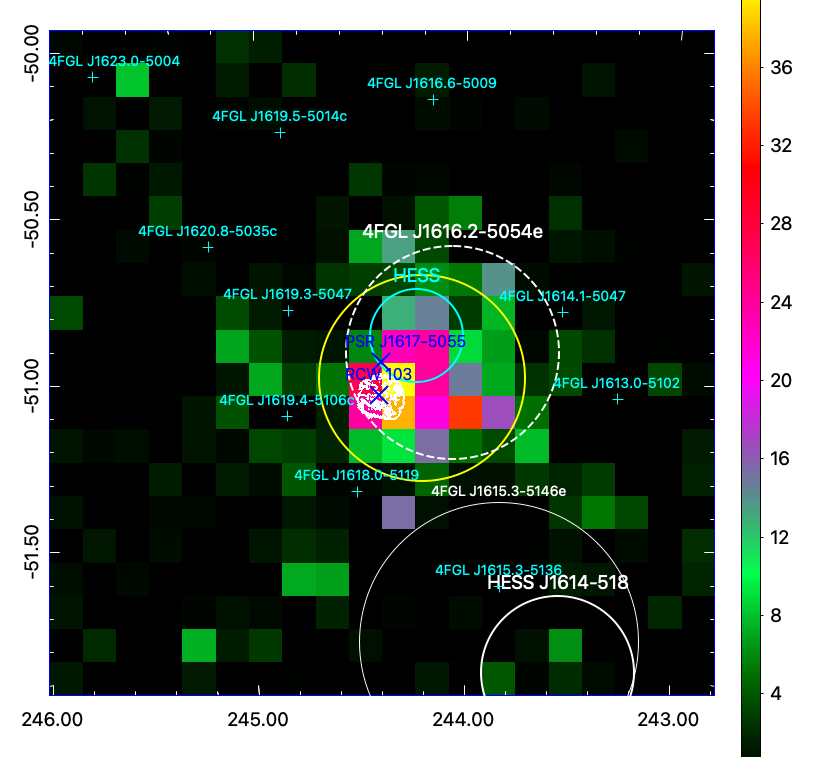}
\end{minipage}
\begin{minipage}[b]{.49\linewidth}
\centering
\includegraphics[width=1.0\linewidth]{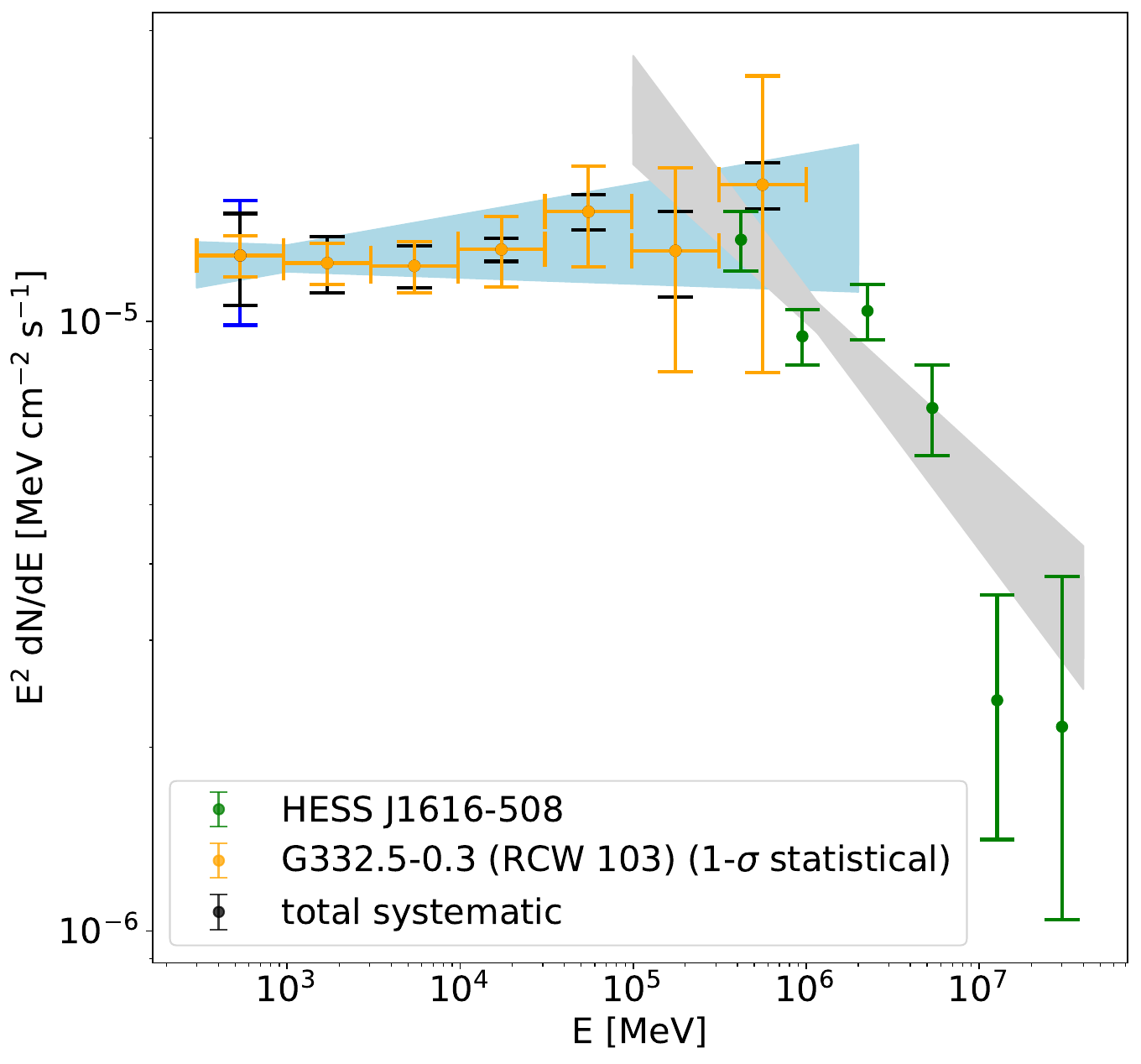}
\end{minipage}
\caption{{\it Left:} A $2\,\degree \times 2\,\degree$ 10\,GeV--2\,TeV TS map of \texttt{PSF3} events for both X-ray PWNe coincident with the \fermi PWN 4FGL~J1616.2--5054e: G332.50--0.30 (RCW~103, with \cha X-ray contours in white) and G332.50--0.28 (PSR~J1617--5055). 4FGL~J1616.2--5054e is indicated as the white dashed circle but is not included in the source model. The best fit for the extended emission is marked as a yellow circle.  HESS~J1616--518 is the cyan circle. The maximum TS occurs between the two X-ray PWNe with value $\sim$ 50 for energies between 1--10\,GeV. {\it Right:} The best-fit \fermi spectral model (blue band) and data in yellow for the extended source coincident with both PWN~G332.50--0.30 (RCW~103) and G332.50--0.28 (J1616.2--5054e) beside the best-fit spectral model (grey band) and data in green of its TeV counterpart HESS~J1616--508 from \citet{hessgps2018}. The blue flux error for $E< 1$\,GeV is the additional systematic error as discussed in Section~\ref{sec:sys}.}\label{fig:hess_j1616-508}
\begin{minipage}[b]{.49\linewidth}
\centering
\includegraphics[width=1.0\linewidth]{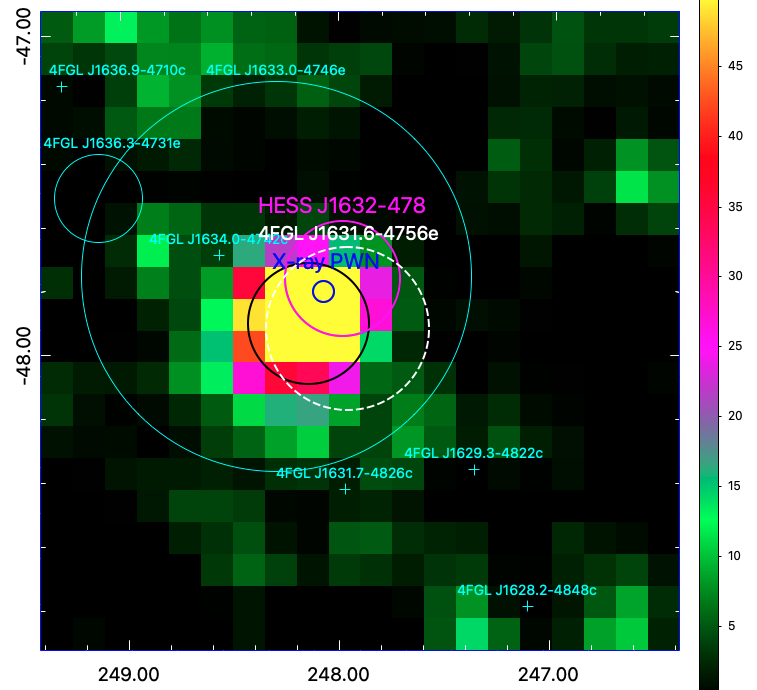}
\end{minipage}
\begin{minipage}[b]{.49\linewidth}
\centering
\includegraphics[width=1.0\linewidth]{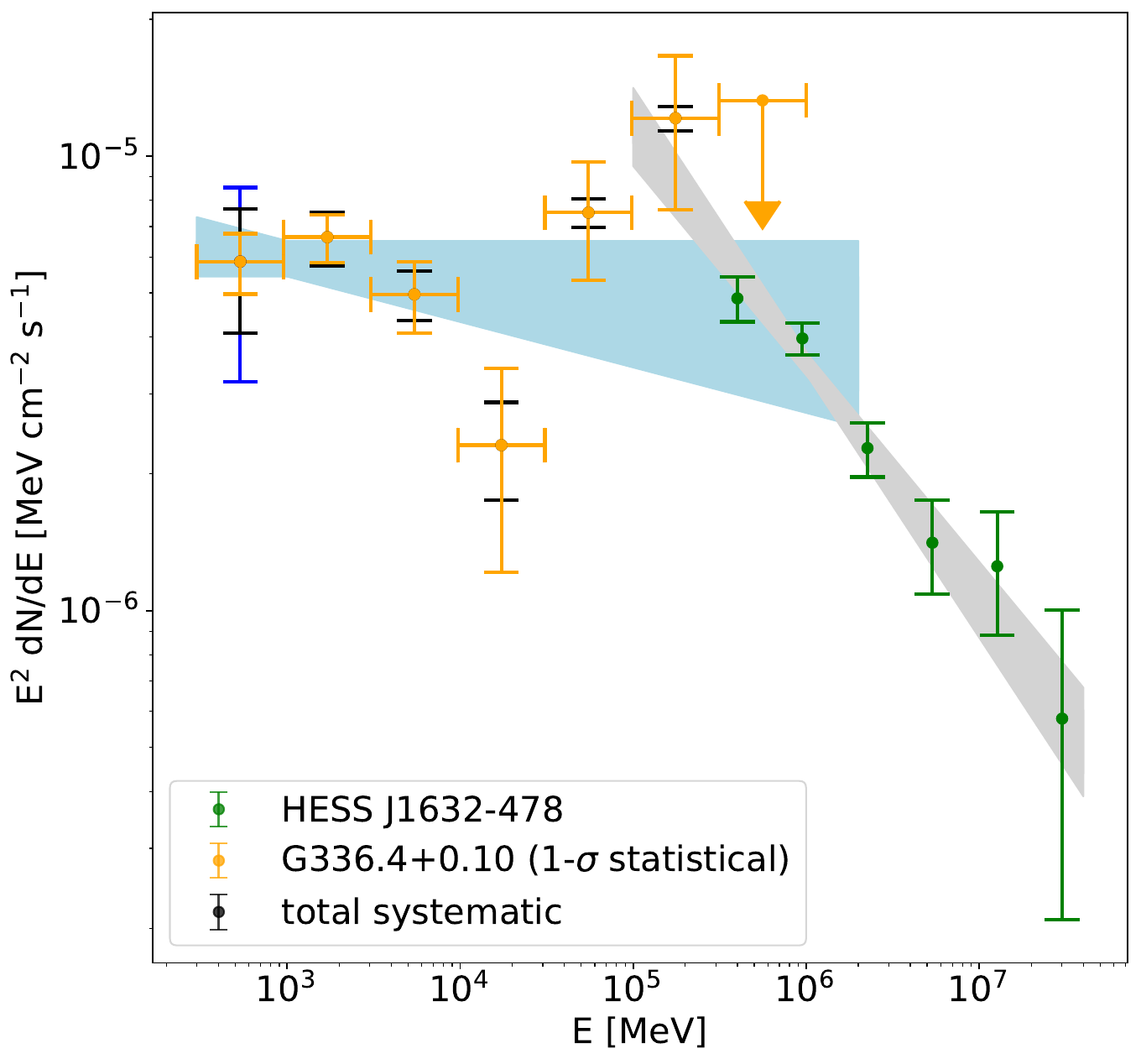}
\end{minipage} 
\caption{{\it Left:} A $2\,\degree \times 2\,\degree$ 1--10\,GeV TS map of \texttt{PSF3} events for PWN~G336.40+0.10. 4FGL~J1631.6--4756e is indicated but not included in the source model. The best-fit radial Gaussian template for the extended emission is the black circle. The X-ray PWN location and extent is marked with a blue circle and the TeV PWN counterpart HESS~J1632--478 in magenta.  The maximum TS at the PWN position is $\sim$ 109 for energies between 1--10\,GeV. Unrelated 4FGL sources are labeled in cyan. {\it Right:} The best-fit spectral model (blue band) and data in yellow for J1631.6--4756e (PWN~G336.40+0.10) beside the best-fit spectral model (grey band) and ata in green for its TeV counterpart HESS~J1632--478 from \citet{hessgps2018}. The blue flux error for $E< 1$\,GeV is the additional systematic error as discussed in Section~\ref{sec:sys}.}\label{fig:hess_fermi_g336.4+0.10_sed}
\end{figure*} 


\smallskip
{\bf G332.50--0.30 (RCW~103) and G332.50--0.28:} 
RCW~103 is observed as a $\sim 5^\prime$ X-ray SNR shell with an inner PWN by \cha \citep{rcw1032015}. A second X-ray PWN powered by the pulsar J1617--5055 is positionally coincident with the \fermi source, 4FGL~J1616.2--5054e, which is characterized as a radial disk with $r = 0.32\,\degree$ in the 4FGL \citep{lande2012,4fgl-dr3}. We report extension results for this source using a radial Gaussian template with $r = 0.31\,\degree$ and an off-set position that is closer to both PWNe, see Figure~\ref{fig:hess_j1616-508}, left panel. The PWN associated with PSR~J1617--5055 is $\sim 1^\prime$ in length observed in X-rays \citep{karg2009}, while RCW~103 is $\sim 10^\prime$ in size \citep{rcw1032015}. The two systems are only $0.02\,\degree$ from one another in location, making it impossible to distinguish the more likely $\gamma$-ray emitter. The TeV counterpart, HESS~J1616--508 is notably smaller and closer to PSR~J1617--5055 than to RCW~103. The best-fit spectral index for the GeV source is $\Gamma = 1.98 \pm 0.03$, which is comparable to the TeV index $\Gamma = 2.32 \pm 0.06$ \citep{hessgps2018}, see Figure~\ref{fig:hess_j1616-508}, right panel. It seems probable that the origin of the GeV and TeV emission is from one or both PWNe \citep{fges2017}, in which case we still consider the extended source 4FGL~J1616.2--5054e a likely \fermi PWN, while noting that there is potentially more than one X-ray counterpart. Future high-energy studies should explore this region further. 

\smallskip
{\bf G336.40+0.10:} HESS~J1632--478 was first discovered as an extended ($\sim 12^\prime$) unidentified TeV source \citep{aha2006}. An X-ray PWN was subsequently identified by \xmm observations as a point source accompanied by faint, diffuse non-thermal X-ray emission extending outward $\sim 32^{\prime\prime}$ in size \citep{balbo2010}. The X-ray nebula is  in positional coincidence with the H.E.S.S. emission in addition to possible extended GeV emission.  The X-ray, GeV, and TeV extended emission favor a PWN scenario. In the \fermi catalogs, the extended GeV emission corresponding to 4FGL~J1631.6--4756e is fit as a radial disk with $r = 0.25\,\degree$ \citep{fges2017,4fgl-dr2}. Re-analysis of the \fermi emission concludes that a radial Gaussian template of size $r = 0.19\,\degree$ can significantly improve the fit. HESS~J1632--478 is $\sim0.21\,\degree$ in size \citep{hessgps2018}, in good agreement with the extension we find. The TeV spectral index $\Gamma = 2.52 \pm 0.06$ and the spectral index measured by \fermi $\Gamma = 2.05 \pm 0.04$ also connect well, encouraging a single origin, see Figure~\ref{fig:hess_fermi_g336.4+0.10_sed}, left panel. Since it is highly likely the GeV source is the counterpart of the TeV PWN, we classify 4FGL~J1631.6--4756e as a likely \fermi PWN. 

\subsubsection{New PWN candidates}\label{sec:pwnc}

\begin{figure*}[hbtp]
\begin{minipage}[b]{.5\linewidth}
\hspace{-0.5cm}
\includegraphics[width=0.97\linewidth]{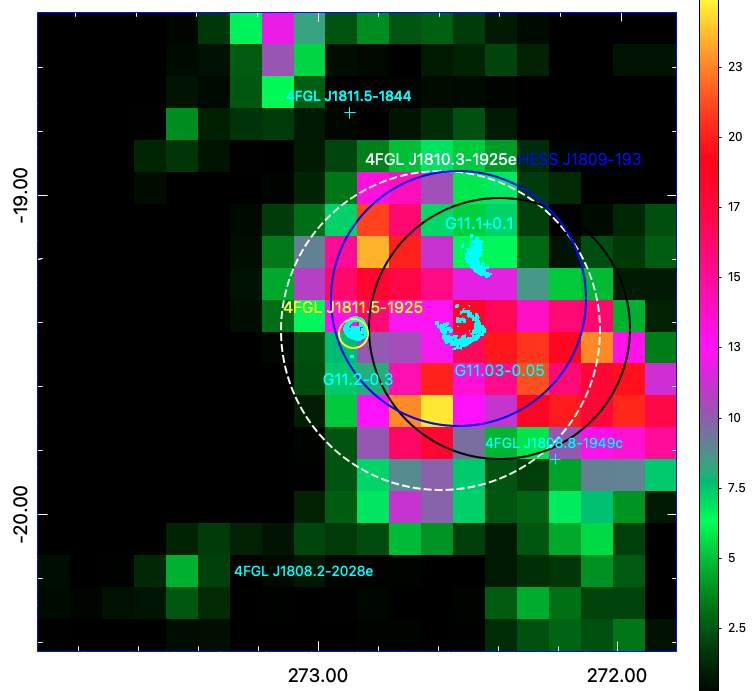}
\end{minipage}
\begin{minipage}[b]{.5\linewidth}
\hspace{-0.5cm}
\includegraphics[width=1.1\linewidth]{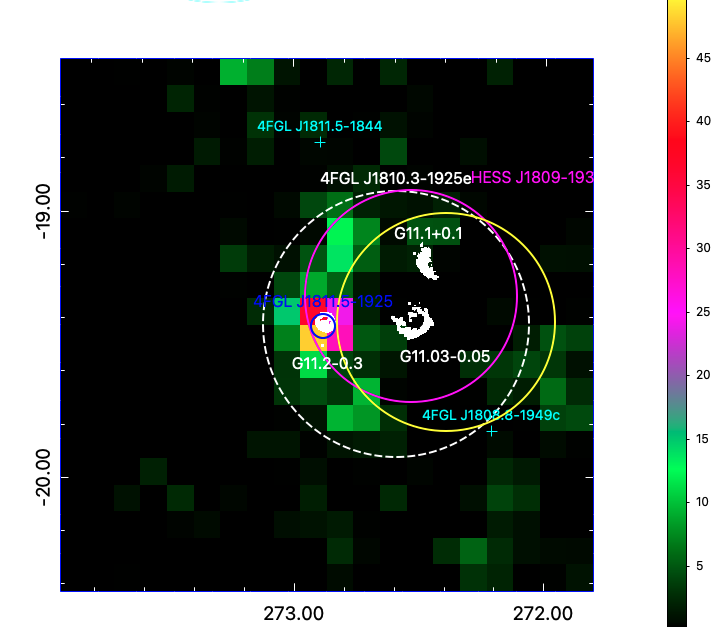}
\end{minipage}
\caption{{\it Left:} A $2\,\degree \times 2\,\degree$ 1--10\,GeV TS map of \texttt{PSF3} events for plerionic SNRs G11.03--0.05 and G11.09+0.08 (denoted by radio contours in cyan). G11.2--0.3 is also plotted with its radio contours and is coincident with 4FGL~J1811.5--1925. An unidentified, extended TeV source HESS~J1809--193 is coincident with all three PWNe, displayed in blue. 4FGL~J1810.3--1925e is indicated but not included in the source model. The best-fit radial Gaussian template for the extended emission is the black circle (yellow in the right panel). The maximum TS is $\sim$25 in the 1--10\,GeV energy range. {\it Right:} A $2\,\degree \times 2\,\degree$ 300\,MeV--2\,TeV TS map for PWN~G11.18--0.35. The 95\% uncertainty region for 4FGL~J1811.5--1925 (blue circle) is coincident with the radio position of SNR~G11.2--0.3 (radio contours in white). Unidentified HESS~J1809--193 is displayed as the magenta circle with radius $r = 0.4\,\degree$. The maximum TS at the PWN/SNR position is TS $\sim$ 48. In both panels, unrelated 4FGL sources in the field of view are labeled in cyan.}\label{fig:g11.0_g11.1_g11.2}
\end{figure*}

{\bf G11.03--0.05, G11.09+0.08, and G11.18--0.35:} G11.03--0.05 and G11.09+0.08 are two uncertain SNR/PWN candidates. 
Both were first identified in \citet{brogan2004} using Very Large Array (VLA) observations at 1465\,MHz along with the Giant Metrewave Radio Telescope (GMRT) at 235\,MHz. G11.18--0.35 belongs to the composite SNR G11.2--0.3 (the ``Turtle'') and is powered by pulsar PSR~J1811--1925. The two SNR candidates, G11.03--0.05 and G11.09+0.08, in addition to their neighbor, SNR~G11.2--0.3, are indicated by plotting their VLA radio contours in both panels of Figure~\ref{fig:g11.0_g11.1_g11.2}. All three plerionic SNRs are coincident in location with an unidentified extended 4FGL source J1810.3--1925e. A second point source is additionally coincident with G11.2--0.35 (see Figure~\ref{fig:g11.0_g11.1_g11.2}, right panel). In our analysis of the \fermi data, we tested for multiple point sources replacing 4FGL~J1810.3--1925e, but one extended \fermi source is clearly required with only one additional point source, 4FGL~J1811.5--1925. The extended source is required to model significant extended $\gamma$-ray emission, though with a location off-set from 4FGL~J1811.5--1925 and G11.2--0.35 and is fit as a radial Gaussian, $r = 0.41\,\degree$. For comparison, the 4FGL catalogs model the extended emission as a radial disk with $r = 0.5\,\degree$, such that the extension encapsulates all three SNRs \citep{araya2018}. With the new template and location, 4FGL~J1810.3--1925e is found to overlap with the PWN candidates G11.09+0.08 and G11.03--0.05 and the unidentified TeV source HESS~J1809--193 that has a Gaussian extension $r = 0.40\,\degree$ \citep{hessgps2018}. 


It is possible that the coincident \fermi and TeV extended emission originate from one or both G11.03--0.05 and G11.09+0.08 PWN candidates. It is probable that 4FGL~J1811.5--1925 is associated with G11.18--0.35. HESS~J1809--193 may be unidentified, but it has spectral properties consistent with a PWN origin \citep[$\Gamma = 2.38 \pm 0.07$,][]{hessgps2018}. 
Both G11.03--0.05 and G11.09+0.08 are therefore considered possible PWN radio counterparts to the extended GeV emission reported here. 

Due to the crowded region, it is difficult to systematically characterize the J1810.3--1925e spectrum (see the online figure set). The uncertain nature of 4FGL~J1810.3--1925e is indicated with the flags 3, 5, and 6 in the 4FGL \citep{4fgl-dr4}. Flag 3 corresponds to the source flux changing with another model or analysis by more than $3\,\sigma$. Flag 5 indicates that the region is confused with a nearby brighter neighbor. Flag 6 indicates an interstellar gas clump may be coincident, and thus is labeled with the ``c'' identifier, signifying that the 4FGL source is a candidate in nature, and may actually be part of the Galactic diffuse background. For this reason, the systematic study performed in this work cannot be applied to this source, as it relies on the use of alternative background models. Understanding the true nature of the \fermi extended source emission with respect to the background is required.

4FGL~J1811.5--1925 is best fit as a power-law with $\Gamma = 2.03 \pm 0.01$ and is coincident with the position of the young and energetic SNR~G11.2--0.3. The radio-quiet central pulsar, PWN, and host SNR are detected in X-ray \citep{madsen2020}. The SNR shell exhibits non-thermal X-ray emission which suggests evidence for efficient particle acceleration of electrons to multi-TeV energies \citep{madsen2020}. Both the 4FGL catalogs \citep{4fgl-dr4} and \citet{hessgps2018} classify the 4FGL source a pulsar candidate. The 4FGL finds significant curvature of the source spectrum, but this may depend on the location of J1810.3--1925e with respect to J1811.5--1925. 
The young age and the energetic nature of the pulsar, PWN, and SNR shell require a broadband investigation to determine the most probable scenario of the observed $\gamma$-ray emission.

\begin{figure*}[ht]
\begin{minipage}[b]{.5\linewidth}
\centering
\includegraphics[width=1.0\linewidth]{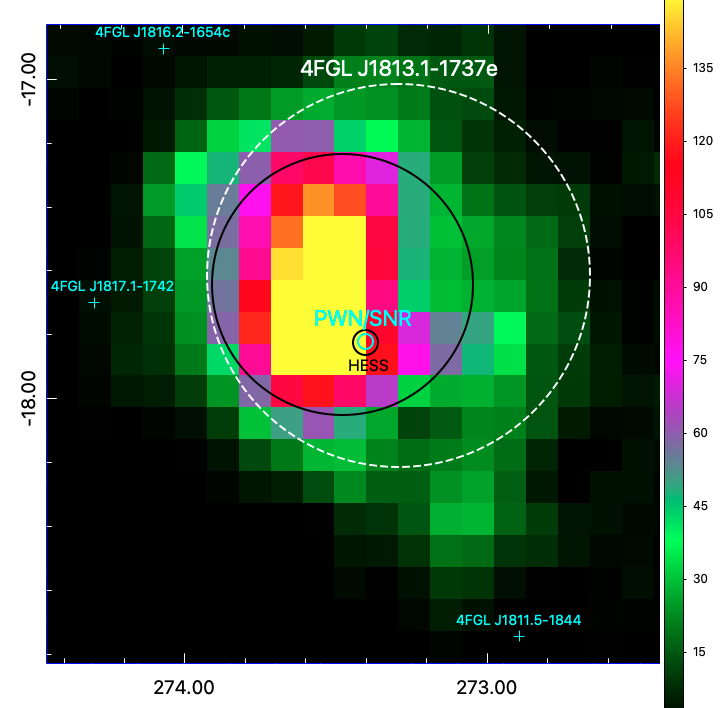}
\end{minipage}
\begin{minipage}[b]{.5\linewidth}
\centering
\includegraphics[width=1.0\linewidth]{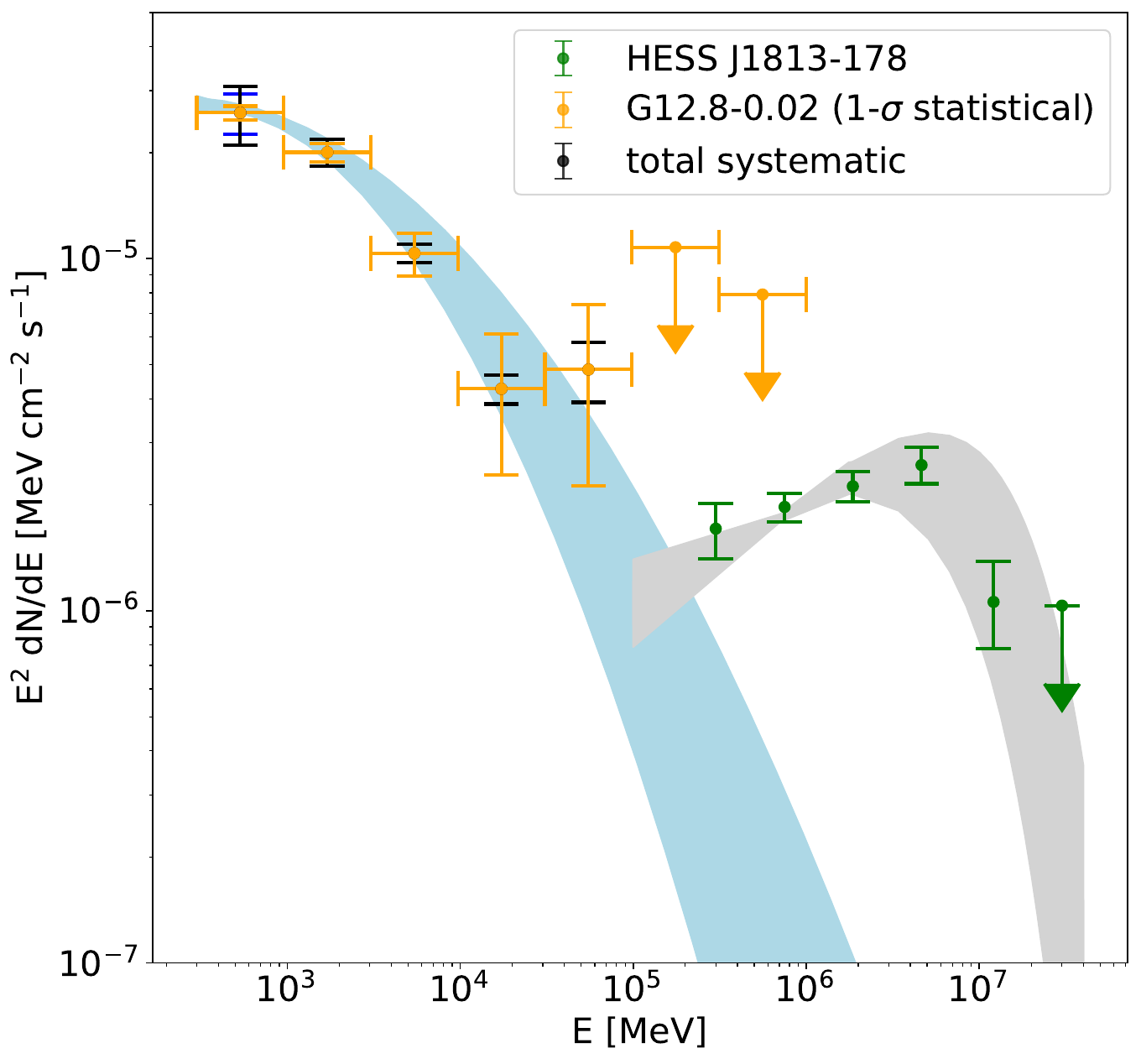}
\end{minipage}
\caption{{\it Left:} A $2\,\degree \times 2\,\degree$ 1--10\,GeV TS map of \texttt{PSF3} events for PWN~G12.82--0.02. The SNR shell is $\sim 3^\prime$ in diameter (cyan circle) which embodies PSR~J1813--1749 and the X-ray PWN. The TeV PWN HESS~J1813--178 location and size is marked in black. The 4FGL~J1813.1--1737e position and extent is indicated by the white dashed circle but is not included in the source model. The best-fit position and extent of the radial Gaussian template is indicated by the larger black circle. The maximum TS is $\sim$ 194. {\it Right:} The best-fit \fermi spectral model (blue band) and data in yellow for the extended source coincident with PWN~G12.82--0.2 (J1813.1--1737e) beside the best-fit spectral model (grey band) and data in green of its TeV counterpart HESS~J1813--178 from \citet{hessgps2018}. The blue flux error for $E< 1$\,GeV is the additional systematic error as discussed in Section~\ref{sec:sys}.}\label{fig:hess_j1813-178}
\begin{minipage}[b]{.5\linewidth}
\centering
\includegraphics[width=1.0\linewidth]{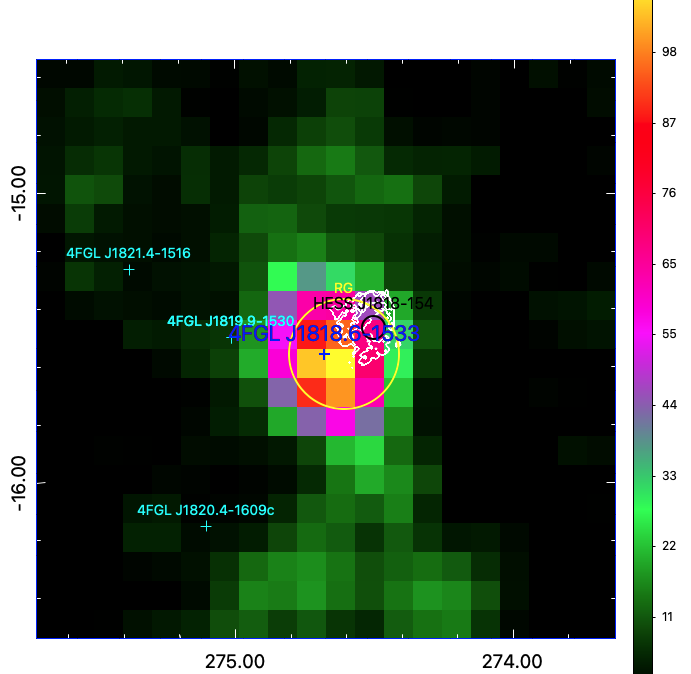}
\end{minipage}
\begin{minipage}[b]{.5\linewidth}
\centering
\includegraphics[width=0.97\linewidth]{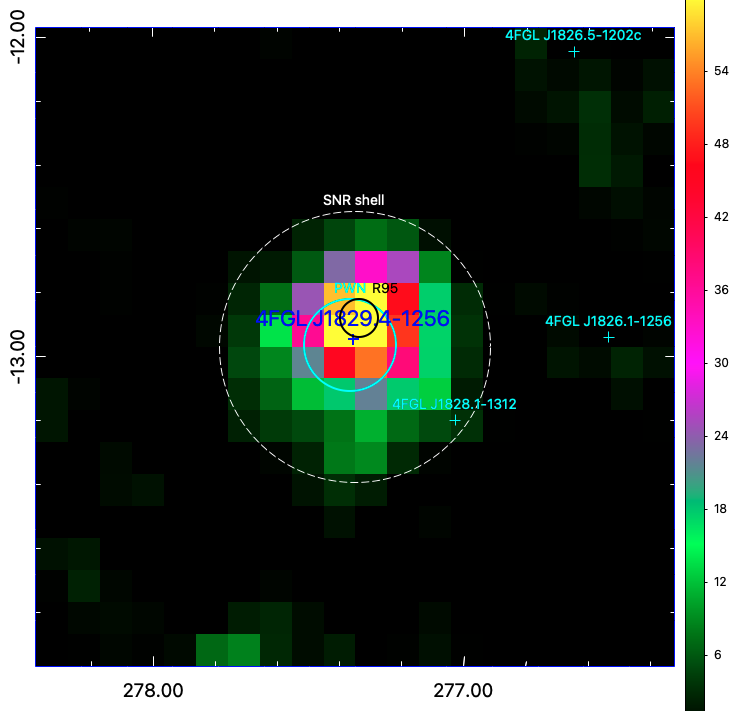}
\end{minipage}
\caption{{\it Left:} A $2\,\degree \times 2\,\degree$ 1--10\,GeV TS map of \texttt{PSF3} events for PWN~G15.40+0.10. The Gaussian extent of 4FGL~J1818.6--1533 is in yellow. The white contours represent the radio SNR. HESS~J1818--154 is located at the PWN near the core of the SNR shell in black. The maximum TS at the PWN/SNR position is $\sim$ 110. {\it Right:} A $2\,\degree \times 2\,\degree$ 1--10\,GeV TS map for PWN~G18.90--1.10. The location and size of the radio PWN is represented as a cyan circle. The white dashed circle corresponds to the size of the SNR shell in radio. The 95\% positional uncertainty region for a point source modeling emission associated with 4FGL~J1829.4--1256 is marked as a black circle.  The maximum TS at the PWN position is $\sim$ 78. Unrelated 4FGL sources are indicated in cyan.}\label{fig:g15.4_g18.9}
\end{figure*} 

\smallskip
{\bf G12.82--0.02:} G12.82--0.02 is a composite SNR housing the central pulsar J1813--1749 that powers the nebula. The source was first detected in the TeV band as HESS~J1813--178 \citep{hess2005} and later identified as a PWN after being identified in radio and X-ray \citep{g12_2005}. The TeV PWN is extended $r = 0.04\,\degree$ and is consistent with the full extent of the SNR shell $r \sim 0.025\,\degree$ in radio \citep{hessgps2018}. An extended \fermi source, 4FGL~J1813.1--1737e overlaps in location, but exhibits a much larger extension $r = 0.6\,\degree$ as a radial disk in the 4FGL catalogs \citep{araya2018,4fgl-dr4}. The extended emission can be better fit with a source closer to the SNR position and modeled as a radial Gaussian and similar extent, $r = 0.41\,\degree$, see Figure~\ref{fig:hess_j1813-178}, left panel. The larger extension in the GeV band compared to the X-ray and TeV morphologies is notably different, as well as the GeV and TeV spectral shapes, see Figure~\ref{fig:hess_j1813-178}, right panel, indicating multiple source contributions are present. 
Recently, \citet{wach2023joint} found compelling evidence that the H.E.S.S. emission has more than one contributing component, such that it can be decomposed into a compact Gaussian component ({\it A}) and a halo-like diffuse component ({\it B}). The authors concluded that the compact source should be the TeV PWN counterpart, while the halo-like diffuse component is reasonably associated with 4FGL~J1813.1--1737e.


\smallskip
{\bf G15.40+0.10:} The radio shell of SNR G15.40+0.10 is coincident with the X-ray position and size ($\sim 0.1\,\degree$) of a PWN, making it a composite SNR. The source is also detected in TeV as point-like HESS~J1818--154 and HAWC~J1819--150\footnote{The HAWC counterpart is possibly contaminated by a close neighbor, see \citet{abeysekara2017}.}. The H.E.S.S. emission is classified as a composite SNR and may have contribution from both the SNR shell and the PWN \citep{supan2015}. An unknown 4FGL source J1818.6--1533 is coincident with G15.40+0.10. A re-analysis of the $\gamma$-ray emission finds evidence for extension TS$_{\text{ext}} = 20.6$ using a radial Gaussian template, $r = 0.19\,\degree$, see Figure~\ref{fig:g15.4_g18.9}, left panel. The TeV spectral index $\Gamma = 2.21 \pm 0.15$ is similar to the GeV spectral index $2.65 \pm 0.07$ \citep{hessgps2018}, though the GeV spectrum indicates a softer spectrum with curvature (Table~\ref{tab:tev_table}). No pulsar is yet identified, but there is observational evidence suggesting molecular interactions with the SNR shell that may at least partially explain the high-energy emission \citep{hess2014,supan2015}.

\begin{figure*}[htbp]
\begin{minipage}[b]{0.5\linewidth}
\centering
\includegraphics[width=1.0\linewidth]{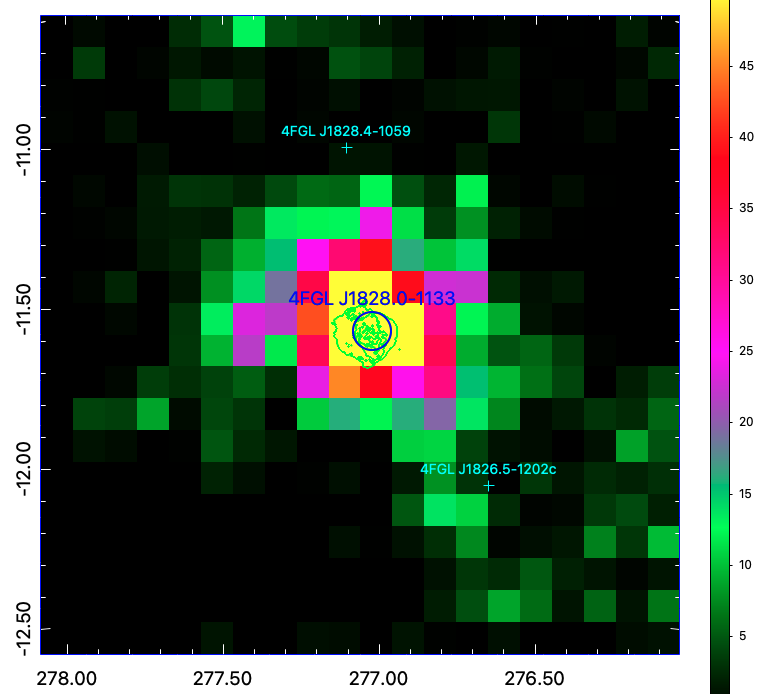}
\end{minipage}
\begin{minipage}[b]{.5\linewidth}
\centering
\includegraphics[width=0.97\linewidth]{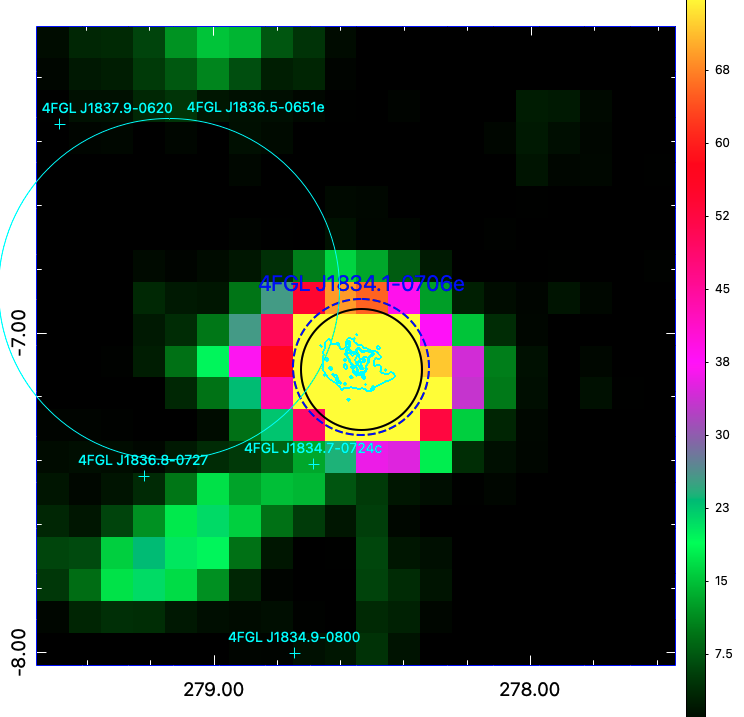}
\end{minipage}
\caption{{\it Left:} A $2\,\degree \times 2\,\degree$ 300\,MeV--2\,TeV TS map of \texttt{PSF3} events for PWN~G20.20--0.20. The 95\% uncertainty region for the coincident source 4FGL~J1828.0--1133 is in blue. The green contours represent the radio SNR and the central peak corresponds to the PWN. The maximum TS at the PWN/SNR position is $\sim$ 88. {\it Right:} A $2\,\degree \times 2\,\degree$ 1--10\,GeV TS map for PWN~G24.70+0.60. The 4FGL~J1834.1--0706e position and extent is indicated by the blue dashed circle but is not included in the source model. The best-fit position and extent of the radial Gaussian template is indicated by the black circle.  The cyan contours represent the Crab-like SNR in radio. The maximum TS at the PWN/SNR position is $\sim$ 204. Unrelated 4FGL sources are indicated in cyan.}\label{fig:g20.2_24.7}
\end{figure*}

\smallskip
{\bf G18.90--1.10:} G18.90--1.10 is a likely PWN based on \cha X-ray observations of a possible point source accompanied by an extended nebula that is $\sim 8^\prime$ in size \citep{tullmann2010}. The central pulsar has not been detected. The larger SNR shell is nearly $0.5\,\degree$ in size in radio \citep{tullmann2010}. An unidentified source 4FGL~J1829.4--1256 is coincident in position to the X-ray PWN, see Figure~\ref{fig:g15.4_g18.9}, right panel. Given the observed $\gamma$-ray emission is point-like, localized well within the SNR radio shell, and has a Log Parabola spectral index $\Gamma = 1.46 \pm 0.24$, we classify this source as a PWN candidate, depending on confirming the X-ray counterpart as a PWN, though it is possible the central pulsar may have some contribution to the observed $\gamma$-ray emission.

\smallskip
{\bf G20.20--0.20:} While the central pulsar has not been identified, the SNR shell and PWN are detected in X-ray, where the SNR morphology is probably the result of the shell interacting with ambient molecular clouds \citep{petriella2013}. A thermal component to the diffuse X-ray emission in the center of the SNR shell is not identified, instead the overlapping nonthermal radio nebula favors a synchrotron origin such as a PWN \citep{becker1985,petriella2013}. An unidentified 4FGL source 4FGL~J1828--1133 coincides with the entire SNR system, see Figure~\ref{fig:g20.2_24.7}, left panel. It is likely the source is associated with the PWN/SNR system in some way, but because the SNR may be interacting with its surroundings, a detailed analysis considering the broadband properties will be required to determine the most likely origin between the central pulsar, PWN, and SNR. The best-fit spectrum is curved with a spectral index $\Gamma = 2.27 \pm 0.13$ at 1\,GeV, favoring a SNR or pulsar origin.  

\smallskip
{\bf G24.70+0.60:} This Crab-like SNR is observed as a bright, centrally peaked core roughly $\sim 3^{\prime}$ in size and is encompassed by an incomplete shell with size $\sim 30^\prime$ in radio at 20\,cm \citep{becker1987}. The central pulsar is not known, but the radio morphology and spectral properties strongly suggest a PWN origin given the flat radio spectrum and linear polarization from the core \citep{becker1987}. A \fermi source 4FGL~J1834.1--0706e is positionally coincident with G24.70+0.60 of similar size and is classified as an SNR in the 4FGL catalogs \citep{4fgl-dr4}. Prior work has argued an SNR origin based on the possible association of a nearby extended TeV source, MAGIC~J1835--069 \citep{fges2017, g24.72019}, motivated by similar spectral indices and the possibility of the SNR shell interacting with denser material. The best-fit spectral index characterizing the $\gamma$-ray emission is $\Gamma = 2.02 \pm 0.04$ for 4FGL~J1834.1--0706e and $\Gamma \sim 2.75$ for MAGIC~J1835--069 \citep{g24.72019}. It is argued that CRs accelerated by the SNR shell diffuse and illuminate a nearby cloud coincident with the TeV emission. Other investigations into the region favor a star forming region origin \citep[e.g.,][]{katsuta2017}.  While it is possible the GeV and TeV sources are associated, we question an SNR origin since the SNR shell is not firmly detected at any wavelength, whereas the polarized, centrally-peaked radio core is and possibly suggests a PWN origin instead. The source 4FGL~J1834.1--0706e is modeled as a radial disk with $r = 0.21\,\degree$ in 4FGL and we find the extended emission is fit marginally better as a radial Gaussian with $r = 0.19\,\degree$, see Figure~\ref{fig:g20.2_24.7}, right panel. The region is complex in more than just the \fermi band, which prevents a reliable classification of this source. It seems plausible that the PWN is at least partially contributing to the observed $\gamma$-ray emission, but a deeper analysis considering multiwavelength data of the region is needed. 

\begin{figure}[htbp]
\centering
\includegraphics[width=1.0\linewidth]{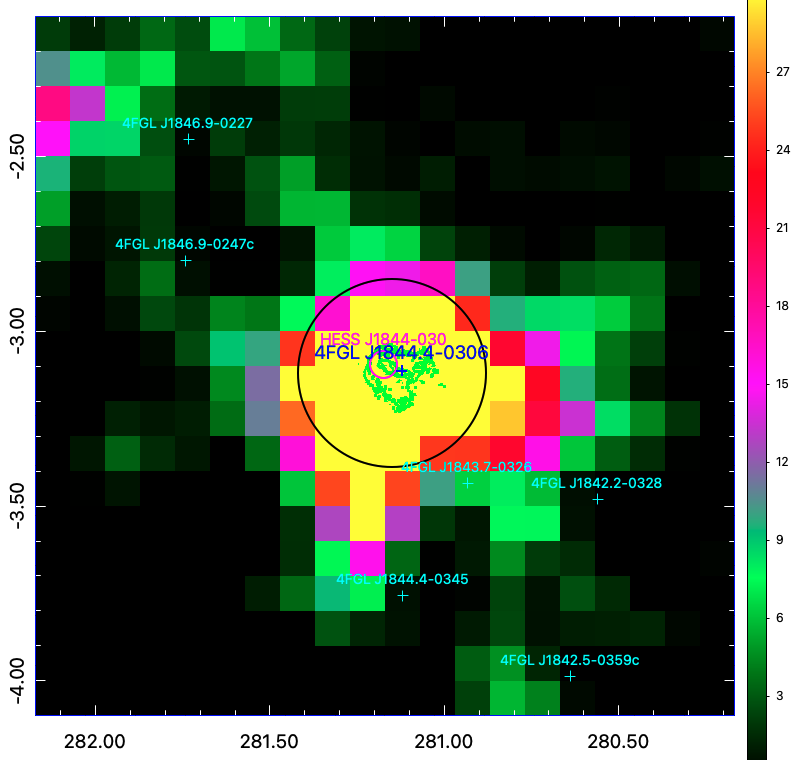}
\caption{A $2\,\degree \times 2\,\degree$ 1--10\,GeV TS map of \texttt{PSF3} events for PWN~G29.40+0.10. The best-fit position and extent of the Gaussian source 4FGL~J1844.4--0306 is in black.  The green contours represent the SNR in radio. The UNID HESS~J1844--030 is marked in magenta. The maximum TS at the PWN/SNR position is $\sim$ 67. Unrelated 4FGL sources are indicated in cyan.}\label{fig:g29.4}
\end{figure}

\smallskip
{\bf G29.40+0.10:} This is a possible composite SNR identified in radio, X-ray, GeV, and TeV energies \citep{petriella2019}. The radio observations suggest the SNR is expanding into a non-uniform medium and the X-ray observations show a morphology and spectrum suggestive of a PWN origin with an embedded point source that may be the central pulsar. An unknown 4FGL source J1844.4--0306 is found in the location of the SNR and PWN. We find evidence for extension TS$_{\text{ext}} = 18.8$ for a Gaussian template with $r = 0.27\,\degree$, see Figure~\ref{fig:g29.4}. The TeV counterpart HESS~J1844--030 is unidentified and point-like, having a spectral index $\Gamma = 2.48 \pm 0.12$ \citep{hessgps2018}. The GeV spectral index is similarly soft, $\Gamma = 2.55 \pm 0.06$, indicating a common origin. A PWN scenario can plausibly explain the TeV emission. Recent work \citep{zheng2023} reports either a hadronic or a leptonic scenario for the TeV emission is possible while a hadronic component is most likely for the GeV.

\begin{figure*}[htbp]
\begin{minipage}[b]{.5\linewidth}
\centering
\includegraphics[width=1.0\linewidth]{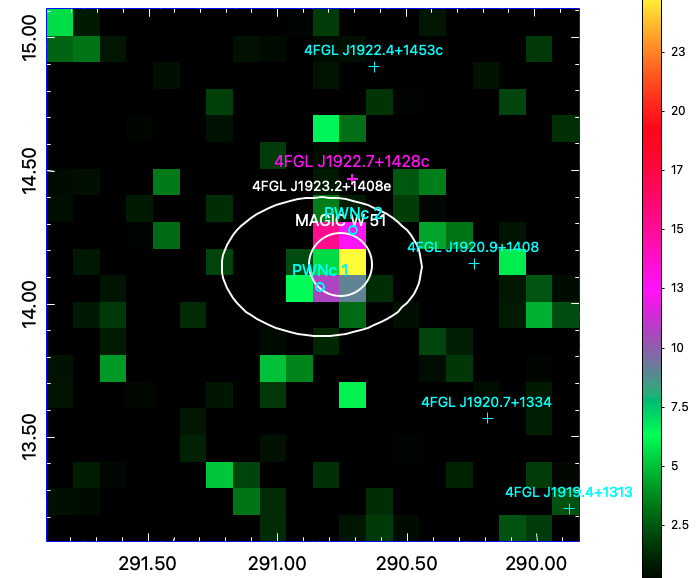}
\end{minipage}
\begin{minipage}[b]{0.5\linewidth}
\centering
\includegraphics[width=0.9\linewidth]{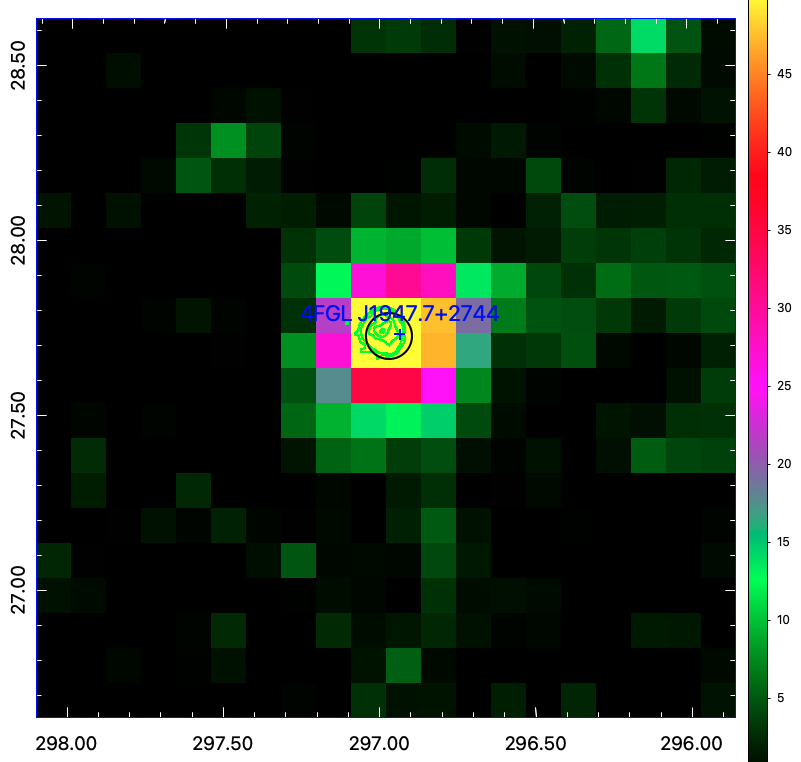}
\end{minipage}
\caption{{\it Left:} A $2\,\degree \times 2\,\degree$ 10\,GeV--2\,TeV TS map of \texttt{PSF3} events for PWN candidates G49.20--0.70 (``PWNc 1'') and G49.20--0.30 (``PWNc 2''), both overlapping SNR~W~51C. The SNR is the GeV emitter 4FGL~J1923.2+1408e. 4FGL~J1922.7+1428c is the \fermi counterpart, replaced by the two point-sources at the PWN locations. The TeV SNR is the inner white circle, which overlaps spatially with both PWN candidates. The maximum TS is $\sim$25 for $E >10$\,GeV. {\it Right:} A $2\,\degree \times 2\,\degree$ 300\,MeV--2\,TeV TS map of \texttt{PSF3} events for PWN~G63.70+1.10. There is one associated \fermi source 4FGL~J1947.7+2744. Radio contours of the SNR shell and the central PWN are indicated in green. The 95\% positional uncertainty for the point source is indicated in black. The maximum TS at the PWN/SNR position is $\sim$ 66. 
}\label{fig:w51c_g49.2_g63.7}
\end{figure*}

\smallskip
{\bf G49.20--0.30 and G49.20--0.70 (W51~C):} The SNR W51~C or G49.2--0.7 is located near a star forming region and houses a pulsar candidate CXO~J192318.5+140335 with compact X-ray emission observed by \cha \citep[core $\sim 1^\prime$,][]{chaw51c2005}. Observations performed by \xmm reveal a second PWN candidate in the SNR which has a similar extent to the other PWN, G49.20--0.30 \citep{xmmw51c2014}. The W51~C SNR has a radio diameter $\sim 1.0 \,\degree$ and is the basis for modeling \fermi extended $\gamma$-ray emission in the region coincident with this system. It was shown that the SNR is likely the $\gamma$-ray emitter with compelling evidence for hadronic CR acceleration and an extension roughly consistent to the radio SNR size \citep{fermiw51c2016}. 4FGL~J1923.2+1408e represents the SNR emission as an elliptical disk with radii 0.375\,$\degree$ and 0.26\,$\degree$, see the left panel of Figure~\ref{fig:w51c_g49.2_g63.7} \citep{abdo2009w51c,4fgl-dr2}. The extended TeV source first discovered by H.E.S.S. \citep{feinstein2009} and subsequently detected by MAGIC \citep{magic2012} is identified as an interaction between the SNR and the surrounding molecular clouds (MCs), similar to the GeV emission \citep{fermiw51c2016}. The spectrum for the \fermi and MAGIC SNR counterparts show agreement above $\sim 3$\,GeV, with a spectral index $\Gamma_\gamma \sim 2.5$. The spectral index below $\sim 3$\,GeV is harder $\Gamma_\gamma \sim 2.1$. In the 4FGL--DR4, the SNR is characterized with a Log Parabola spectrum with an index $\sim 2.2$ at $2.7$\,GeV.

Based on the smaller TeV extension $0.12 \,\degree$ compared to the SNR size observed in the GeV and radio bands, as well as its compelling overlap with both of the PWN candidates G49.20--0.70 (``PWNc 1'') and G49.20--0.30 (``PWNc 2''), a PWN contribution is plausible. Furthermore, as shown in the left panel of Figure~\ref{fig:w51c_g49.2_g63.7}, there is significant TS $\sim 25$ residual emission coincident with both PWN candidates not accounted for by 4FGL~J1923.2+1408e or the background components. We find that 4FGL~J1923.2+1408e is required to model extended emission in the region, but that 4FGL~J1922.7+1428c is the likely counterpart to G49.20--0.30, and is therefore removed from the global source model. Two point sources are tested, each fixed at the PWN X-ray positions. The addition of these two sources significantly improves the fit over the 4FGL model (TS $ = 2\log{\frac{L_{\text{2ps}}}{L_{\text{4FGL}}}} = 166$) and over a single additional extended source (TS $ = 2\log{\frac{L_{\text{2ps}}}{L_{\text{1ext}}}} = 99$) that is localized. G49.20--0.30 (``PWNc 2'')  results in a point source detection TS $= 157.2$ with a Log Parabola spectral index $\Gamma = 2.35 \pm 0.06$. ``PWNc 1'' or G49.20--0.70 is detected with TS $= 28.0$ and has a power-law spectral index $\Gamma = 2.38 \pm 0.09$. We therefore classify the two new point sources (replacing 4FGL~J1922.7+1428c) as tentative PWN detections. A deeper analysis considering multiwavelength information is needed to determine the capacity for either PWN candidate to emit $\gamma$-rays, as it is possible that one or both sources may be components of the SNR.

\smallskip
{\bf G63.70+1.10:} G63.70+1.10 is the central PWN to SNR~G63.7+1.1, though the central pulsar remains unknown. X-ray observations revealed a point source embedded within a diffuse non-thermal X-ray nebula, which coincides with the bright radio core of the SNR $\sim 8^\prime$ in diameter \citep{matheson2016}. The PWN has a possible \fermi association 4FGL~J1947.7+2744. Re-analysis of the region confirms that there is point-like $\gamma$-ray emission in the vicinity of the PWN/SNR system, see Figure~\ref{fig:w51c_g49.2_g63.7}, right panel. We consider the source association tentative for the PWN. This is motivated by the SNR being located among dense material and possibly interacting with local clouds \citep{matheson2016}. Considering the spectral properties of the $\gamma$-ray source such as its best-fit photon index $\Gamma = 1.84 \pm 0.22 $, and the age of the system \citep[$\tau \gtrsim 8\,$kyr,][]{matheson2016}, it seems equally likely for the pulsar, PWN, or the SNR to be the \fermi counterpart. 

\begin{figure*}[htbp]
\begin{minipage}[b]{.5\linewidth}
\centering
\includegraphics[width=0.9\linewidth]{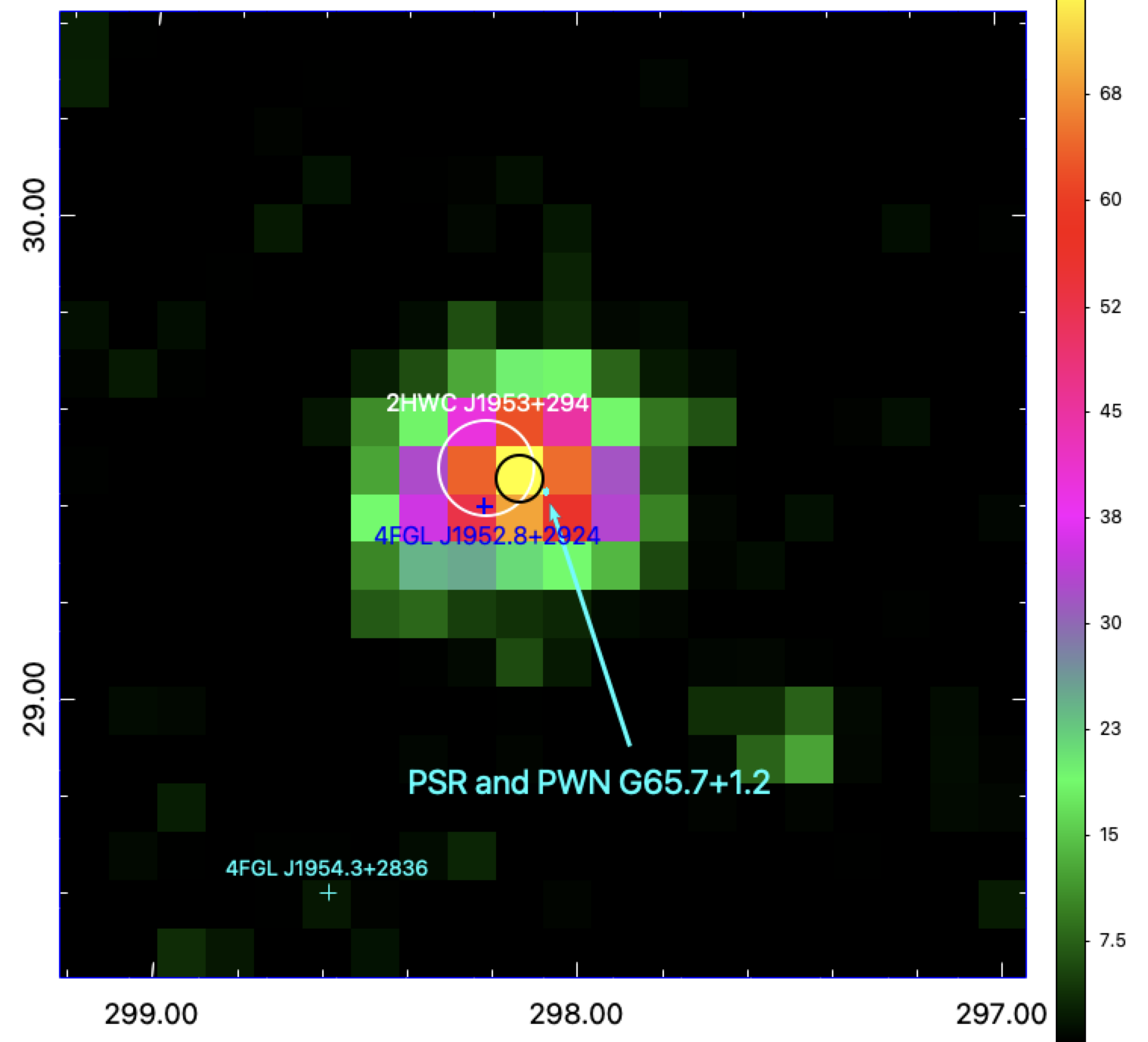}
\end{minipage}
\begin{minipage}[b]{.5\linewidth}
\centering
\includegraphics[width=1.0\linewidth]{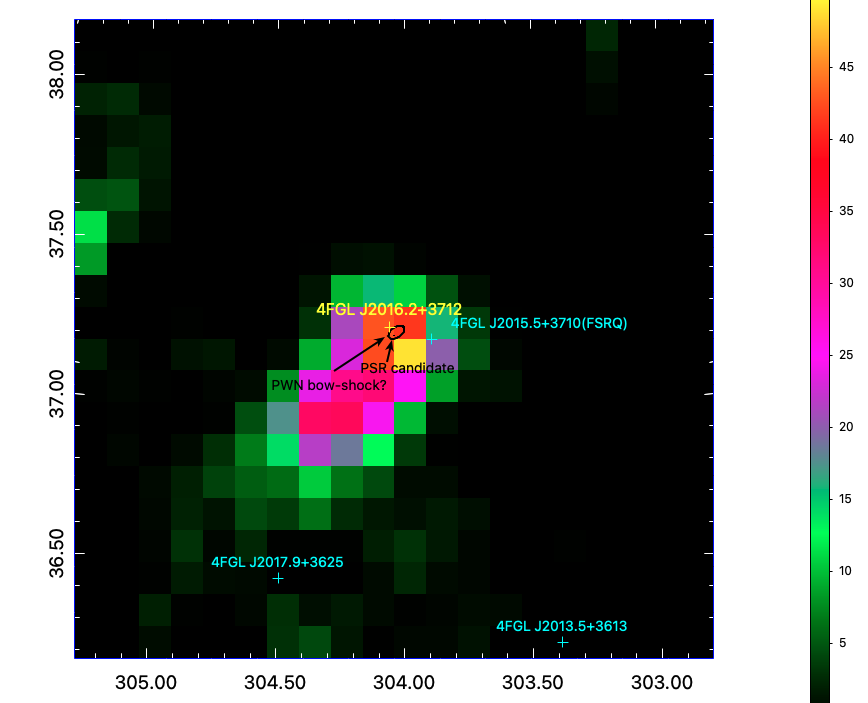}
\end{minipage}
\caption{{\it Left:} A $2\,\degree \times 2\,\degree$ 300\,MeV--2\,TeV TS map of \texttt{PSF3} events for PWN~G65.73+1.18. The pulsar and PWN in X-ray as observed by \cha are denoted with cyan contours and are highlighted using the cyan arrow and label. The 95\% uncertainty region of a point source at the PWN position is indicated in black and corresponds to the best-fit position of 4FGL~J1952.8+2924. The maximum observed extension of the TeV PWN 2HWC~J1953+294 is indicated in white. The maximum TS at the PSR/PWN position is $\sim$ 89. {\it Right:} A $2\,\degree \times 2\,\degree$ 1--10\,GeV TS map of \texttt{PSF3} events for PWN~G74.94+1.11. The \fermi counterpart 4FGL~J2016.2+3712 is shown in yellow. The X-ray PSR and PWN position and extent are shown in black and highlighted using the black arrow. The maximum TS at the source position is $\sim$ 50. Unrelated 4FGL sources are indicated in cyan.}\label{fig:65.7_ctb87}
\begin{minipage}[b]{.5\linewidth}
\centering
\includegraphics[width=0.97\linewidth]{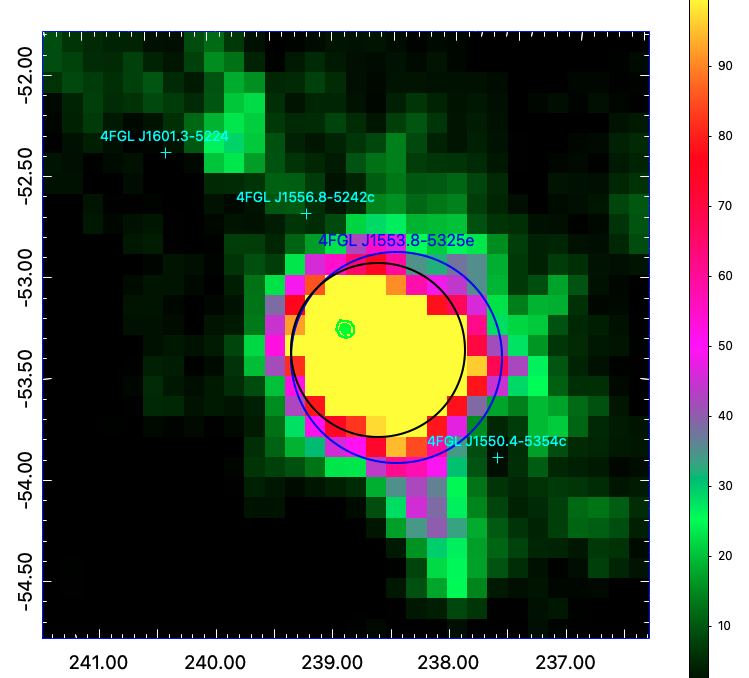}
\end{minipage}
\begin{minipage}[b]{.5\linewidth}
\centering
\includegraphics[width=1.0\linewidth]{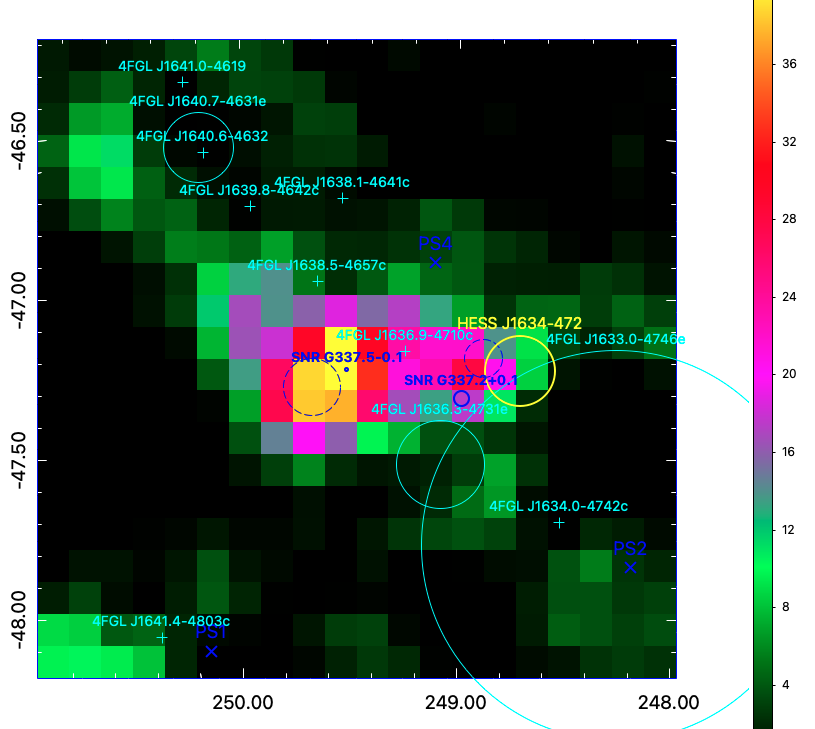}
\end{minipage}
\caption{{\it Left:} A $2\,\degree \times 2\,\degree$ 1--10\,GeV TS map of \texttt{PSF3} events for PWN~G328.40+0.20. The Crab-like radio SNR is shown as the green contours. 4FGL~J1553.8--5352e is shown in blue but is not included in the source model. The best-fit Gaussian source is marked in black. The maximum TS at the PWN/SNR position is $\sim$ 300. {\it Right:} A $2\,\degree \times 2\,\degree$ 1--10\,GeV TS map of \texttt{PSF3} events for PWN candidates G337.20+0.10 and G337.50--0.10. The PWN positions and approximate sizes are indicated with solid blue circles. The best-fit positions with 95\% positional uncertainty are shown as the dashed blue circles. 4FGL~J1636.3--4731e is the unrelated SNR~G337.1--0.1. The unidentified TeV source HESS~J1634--472 corresponds to the yellow circle. The maximum TS at the PWN position is TS $\sim 45$ for 1--10\,GeV. 
Unrelated 4FGL sources are labeled in cyan.}\label{fig:g328_g337.5}
\end{figure*}

\begin{figure}[ht]
\centering
\includegraphics[width=1.0\linewidth]{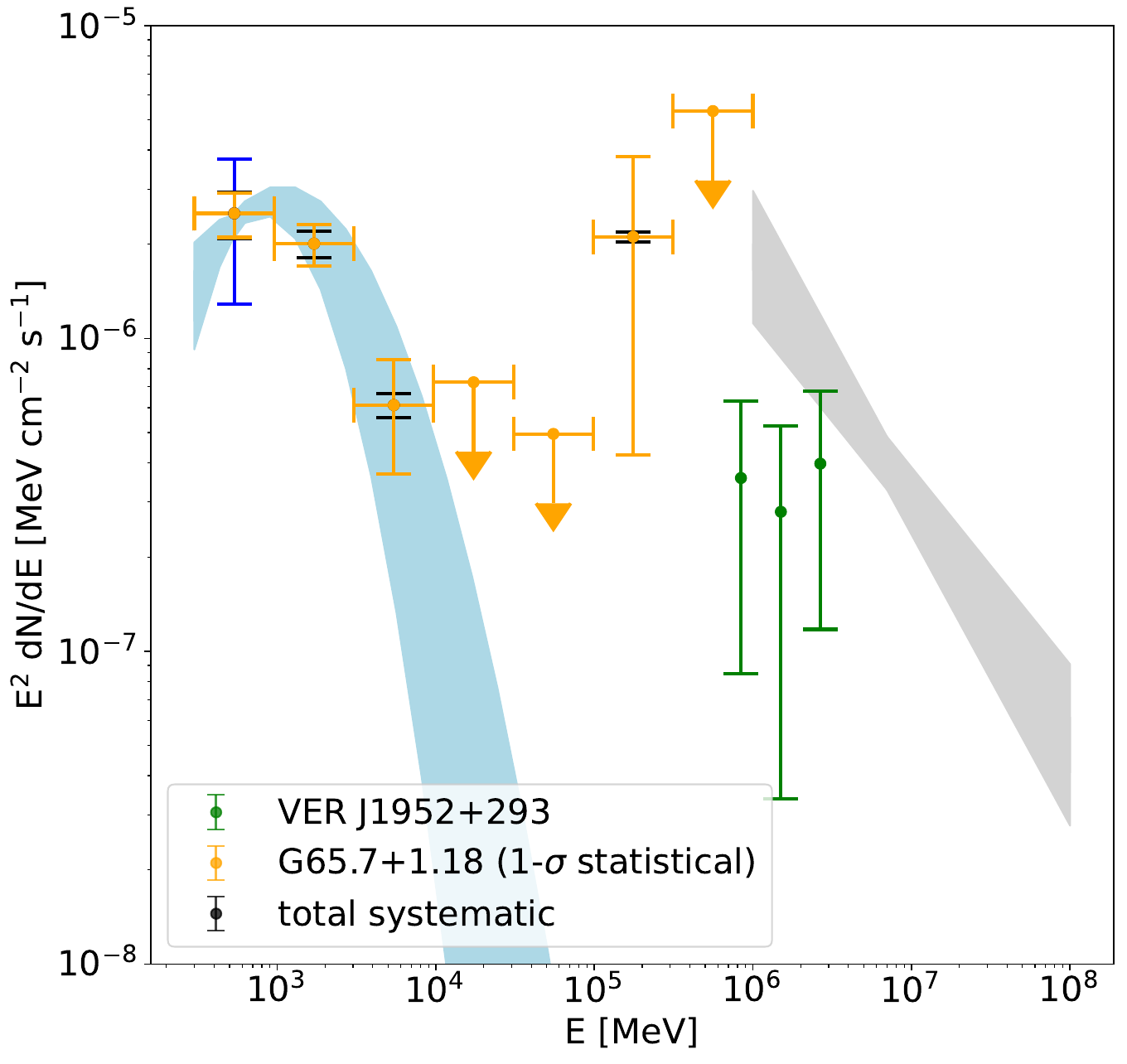}
\caption{The best-fit \fermi spectral model (blue band) and data in yellow for the source coincident with PWN~G65.73+1.18 (J1952.8+2924) beside the best-fit spectral model of 2HWC~J1953+294 in grey and the data of VER~J1952+293 is in green \citep{abeysekara2018}. The blue flux error for $E< 1$\,GeV is the additional systematic error as discussed in Section~\ref{sec:sys}.}\label{fig:2hwc_j1953+294}
\end{figure}

\smallskip
{\bf G65.73+1.18:} This source is another Crab-like PWN with an identified TeV counterpart. {\it NuSTAR} and \cha observations revealed a presumed pulsar located at the emission peak of a compact X-ray nebula within a radius $r \sim 20^{\prime\prime}$, overlapping in location with the previously unknown TeV source 2HWC~J1953+294 \citep[shown in Figure~\ref{fig:65.7_ctb87}, left panel,][]{cover2019}. A spectral index $\sim 2.0$ characterizes X-ray emission from 2--20\,keV with no apparent spectral cutoff and the 2--10\,keV luminosity is $\approx 10^{31}\,$erg s$^{-1}$, similar to the 1--30\,TeV luminosity. The TeV emission has no evidence for extension observed by HAWC, but the VERITAS counterpart VER~J1952+293 has a Gaussian extension $r = 0.14\,\degree$ \citep{abeysekara2018}. 
The observed $\gamma$-rays from the {\it Fermi}--LAT, HAWC, and VERITAS suggest a PWN origin \citep{abeysekara2018}. An unidentified \fermi source 4FGL~J1952.8+2924 is plausibly the GeV counterpart due to positional coincidence (see Figure~\ref{fig:65.7_ctb87}, left panel). The TeV counterpart's spectral index $\Gamma = 2.78 \pm 0.15$ \citep{hawc2017} is somewhat softer than the Log Parabola GeV spectral index $\Gamma = 2.13 \pm 0.19$. VER~J1952+293 has a spectral index $\Gamma = 2.65 \pm 0.49$, but with fluxes that are lower than what is observed by HAWC, see Figure~\ref{fig:2hwc_j1953+294}. This may be in part due to the different source sizes observed \citep[for more details see][]{abeysekara2018}. We hence classify 4FGL~J1952.8+2924 as a PWN candidate based on the similar position and energetics of the Crab-like SNR observed in each waveband, however, we cannot rule out a pulsar contribution.

\smallskip
{\bf G74.94+1.11:} Also known as CTB~87, the SNR~G74.9+1.2 has a filled center in X-ray with no identified shell, hosting PWN G74.94+1.11 and pulsar candidate CXOU~J201609.2+371110 \citep{ctb872020}. 4FGL~J2016.2+3712 is coincident, see Figure~\ref{fig:65.7_ctb87}, right panel. Point-like TeV emission VER~J2016+371 is a possible counterpart with a power-law spectral index $\Gamma_\gamma = 2.3 \pm 0.4$ \citep{verj20162014}, similar to the \fermi index $\Gamma_\gamma = 2.24 \pm 0.41$. An SNR interaction with molecular material is possible \citep{ctb872018} and recently pulsations from CXOU~J201609.2+371110 (now PSR~J2016+3711) are detected in radio \citep{ctb872024}. No pulsations in the $\gamma$-rays are found in \citet{ctb872024}, but it is possible the \fermi emission may be contributions from both a pulsar at low-energies, and a PWN or SNR at high-energies.


\smallskip
{\bf G328.40+0.20:} This is a radio-bright Crab-like SNR having a diameter $\sim 5^\prime$ \citep{gaensler2000}. The PWN is also detected in X-ray with a size $\sim 1^\prime$ in diameter \citep{gelfand2007}. Bright, extended GeV emission 4FGL~J1553.8--5352e overlaps with the source region \citep{fges2017}, characterized as a Gaussian, $r = 0.43\,\degree$, see Figure~\ref{fig:g328_g337.5}, left panel. An additional point source at the PWN location is tested, yielding TS $\sim 9$. The extension and significant emission present above 20\,GeV favors a PWN origin. A pulsar contribution is possible but not dominant. The large GeV extension compared to the radio PWN/SNR size, however, prevents a confident classification.

\smallskip
{\bf G337.20+0.10:} Another crowded region in the \fermi sky, the plerionic SNR~G337.2+0.1 may be contributing to the observed $\gamma$-ray emission. In X-ray the SNR appears Crab-like with non-thermal emission forming a central peak, presumably the pulsar and PWN, accompanied by an uncertain shell \citep{combi2006}. The diffuse X-ray emission attributed to a nebula is $\sim 1.5^\prime$ in size. An unidentified extended TeV source HESS~J1634--472 is in the vicinity and possibly associated \citep{hessgps2018}, see Figure~\ref{fig:g328_g337.5}, right panel.  A GeV extended source 4FGL~J1636.3--4731e is also closeby, but likely unrelated as the source coincides better to  SNR~G337.0--0.1, both in size and location \citep{fges2017}. 
Adding an additional point source at the G337.20+0.10 position 
is detected at TS $ = 18.4$ (see Figure~\ref{fig:g328_g337.5}, right panel). The best-fit photon index is $\Gamma = 2.26 \pm 0.12$. We note there is no 4FGL counterpart to this source in the 4FGL--DR4 catalog. 

\smallskip
{\bf G337.50--0.10:} The \cha X-ray Survey unveiled a probable bow-shock PWN powered by central pulsar candidate CXOU~J163802.6--471358 \citep{jakobsen2014}. The X-ray nebula extends on the sub-arcminute scale ($\sim40^{\prime\prime}$). The position and approximate size of the nebula is displayed in Figure~\ref{fig:g328_g337.5}, right panel. There is a close candidate source in the 4FGL--DR3, 4FGL~J1638.4--4715c, which we find is the likely GeV counterpart to this system and yields TS $= 42.7$ and power-law spectral index $\Gamma = 2.69 \pm 0.10$ when localized to the PWN, a significant improvement to the 4FGL--DR2 fit. 
We consider the $\gamma$-ray source 4FGL~J1638.4--4715c a possible counterpart to G337.50--0.10, but it is a tentative detection due to it being a heavily crowded region in the Galactic plane that is among uncertain diffuse residual $\gamma$-ray emission.

\subsubsection{Nondetections of PWNe}\label{sec:nondetect}
There are 19 PWNe and PWN candidates that have been identified at radio, X-ray, and TeV bands that are not significantly detected in the \fermi data.  All of the undetected PWNe are listed in Table~\ref{tab:nondetect} along with their measured TS and the 95\% C.L. flux upper limit for the 300\,MeV--2TeV energy range. Another source is listed in Table~\ref{tab:nondetect}: G25.10+0.02. The reported nondetection refers to the removal of a second extended PWN candidate (4FGL~J1838.0--0704e) in the region associated with TeV PWN HESS~J1837--069 (see also HESS~J1837--069 in Section~\ref{sec:new}).

Identifying and characterizing PWNe inside $\gamma$-ray bright SNRs is challenging. 
The PWN G266.97--1.00 is powered by the pulsar PSR~J0855--4644 and together they lie on the SW edge of the Vela Jr. SNR \citep{acero2013b}. 
The Vela Jr. SNR is a GeV emitter, 4FGL~J0851.9--4620e and is nearly $1\,\degree$ in radius.
A TS signal is exclusively detected at $E < 1\,$GeV in the nearby region of G266.97--1.00. This would not be unusual for \fermi pulsars, but may also be explained as structure associated with the Galactic diffuse emission that is not being accounted for properly by the background. 
The extended TeV counterpart RX~J0852.0--4622 to the SNR was investigated for a PWN or pulsar contribution, but disentangling TeV emission components was not feasible \citep{arribas2012}. A deeper analysis exploring both the GeV and TeV extended emission may provide better insight to any additional significant emission in the region \citep[e.g.,][]{donath2024}. 

A similar case is seen in the ROI for IC~443. 
There is a PWN candidate G189.10+3.00 observed in X-ray that overlaps with two SNRs including IC~443 and the unconfirmed SNR candidate G189.60+3.30 \citep{ic4432004,ic4432018}. 
We place an additional point source at the PWN position, which yields TS $= 26$ and spectral index $\Gamma = 2.02 \pm 0.09$ for energies 300\,MeV--2\,TeV, 
but we do not attempt to analyze this region further due to the complexity. A dedicated study is required to reliably determine any potential PWN presence within IC~443 in the \fermi data.



\fermi data analysis is even more complicated within $\sim 1.0\,\degree$ of the Galactic center, where at least four PWN candidates are located: SNR~G00.00+0.00, PWN~G0.13--0.11, PWN~G359.90--0.04, and PWN~G358.50--0.96. The last source may have a possible extended GeV counterpart 4FGL~J1745.8--3028e \citep{marchesi2024}. The ROIs are heavily crowded due to the proximity to the Galactic center, introducing convergence problems in the global fits as well as making source identification extremely difficult if not impossible. Hence we do not analyze or discuss any of the PWN candidates located within $\sim 1.0\,\degree$ from the Galactic center. 



Aside from the uncertainties in the ROIs just described, the regions for undetected PWNe and PWN candidates are relatively sparse, making any source detection fairly straightforward. In all cases, a point source is added to the PWN position assuming a power-law spectrum. The TS results for each source are provided in Table~\ref{tab:nondetect}. In the Appendix (Table~\ref{tab:spectral_ul_fluxes}), we provide the 95\% C.L. upper limits on the flux for nine energy bins of each undetected PWN. 

\subsection{Summary of Results}
In the 58 regions analyzed, we detect 9 unidentified $\gamma$-ray sources that we classify as likely PWNe and present them in the top panel of Table~\ref{tab:points} (6 point-like detections) and in the middle panel of Table~\ref{tab:extent} (3 extended detections, excluding RCW~103). 
These sources 
likely have a PWN contribution, particularly at high energies ($E>10$ GeV), but need follow-up multiwavelength studies in order to determine and characterize the $\gamma$-ray origin. The three likely extended $\gamma$-ray PWNe would bring the total number of extended \fermi PWNe with no detectable $\gamma$-ray pulsar from 6 to 9, and 
the entire extended \fermi PWN population from 12 to 15\footnote{The \fermi PWN 3C~58 is extended in the 4FGL--DR4 catalog, based on the work of \citet{li2018}.}. 
There are currently no identified point-like \fermi PWNe in the 4FGL catalogs \citep{4fgl-dr4}. 
%

We classify another 21 $\gamma$-ray sources as PWN candidates and list them in the bottom panels of Tables~\ref{tab:points} and \ref{tab:extent} based on whether they are observed as point-like (15/21) or extended (6/21). The dominant $\gamma$-ray origin for the 21 weaker PWN candidates is less clear, either due to the lack of a TeV PWN counterpart and(or) being associated with a PWN/SNR system that is energetic in nature, where the pulsar, PWN, or SNR shell could plausibly explain the high-energy emission. The most interesting cases in both the likely and weak PWN candidate classes are described in more detail in Section~\ref{sec:results_by_roi}. There are 19 PWNe where no significant residual emission is detected. They are listed in Table~\ref{tab:nondetect} and discussed in Section~\ref{sec:nondetect}. In total, there are 36 detected $\gamma$-ray sources of which three coincide with two lower-energy PWN counterparts each (see Table~\ref{tab:extent}). 

\begingroup
\begin{table}[h]
\scalebox{0.9}{
\hspace{-1.75cm}
\begin{tabular}{lcc}
\hline
\textbf{Criterion} & \textbf{Likely PWN} & \textbf{Candidate PWN} \\
 & (Total = 9) & (Total = 21) \\
\hline
TeV counterpart present & 7 & 8 \\
TeV source identified as PWN & 6 & 2 \\
MWL$^a$ study favors PWN origin & 9 & $\sim$3 \\
Confirmed MWL PWN counterpart & 9 & 8 \\
Location in LAT sky & Generally & Some in complex \\
& not complex & or uncertain regions \\
\hline
\end{tabular}}
\caption{Summary of Classification Criteria. The numbers indicate how many sources meet each criterion within the classification group. Because a single source can satisfy multiple criteria, the values do not sum to the total number of sources. \footnotesize{$^a$ multiwavelength (MWL).}}\label{tab:criteria}
\end{table}
\endgroup

We point out three important distinctions that separate the likely from the weak candidate class. The first distinction is that the majority of likely PWNe are coincident with TeV counterparts that are classified as PWNe based on multiwavelength investigations of the lower-energy counterparts known to be PWNe (6/9, see Table~\ref{tab:tev_table}). The second applies to the \fermi sources (2/9) that do not have identified TeV counterparts (G8.40+0.15 and B0453--685), but which have prior detailed multiwavelength analyses where a PWN origin of the \fermi $\gamma$-ray emission is preferred. The third is for the final likely PWN, G315.78--0.23, an energetic bow-shock PWN fully displaced from its SNR shell and powered by the pulsar now traveling supersonically in the ISM beyond the SNR. These distinctions are summarized in Table~\ref{tab:criteria}. An identified multiwavelength counterpart with a detailed broadband study are the primary qualifiers for likely PWNe. As a final note, in general, a confident PWN classification requires either a morphological study showing compelling correspondence \citep[e.g.,][]{li2018} or correlated variability, verified only by the flaring Crab Nebula.

\begin{figure*}[htb]
\begin{minipage}[b]{.5\linewidth}
\centering
\includegraphics[width=1.0\linewidth]{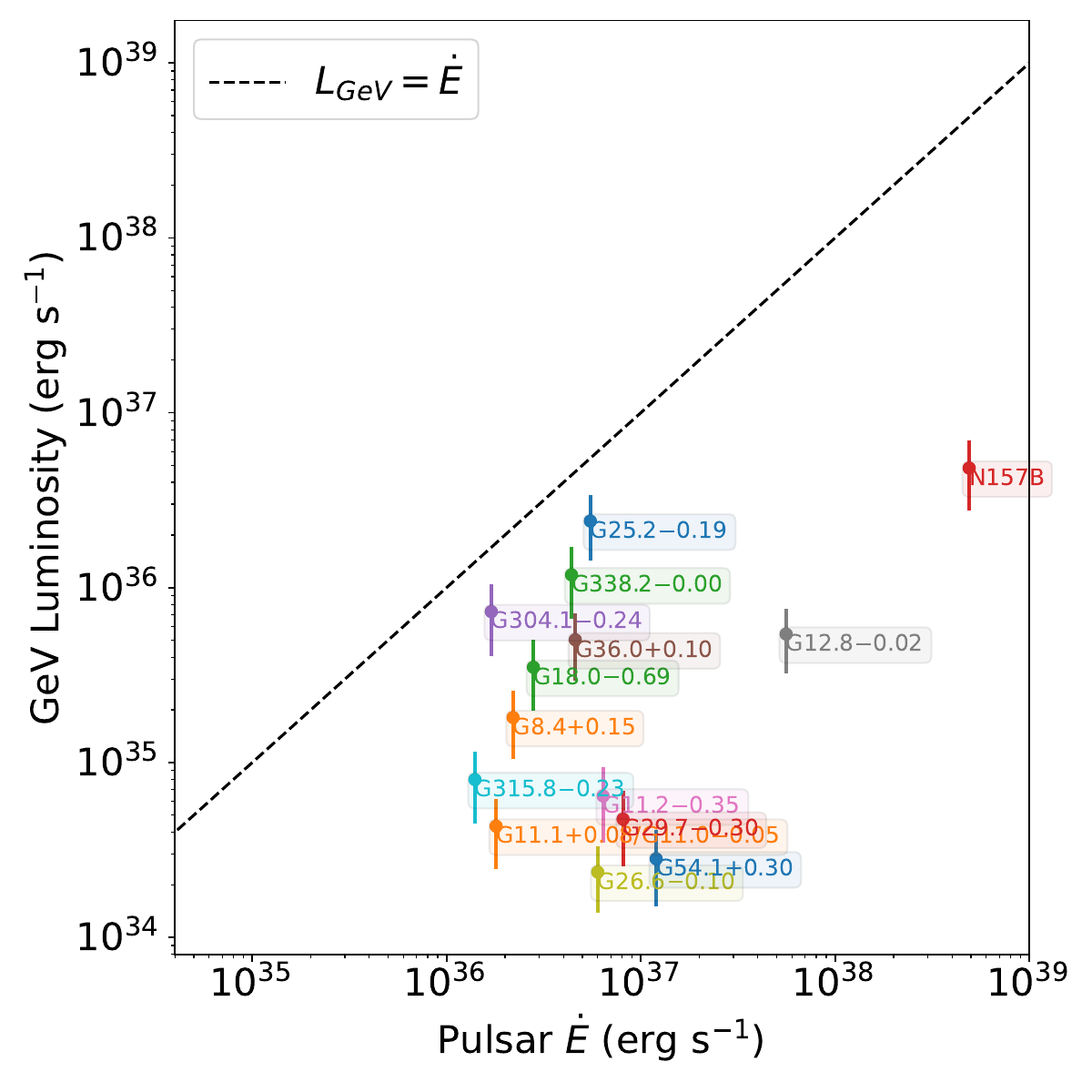}
\end{minipage}
\begin{minipage}[b]{.5\linewidth}
\centering
\includegraphics[width=1.0\linewidth]{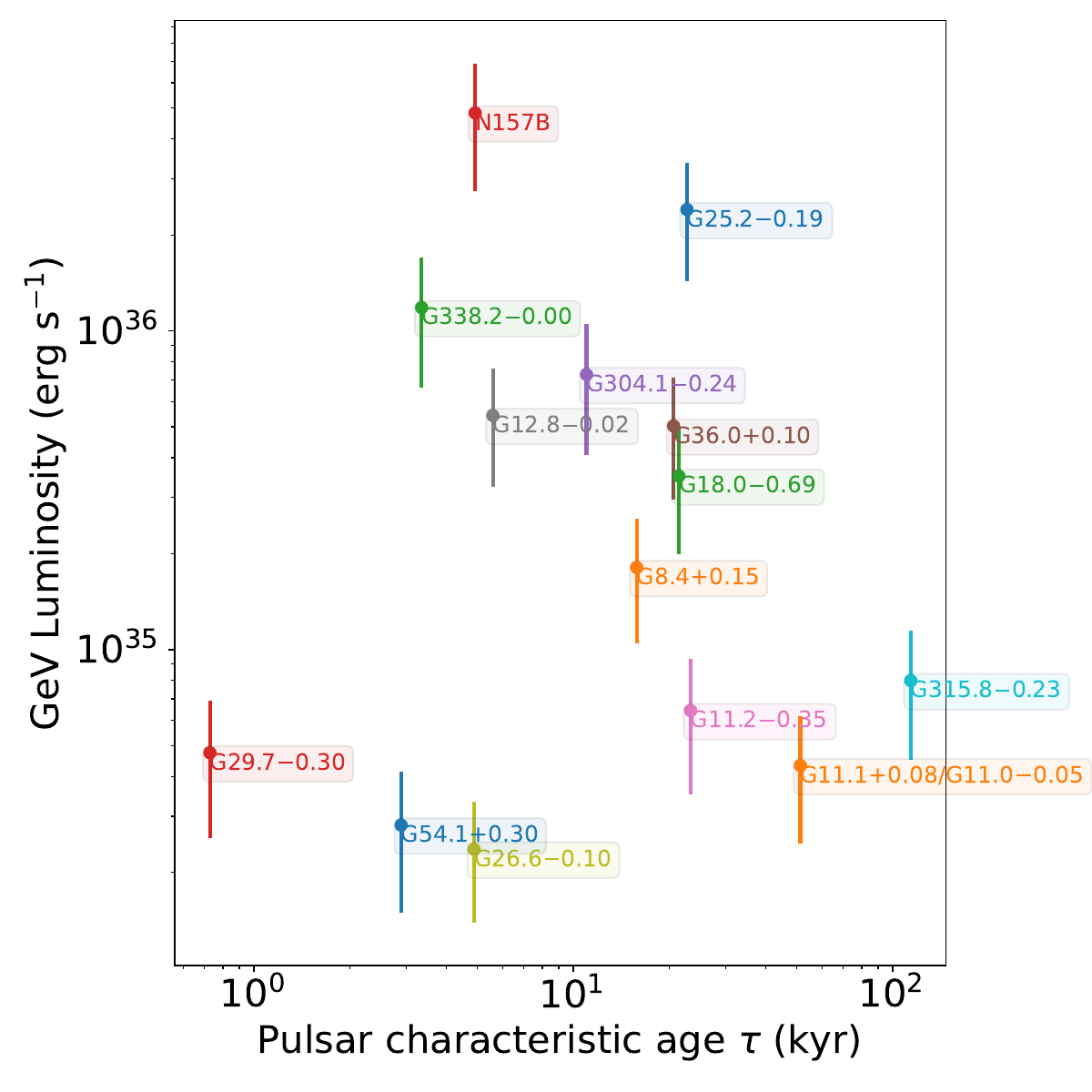}
\end{minipage}
\caption{{\it Left:} 300\,MeV--2\,TeV luminosity of the PWNe as a function of pulsar spin-down power. {\it Right:} 300\,MeV--2\,TeV GeV luminosity as a function of pulsar characteristic age.}\label{fig:gev_lum_edot_charage_plots}
\end{figure*} 

%
\section{Discussion}\label{sec:discuss}

We have systematically characterized the 300\,MeV--2\,TeV
emission from 11.5\,years of \fermi data in 58 ROIs containing known
PWNe and PWN candidates identified at other wavelengths, and the source sample analyzed includes 6 previously identified \fermi PWNe and 8 PWN associations.

Approximately $\sim$ 40\% of the source detections reported here are found to be extended. Two sources are reported as extended that are not considered extended in previous \fermi catalogs: 4FGL~J1818.6--1533 and 4FGL~J1844.4--0306. 
One new $\gamma$-ray source considered a likely PWN is reported here that was not reported in any previous \fermi catalogs, corresponding to the PWN in the LMC, B0453--685 (Figure~\ref{fig:n157b_b0453}, right panel). If a PWN origin can be identified, B0453--685 would represent the second extragalactic $\gamma$-ray PWN to be detected at such high energies after N~157B, noting that the PWN origin of N~157B still needs to be confidently known as well. Only one candidate $\gamma$-ray source is first reported here that has no prior \fermi counterpart: 
G337.20+0.10.

In Figure~\ref{fig:gev_lum_edot_charage_plots}, left panel, is the 300\,MeV--2\,TeV GeV luminosity as a function of the associated pulsar's spin-down power $\dot{E}$ (if known). The properties of firm pulsar associations are listed in the Appendix, Table~\ref{table:psr_table}. The GeV luminosity accounts for the uncertainty on the 300\,MeV -- 2\,TeV flux in addition to assuming 20\% uncertainty on the pulsar distance, following \citet{acero2013}, which we collect from the Australia Telescope National Facility (ATNF) pulsar catalog\footnote{\url{https://www.atnf.csiro.au/research/pulsar/psrcat/}}, adopting the distances measured from dispersion measure. As what is reported in \citet{acero2013}, there is no clear correlation between the GeV luminosity and the pulsar spin-down power. In the right panel of Figure~\ref{fig:gev_lum_edot_charage_plots}, we plot the GeV luminosity as a function of pulsar characteristic age, which we also gather from the ATNF pulsar catalog. No clear correlation is found between the two properties, similar to \citet{acero2013}. 

Finally, we measure the ratio between both the 300\,MeV--2\,TeV and 1--10\,TeV $\gamma$-ray luminosity to the 2--10\,keV X-ray luminosity for those in Tables~\ref{tab:tev_table} and \ref{tab:nondetect} that have an X-ray and TeV counterpart and plot the ratio as a function of the pulsar spin-down power $\dot{E}$ to compare to the derived relation in \citet{mattana2009}. We find a similar trend as \citet{mattana2009} with one outlier from the nondetected group, G358.3+0.24, which is explained by the low $\dot{E}$ of the associated pulsar, $\dot{E} \sim 10^{34}\,$erg s$^{-1}$.

\begingroup
\begin{table*}[!htbp]
\centering
 \begin{tabular}{c|c|c|c|c|c|c}
 \hline
 \hline
 PWN Name & Type of Candidate & $\dot{E}$ & $\tau_c$ & $d$ & PWN size & References \\
 & Likely (L) or Weak (W) & (erg s$^{-1}$) & (kyr) & (kpc) & ($\degree$) & \\
  \hline    
 
  G11.18--0.35 & W & 6.4$\times10^{36}$& 29.88& 3.7& 0.011 & \citet{madsen2020}  \\
      & & & & & & \citet{ranasinghe2022distances}  \\
      \hline
  G12.82--0.02 & W & 5.6$\times10^{37}$& 5.6 & 6.2& 0.05 & \citet{joshi2023}  \\
    & & & & & & \citet{hessgps2018}\\
  \hline
  G15.40+0.10 & W & 7.0$\times10^{36}$ & 17 & 9.3 & 0.14 & \citet{hess2014} \\
  & & & & & & \citet{su2017revised} \\
  \hline
  G29.70--0.30 & L & 8.1$\times10^{36}$ & 0.723 & 5.8 & 0.0083 & \citet{straal2022} \\
  \hline
  G327.15--1.04 & L & 3.1$\times10^{36}$ & 17.4 & 9 & 0.02 & \citet{temim2015} \\
   & & & & & & \citet{hessgps2018} \\
\hline
\hline
\end{tabular}
\caption{The five ROIs explored using the evolutionary model in \citet{Martin2022} and their 
observational constraints used in the model: pulsar spin-down power, distance, age, and PWN radius. The second column identifies the type of PWN classification: (L) for likely or (W) for weak. See Section~\ref{sec:rad_model} for details.
For G327.15--1.04 and G15.40+0.10, the pulsar is not known and therefore the pulsar spin-down power, age, and distance are fixed to estimates adopted from the listed references. The spin-down power of G15.40+0.10 is assumed as an appropriate value. 
}
\label{tab:torres}
\end{table*}
\endgroup

\begin{figure*}[!htbp]
\centering
\includegraphics[width=0.32\linewidth]{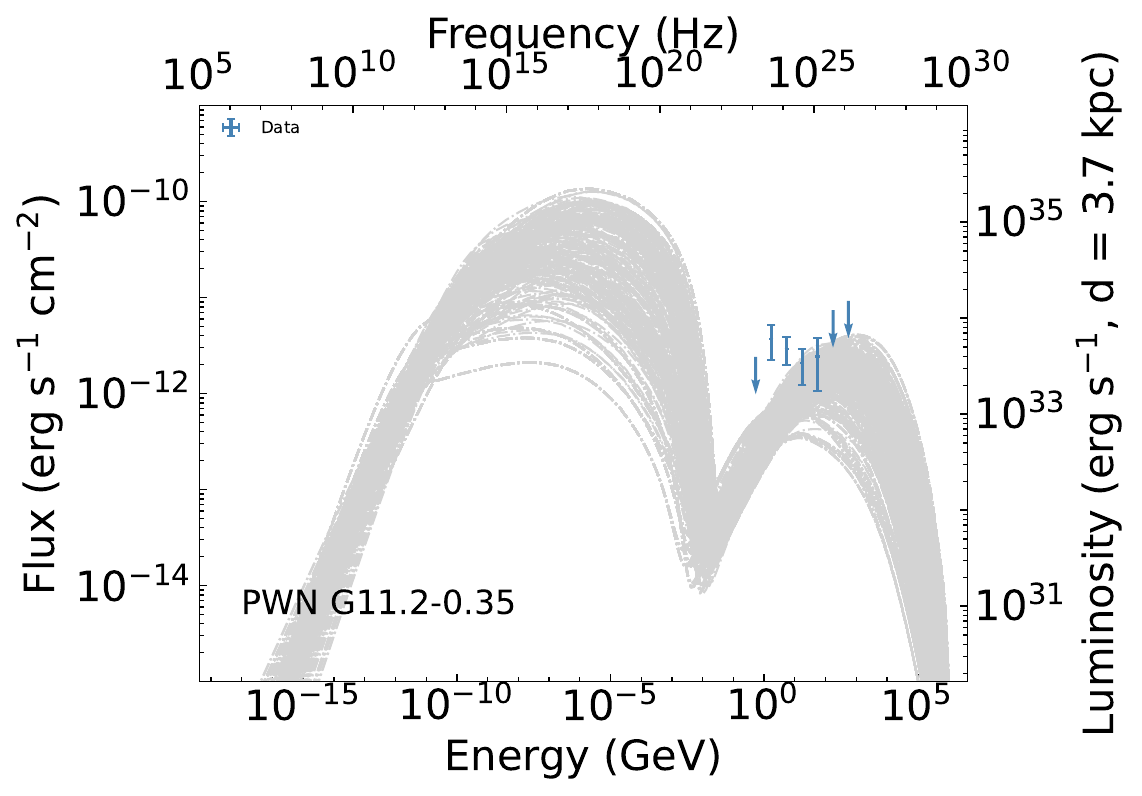}
\includegraphics[width=0.32\linewidth]{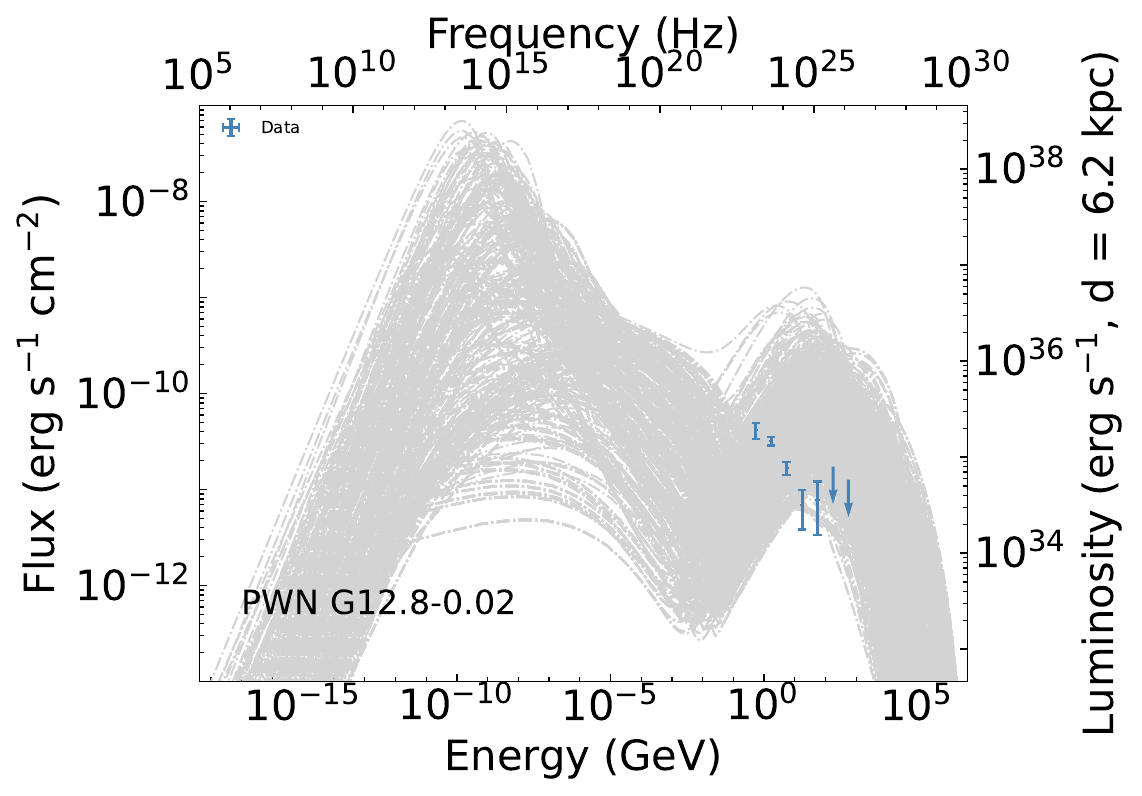}
\includegraphics[width=0.32\linewidth]{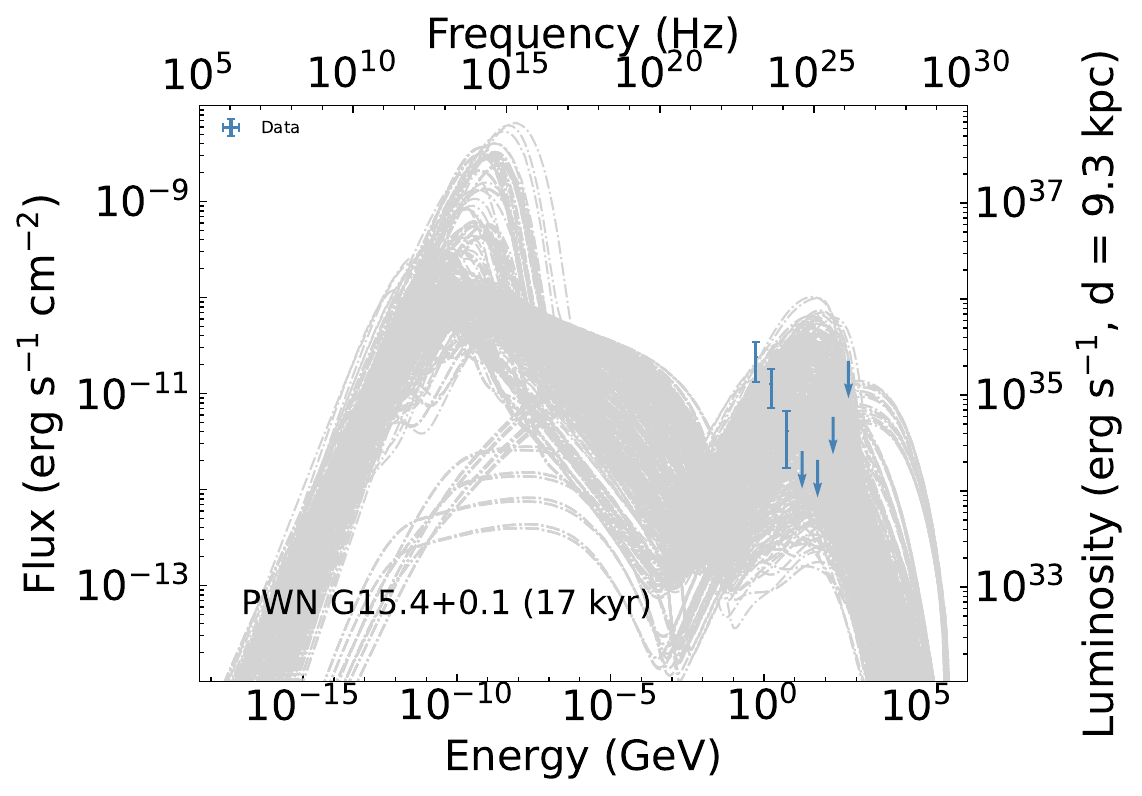}
\includegraphics[width=0.32\linewidth]{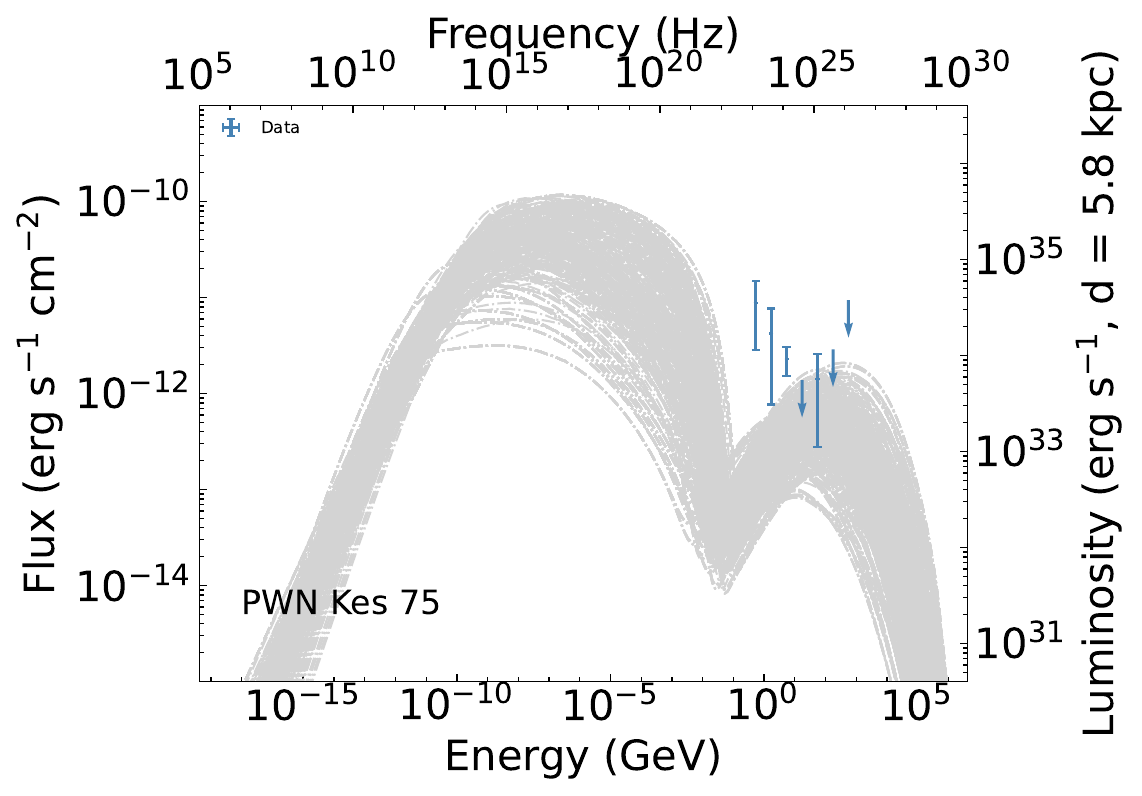}
\includegraphics[width=.32\linewidth]{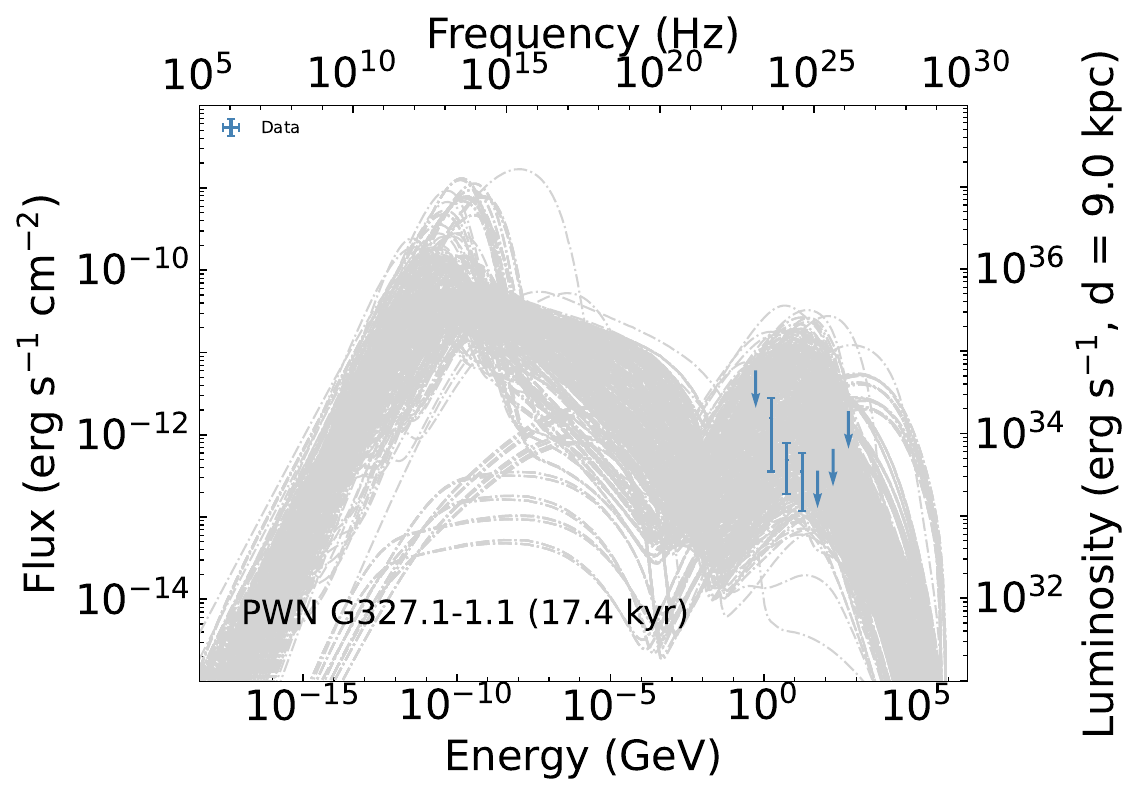}
\caption{712 (456 for G11.2--0.3) predicted SEDs from an evolutionary model \citep{Martin2022} for several of the sources studied as examples. See Table~\ref{tab:torres} for relevant parameters \citep[see also][]{zhang2024}}.
\label{fig:diego_models1}
\end{figure*}

\subsection{PWN Origin through Radiative Modeling}\label{sec:rad_model}

For some of the best candidates and a few more as examples of young PWNe,
listed in Table~\ref{tab:torres}, 
we investigate 
their possible multi-band spectra in the context of a PWN evolutionary model, as outlined in \citet{torres2014,Martin2022}.
The aim is not to fit the GeV data (for there is too scarce a number of data points) but rather to explore whether our data are consistent with model predictions using conservative ranges of free parameters. 
The predicted SEDs are obtained using the observational data detailed in Table~\ref{tab:torres} and a range of reasonable values for 9 unknown parameters \citep{zhang2024}. Assumed ranges to span possible theoretical PWNe spectra include the SN ejecta mass $M_{ej}$ between 8--15\,$M_\odot$, particle index before the break $\alpha_1$ = [1.0,1.6], break $\gamma_b$ = [$10^5$,$10^6$], FIR/NIR energy density $U_{FIR}$ = 1--3 times the GALPROP value, true age between [0.7--1.3]\,$\tau_c$ \citep[except for G11.2--0.35 with a true age of about 1.6 kyr,][]{{clark1977}}, braking index $n = 2, 2.5, 3$ \citep[except for G29.70--0.30 which has measured values before/after the burst where we adopt the post-burst index = 2.16,][]{livingstone2011}, magnetic energy fraction $\eta_B$ = 0.02--0.04, particle index after the break energy $\alpha_2$ = [2.2, 2.8], and ISM particle density $n_{ISM}$ = [0.1, 1.0]\,cm$^{-3}$. The employed parameter ranges are motivated by the observed properties of several PWNe as well as predicted values of the properties from simulations \citep[see][and references therein]{Martin2022}.

We have generated 712 possible models for each source by combining the minimum and maximum values of these 9 parameters to create 2$^{9}=512$ specific SEDs (2$^{8}=256$ for G11.2–0.35), and then adding 200 models by randomly selecting values within these intervals. 
These models span possible SEDs for these $\gamma$-ray sources
under the assumption that they are similar to other PWNe already detected by the \fermi
like the Crab or 3C~58.   
The panels of Figure~\ref{fig:diego_models1} 
show that PWN models are in general 
agreement with the observed fluxes, but the spectral slope is not reproduced for most of the models.
This 
suggests 
that either these PWNe behave differently from the rest or the $\gamma$-ray data suffer from contamination from other members of the complex (pulsar or remnant). 

\subsection{Caveats}\label{sec:caveats}

This \fermi study
is limited by the complexity of the $\gamma$-ray data including the systematic uncertainties from the diffuse Galactic background and the uncertainties from the effective area. We emphasize here that while we attempt to account for these effects, the uncertainties explored here may not fully represent the range of systematics involved \citep[see e.g., Section 2.4.2 in][]{acero2016}.  
The systematic study explored in Section~\ref{sec:sys} adopts the method from \citet{depalma2013,acero2016}. It is found that typically the systematic uncertainties on the flux dominate over the statistical uncertainties for sources that lie along the Galactic plane ($|b| < 1.0\,\degree$) and for energies below 5\,GeV. The systematic impact on source extension is on the same order as the statistical error. These findings are in agreement to prior studies \citep{acero2016,fges2017}. In general, uncertainties from the IEM model remain the most important, while consideration of the location to nearby bright sources especially those of \fermi pulsars can also be critical. Though an incomplete systematic study as mentioned above, this work is among the most detailed investigations on the systematic uncertainties for fluxes and extensions of \fermi detected PWNe.

\section{Conclusion}\label{sec:conclude}

In summary, we have verified the characterization and classification of the 6 \fermi PWNe that lack any detectable $\gamma$-ray pulsar, and show that all of them, with the exception of MSH 15--56, can be characterized using radial Gaussian templates that either provide comparable fits to a radial disk template or better. We note that two likely extended \fermi PWNe, 4FGL~J0836.5--0651e and 4FGL~J1616.2--5054e, have more than one possible PWN counterpart identified in other wavelengths (see Section~\ref{sec:new}). 
Further, we analyze 8 possible PWN associations from the \fermi catalogs and find that most of them (5/8) are likely $\gamma$-ray PWNe based on their multiwavelength properties; they are discussed in detail in Section~\ref{sec:new}. We argue that one of the associations, 4FGL~J1838.9--0704e, a second extended source possibly modeling emission from the PWN~HESS~J1837--069, is not required. 4FGL~J1836.5--0651e, when modeled as a radial Gaussian extended source, can adequately model any extended residual $\gamma$-ray emission coincident with HESS~J1837--069. 

4FGL~J1810.3--1925e overlaps two radio PWN candidates that are plausible counterparts (Section~\ref{sec:pwnc} and Figure~\ref{fig:g11.0_g11.1_g11.2}). 
In many of these less clear cases such as G11.03--0.05, G11.09+0.08, and G18.90--1.10, the source classification depends on the lower-energy counterparts being confirmed as PWNe rather than candidates. For the youngest systems such as G11.18--0.35 and G20.20--0.20, an additional challenge is ruling out contribution from the central pulsar and/or host SNR shell. 

The 9 previously unidentified $\gamma$-ray sources that are likely PWNe, if confirmed, would increase the PWN population detected by the \fermi from 12 \citep{4fgl-dr4} to 21. An additional 21 previously unidentified $\gamma$-ray sources are considered PWN candidates. 
The total number of detected $\gamma$-ray sources is 36 with three extended sources that overlap with more than one lower-energy PWN counterpart as mentioned just above. Among the 58 PWNe and PWN candidates analyzed here, 19 remain undetected. 
36/58 or 62\% of the sample selection is associated with a $\gamma$-ray source.
We have demonstrated that the \fermi data set probably contains a larger PWN population present than is currently known. Nevertheless, many of the source classifications are uncertain and hence require a justification from a thorough broadband analysis. A cursory study characterizing the radiative properties of a selection of PWN candidates (Section~\ref{sec:rad_model}) highlights the need to include broadband information. This work represents a list of ideal targets for future X-ray and TeV observations. A subsequent search will analyze the off-pulse data of \fermi detected pulsars for the presence of a PWN and will be reported in the future. 

\bigskip
The \fermi Collaboration acknowledges generous ongoing support from a number of agencies and institutes that have supported both the development and the operation of the LAT as well as scientific data analysis. These include the National Aeronautics and Space Administration and the Department of Energy in the United States, the Commissariat \a`a l'Energie Atomique and the Centre National de la Recherche Scientifique / Institut National de Physique Nucl\a'eaire et de Physique des Particules in France, the Agenzia Spaziale Italiana and the Istituto Nazionale di Fisica Nucleare in Italy, the Ministry of Education, Culture, Sports, Science and Technology (MEXT), High Energy Accelerator Research Organization (KEK) and Japan Aerospace Exploration Agency (JAXA) in Japan, and the K. A. Wallenberg Foundation, the Swedish Research Council and the Swedish National Space Board in Sweden.

Additional support for science analysis during the operations phase is gratefully acknowledged from the Istituto Nazionale di Astrofisica in Italy and the Centre National d'\'Etudes Spatiales in France. This work performed in part under DOE Contract DE- AC02-76SF00515.

This work has also been partially supported by the grant PID2021-124581OB-I00 funded by MCIN/AEI/10.13039/501100011033, 2021SGR00426, by the Spanish program Unidad de Excelencia María de Maeztu CEX2020-001058-M, and European Union NextGeneration EU funds (PRTR-C17.I1).


\software{FermiPy \citep[v.1.0.1,][]{fermipy2017}, Fermitools: Fermi Science Tools \citep[v2.0.8,][]{fermitools2019}}

\restartappendixnumbering
\appendix

\section{System Checks}\label{sec:system_checks}

\begin{figure*}[htb]
\begin{minipage}[b]{.35\linewidth}
\hspace{-1.25cm}
\includegraphics[width=1.05\linewidth]{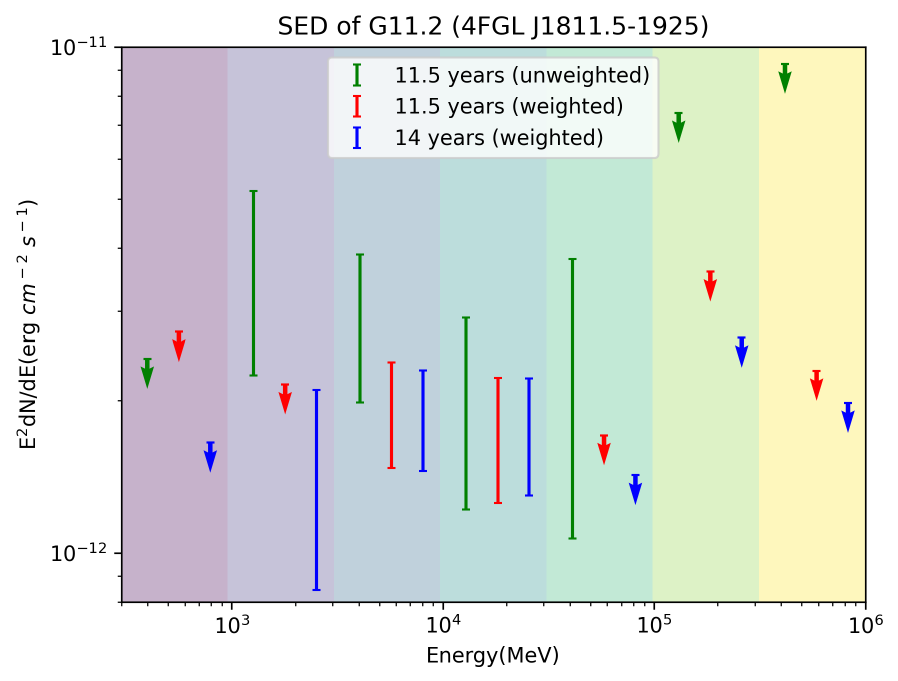}
\end{minipage}
\begin{minipage}[b]{.35\linewidth}
\hspace{-1cm}
\includegraphics[width=1.05\linewidth]{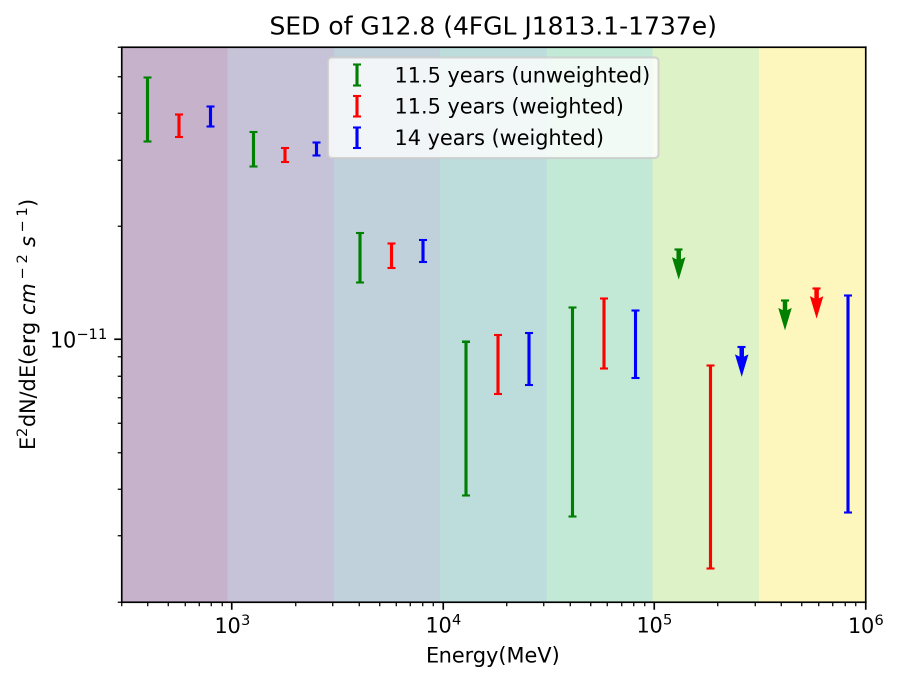}
\end{minipage}
\begin{minipage}[b]{.35\linewidth}
\hspace{-0.75cm}
\includegraphics[width=1.05\linewidth]{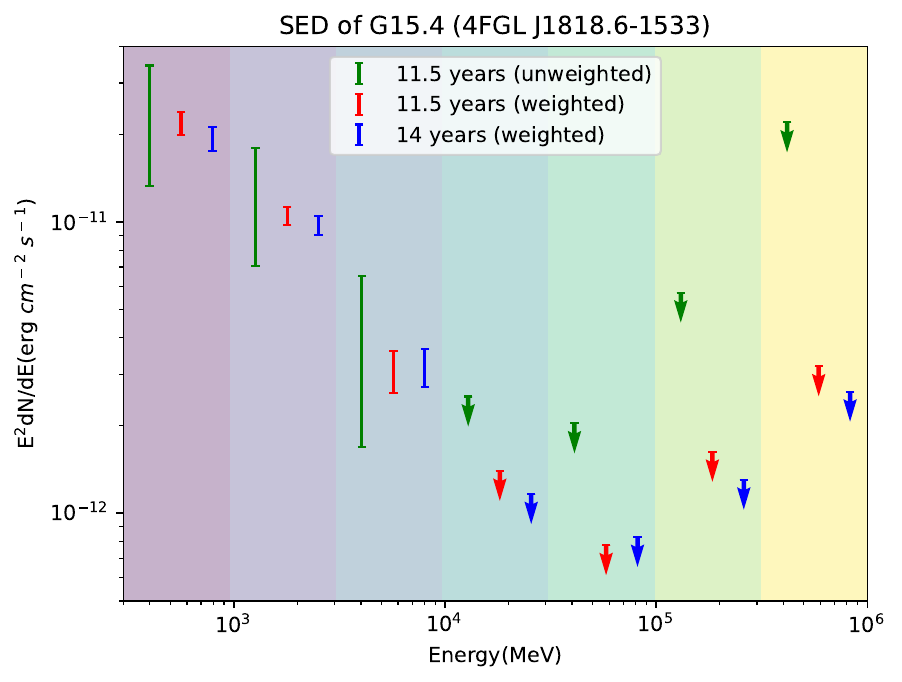}
\end{minipage}
\begin{minipage}[b]{.5\linewidth}
\centering
\includegraphics[width=0.8\linewidth]{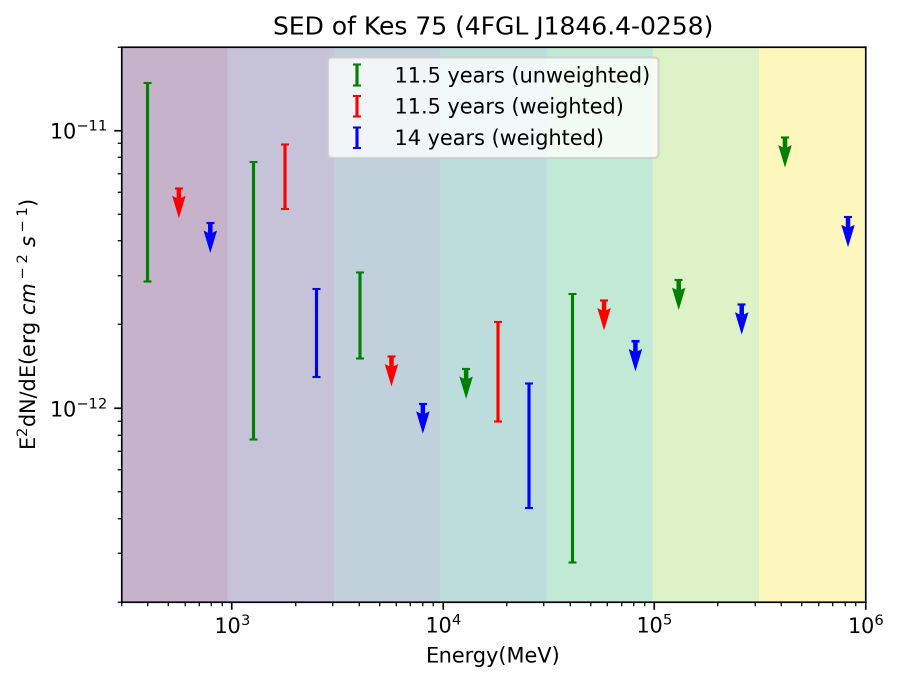}
\end{minipage}
\begin{minipage}[b]{.5\linewidth}
\centering
\includegraphics[width=0.8\linewidth]{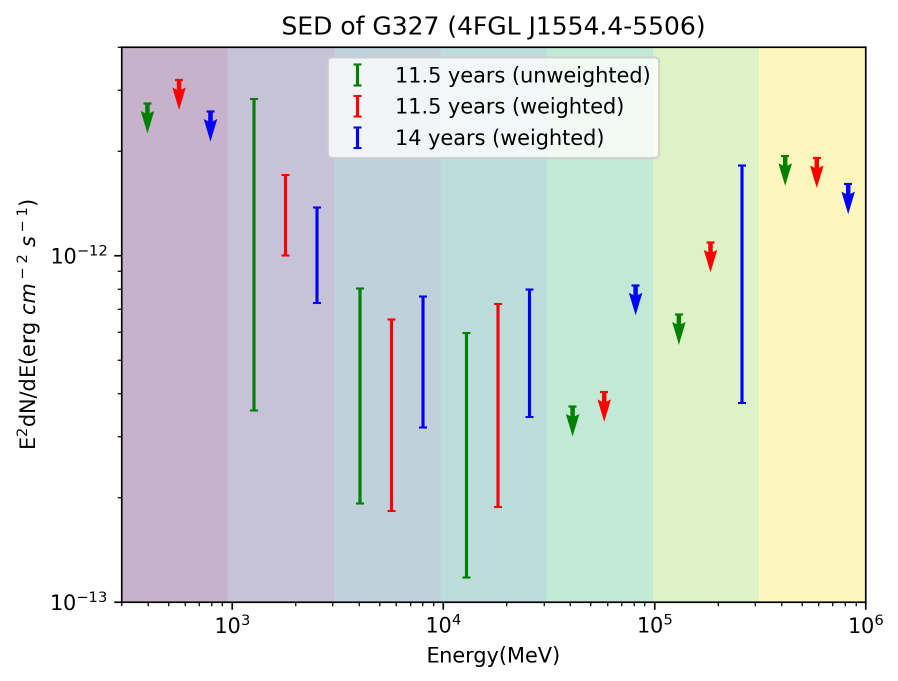}
\end{minipage}
\caption{Comparison of SEDs under different data configurations for 5 sources (Table~\ref{tab:fermipy}). The 11.5-year (unweighted) data points correspond to the main results of this work and include both statistical and systematic errors.}\label{fig:compare_seds}
\end{figure*}

We re-analyze five sample ROIs considering more \fermi data and the latest comprehensive 4FGL--DR4 catalog as a system check for the results in this report. Three of the sample PWNe are considered weak candidate source detections (G11.18--0.35, G12.82--0.02, and G15.40+0.10) and two are considered likely PWN source detections (G29.70--0.30, and G327.15--1.04). We summarize the configuration details of the re-analysis results using 14\,years of \fermi data compared to the analysis using 11.5\,years in Table~\ref{tab:fermipy}. In all five cases, the analysis and re-analysis results are consistent. We compare the SEDs reported in the main paper and those found with the updated data configuration and integration time in Figure~\ref{fig:compare_seds}. 

\begingroup
\begin{table}
\centering
\begin{tabular}{c|ccc}
\hline
\ \fermi Data Configuration & 14\,years (Weighted) & 11.5\,years (Weighted) & 11.5\,years (Unweighted) \\
\hline
\ Time range & 4 Aug. 2008--15 Aug. 2023 & 4 Aug. 2008--1 Jan. 2020 & 4 Aug. 2008--1 Jan. 2020\\
\ Energy range & 300\,MeV--1\,TeV & 300\,MeV--1\,TeV & 300\,MeV--2\,TeV \\
\ Catalog & 4FGL--DR4 & 4FGL--DR2 & 4FGL--DR2 \\
\ Event Type &\makecell*{8/16/32 (0.3--1 GeV) \\4/8/16/32 (1--1000 GeV)} &\makecell*{8/16/32 (0.3--1 GeV) \\4/8/16/32 (1--1000 GeV)} & 32\\
\ Zmax &\makecell*[c]{100$^o$ (0.3--1 GeV) \\105$^o$ (1--1000 GeV)} &\makecell*[c]{100$^o$ (0.3--1 GeV) \\105$^o$ (1--1000 GeV)} &100$^o$ \\
\ Extended Source Template & Extended\_14years & Extended\_8years & Extended\_8years \\
\ Weighted (Y/N) &Y &Y & N \\
\ Fermitools & 2.2.0 & 2.2.0 & 2.0.8 \\
\hline
\hline
\end{tabular}
\caption{Data configuration details for the 11.5-year data analysis of this report compared to a 14-year re-analysis using an updated source model. 
}
\label{tab:fermipy}
\end{table}
\endgroup

\section{Population Studies: High-Energy Characteristics}\label{sec:pop}
\begin{figure*}[htb]
\begin{minipage}[b]{.5\linewidth}
\centering
\includegraphics[width=0.9\linewidth]{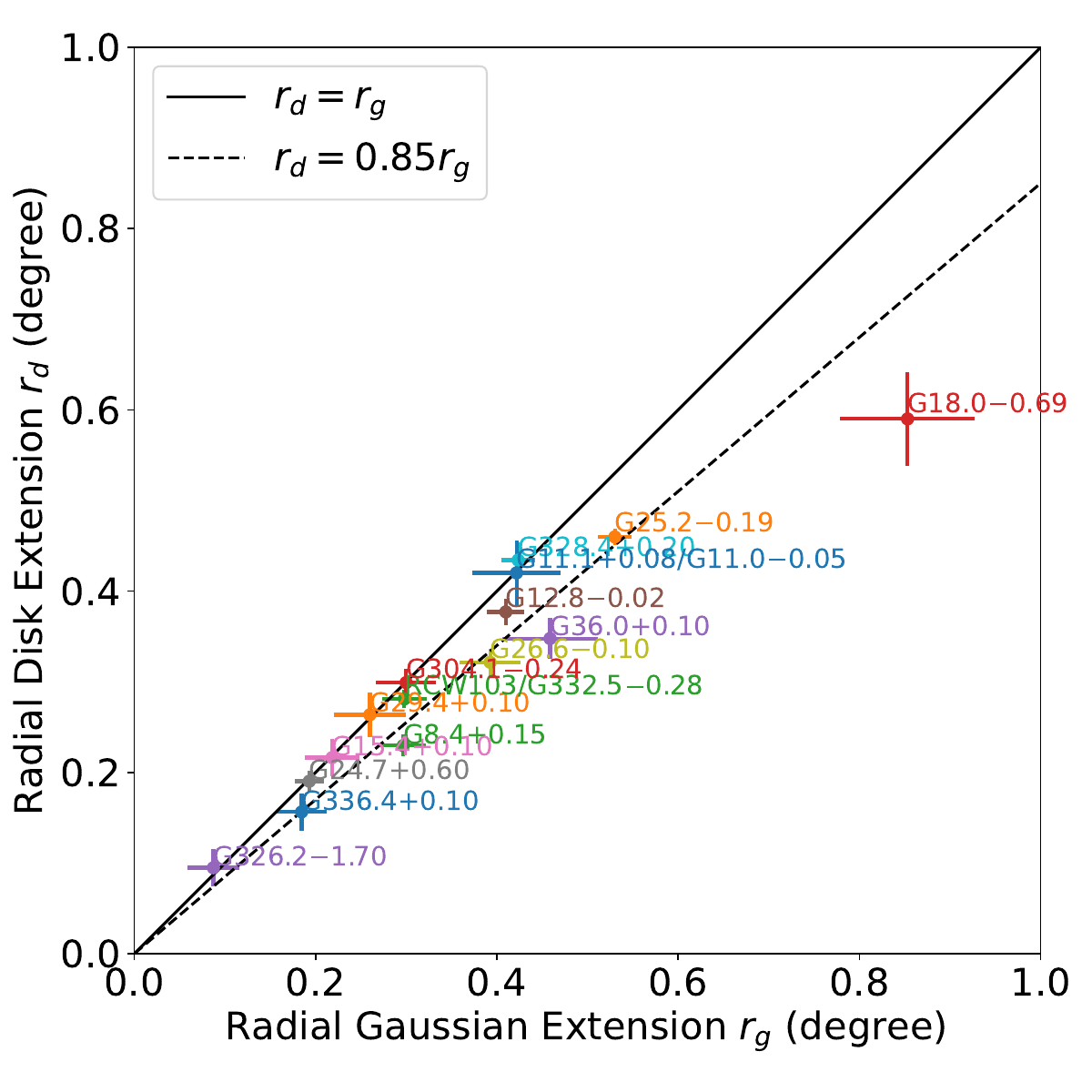}
\end{minipage}
\begin{minipage}[b]{.5\linewidth}
\centering
\includegraphics[width=0.9\linewidth]{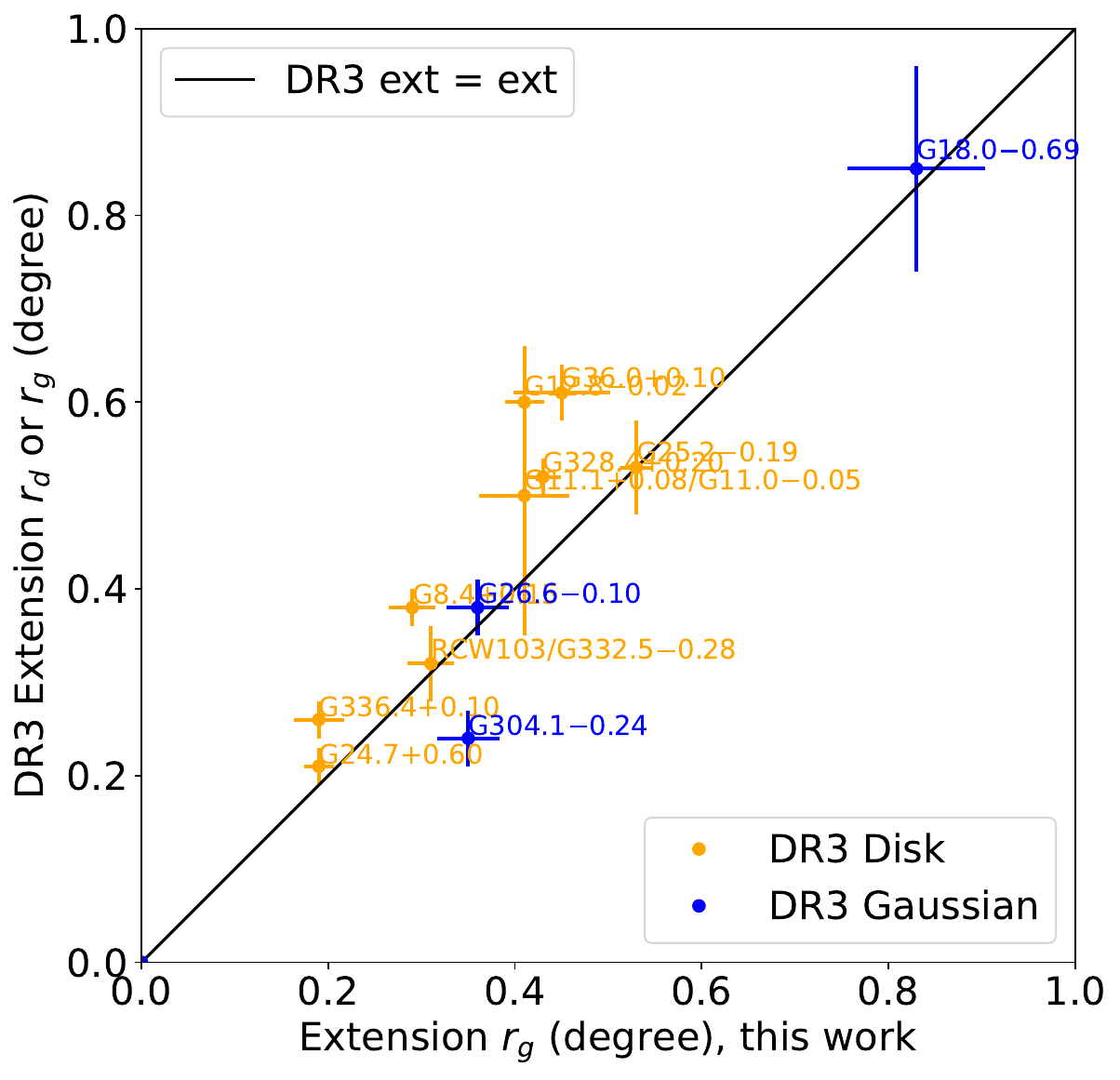}
\end{minipage}
\caption{{\it Left:} Best-fit radii to best-fit Gaussian extensions (68\% containment) from this work. {\it Right:} Best-fit Gaussian extensions from this work to the best-fit radius (yellow) or Gaussian radius (blue) in DR3.}\label{fig:extension_plots}
\end{figure*}

For extended sources, the best-fit radius $r$ when assuming a 2D radial disk template can be up to 1.85 times larger than $\sigma$ when assuming a 2D radial Gaussian such that the $r_{68}$ of either template is approximately the same \citep{lande2012}. The plots in Figure~\ref{fig:extension_plots}, both panels, explore this behavior. In the left panel of Figure~\ref{fig:extension_plots} we plot the best-fit $r_d$ as a function of the best-fit $r_g = 1.51 \sigma$, both measured in this work. On the whole, the extension sizes agree with one another, lying either on the 1:1 solid black line or just below it where $r_d$ is somewhat smaller than $r_g$. The exception is clearly HESS~J1825--137 (G18.00--0.69 in red), where we measure $r_g = 0.83\,\degree$, compared to the radial disk fit that finds $r_d \sim 0.60\,\degree$ ($r_d$ is $\sim 1.4$ times smaller than $r_g$). 
In the right panel of Figure~\ref{fig:extension_plots} we plot the best-fit $r$ that characterizes the extended sources in the 4FGL catalogs \citep{4fgl-dr4} as a function of the best-fit $r_g$ reported in this work. Sources that are fit as a disk in 4FGL--DR4 are plotted in yellow and those fit as a Gaussian in 4FGL--DR4 are plotted in blue. Those that are fit using a disk template in DR4 (yellow) tend to have larger extensions in DR4 than found here but are still reasonably consistent with the effect explored in the left panel. One source, G304.10--0.24, is fit as a radial Gaussian in both works, though we find a slightly larger extension (0.35$\,\degree$) than in DR4 (0.24$\,\degree$).
\begin{figure*}[htbp]
\begin{minipage}[b]{0.5\linewidth}
\centering
\includegraphics[width=0.9\linewidth]{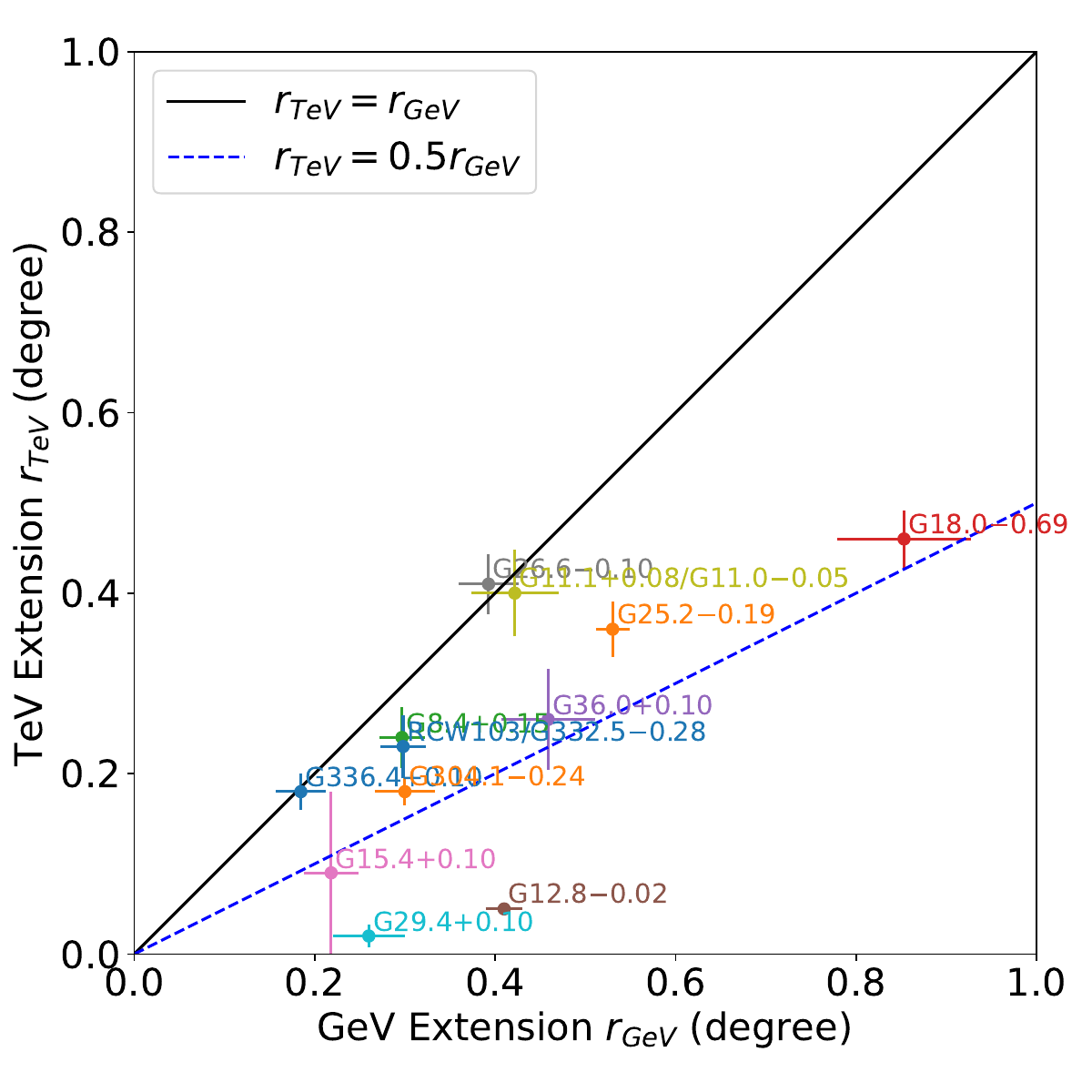}
\end{minipage}
\begin{minipage}[b]{.5\linewidth}
\centering
\includegraphics[width=0.9\linewidth]{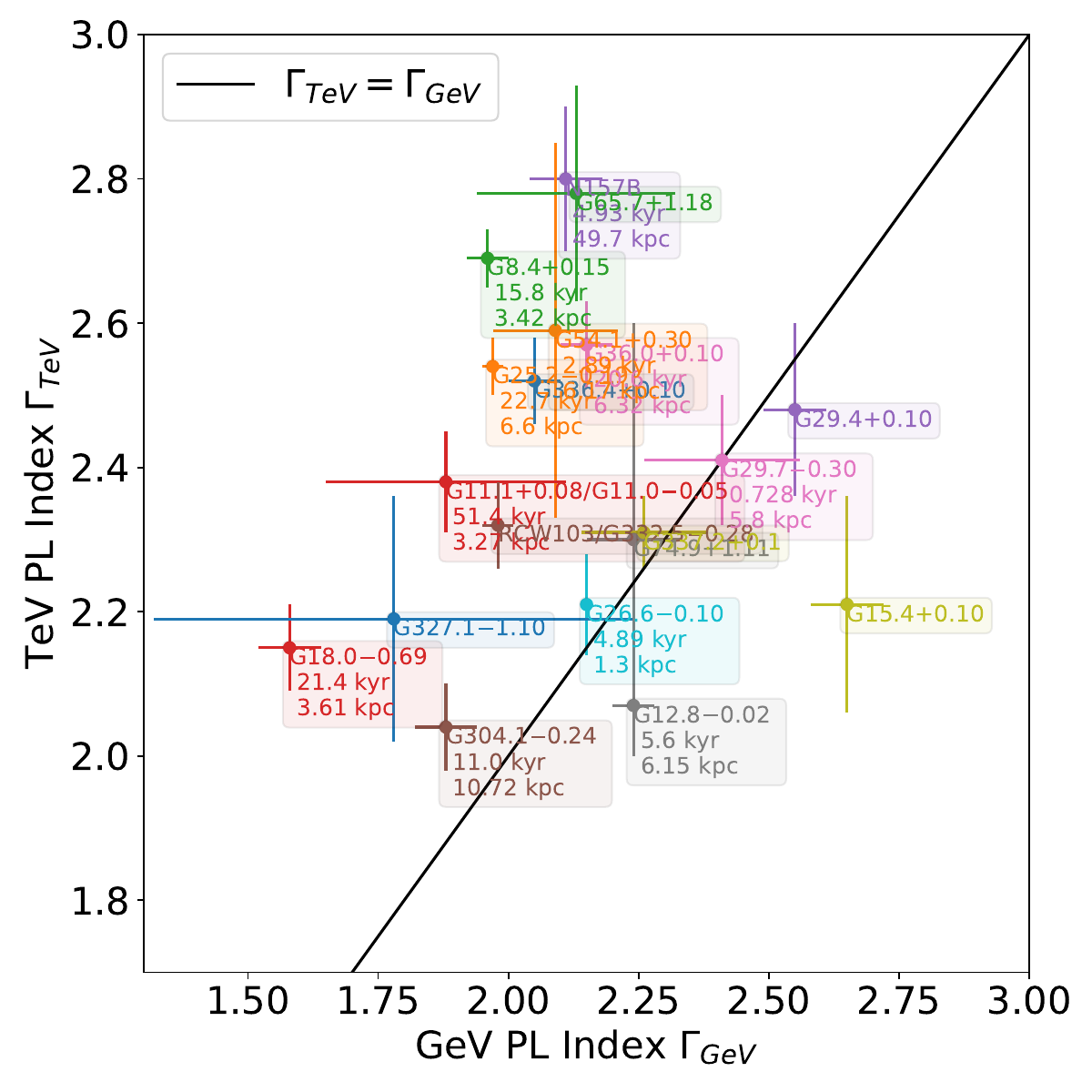}
\end{minipage}
\caption{{\it Left:} Best-fit TeV extensions to the best-fit GeV extensions (68\% containment) found in this work. {\it Right:} 
TeV index as a function of the GeV index.}\label{fig:tev_extension_index_comparison_plot}
\end{figure*}
In Figure~\ref{fig:tev_extension_index_comparison_plot}, left panel, the TeV extension is plotted as a function of the GeV extension found here. Either consistent TeV and GeV extensions or slightly smaller TeV extensions are observed. This is in agreement with what is found in \citet{acero2013}. From an energetics standpoint, this is not surprising considering evolved PWNe may have large differences in TeV and GeV extensions due to the rapid cooling of the highest-energy particles. 


\begingroup
\begin{table*}
\centering
\begin{tabular}{ccccccc}
\hline
ROI \,\, & Galactic PWN Name & 4FGL ID & Associated Pulsar\,\, & $d$ (kpc)\,\, & $\tau_c$ (kyr)\,\, & $\dot{E}$ ($10^{36}$ erg s$^{-1}$)  \\
\hline
\hline
			1 & G8.40+0.15 & J1804.7--2144e & J1803--2137 & 3.42 & 15.8 & 2.2 \\
			2 & G11.09+0.08 & J1810.3--1925e  & J1809--1917 & 3.27 & 51.4 & 1.8 \\
			3 & G11.18--0.35 & J1811.5--1925  & J1811--1925 & 5 & 23.3 & 6.4 \\
			4 & G12.82--0.02 & J1813.1--1737e & J1813--1749 & 6.15 & 5.6 & 56 \\
			5 & G18.00--0.69 & J1824.5--1351e & J1826-1334 & 3.61 & 21.4 & 2.8 \\
			6 & G25.24--0.19 & J1836.5--0651e  & J1838-0655 & 6.6 & 22.7 & 5.5 \\
			7 & G26.60--0.10 & J1840.9--0532e  & J1838-0537 & 1.3 & 4.89 & 6.0 \\
			8 & G29.70--0.30 & J1846.4--0258 (DR4)  & J1846--0258 & 5.8 & 0.728 & 8.10 \\
			9 & G36.01+0.10 & J1857.7+0246e & J1856+0245 & 6.32 & 20.6 & 6.4 \\
			10 & G54.10+0.27 & J1930.5+1853 (DR3) & J1930+1852 & 6.17 & 2.89 & 12 \\
			11 & G279.60--31.70 (N~157B) & J0537.8--6909 & J0537--6910 & 49.7 & 4.93 & 490 \\
			12 & G304.10--0.24 & J1303.0--6312e  & J1301--6305 & 10.7 & 11 & 1.7 \\
			13 & G315.78--0.23 & J1435.8--6018  & J1437--5959 & 8.54 & 114 & 1.4 \\
			14 & G332.50--0.28 & J1616.2--5054e & J1617--5055 & 4.74 & 8.13 & 16 \\
			15 & G338.20--0.00 & J1640.6--4632 (DR1), J1640.7--4631e (DR3) & J1640--4631 & 12.7 & 3.35 & 4.4 \\
			\hline
			1 & G0.87+0.08 & -- & J1747--2809 &	8.1 & 5.3 & 43.0 \\
			2 & G23.50+0.10 & -- & J1833-0827 & 4.4 & 147 & 0.58 \\
			3 & G32.64+0.53 & -- & J1849--0001 & 7 & 43.1 & 9.8 \\
			4 & G34.56--0.50 & -- & J1856+0113 & 2.81 & 20.3 & 0.43 \\
			5 & G47.38--3.88  & -- & J1932+1059 &  0.2 & 3100 & 0.004 \\
			6 & G108.60+6.80  & -- & J2225+6535 & 1.9 & 1120 & 0.001 \\
			7 & G179.72--1.69  & -- & J0538+2817 & 0.95 & 618 & 0.05 \\
			8 & G266.97--1.00 & -- & J0855--4644 & 5.6 & 141 & 1.1 \\
			9 & G290.00--0.93 & -- & J1101--6101 & 8$^\dag$ & 116 & 1.4 \\
			10 & G310.60--1.60  & -- & J1400--6325 & 9.1 & 12.7 & 51 \\
			11 & G341.20+0.90  & -- & J1646-4346 & 6.2 & 32.5 & 0.36 \\
			12 & G358.29+0.24  & -- & J1740--3015 & 2.9 & 20.6 & 0.008 \\
\hline
\hline
\end{tabular}
\caption{The associated pulsar name and relevant properties for the detected sources (top) that have confirmed pulsar associations and for nondetected sources (bottom). The pulsar properties are from the ATNF pulsar catalog \citep{atnf2005}. $^\dag$ The distance estimate listed for PSR~J1101--6101 comes instead from \citet{halpern2014}.}\label{table:psr_table}
\end{table*}
\endgroup

We investigate the spectral trends expected of $\gamma$-ray PWNe, shown in Figure~\ref{fig:tev_extension_index_comparison_plot}, right panel, plotting the TeV spectral index as a function of GeV index. For the systems whose associated pulsars are known, we include this information in the plot. There is a broad spread, but the majority (14/19) of sources fall above the 1:1 slope plotted in black, suggesting that most have a softer (larger) TeV spectral index than GeV spectral index, which is consistent with the predicted high-energy emission characteristics from \citet{acero2013}, but may have important contributions from other physical properties of the systems such as the characteristic pulsar age and(or) the distance. 
\begin{figure*}[htb]
\begin{minipage}[b]{.5\linewidth}
\centering
\includegraphics[width=0.8\linewidth]{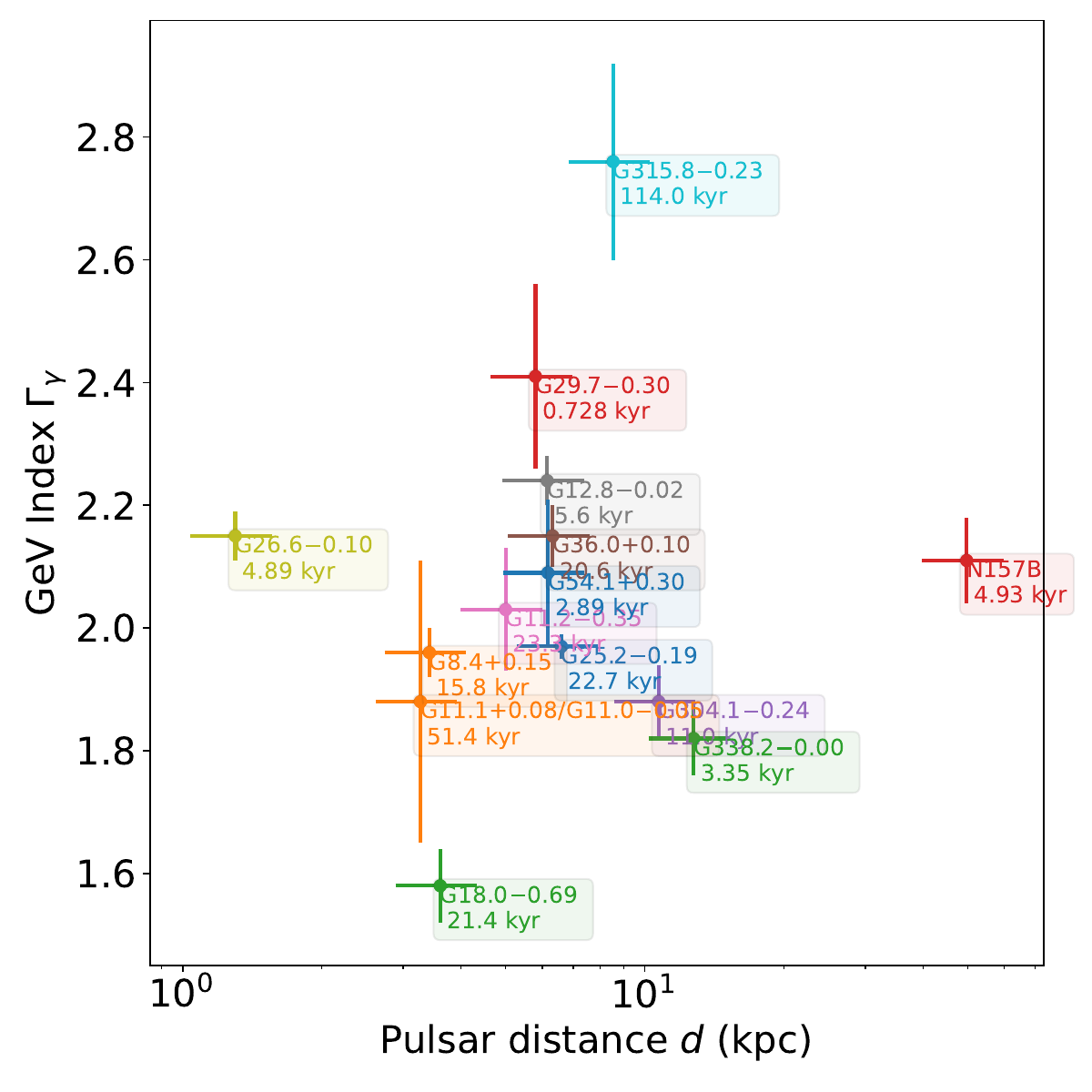}
\end{minipage}
\begin{minipage}[b]{.5\linewidth}
\centering
\includegraphics[width=0.8\linewidth]{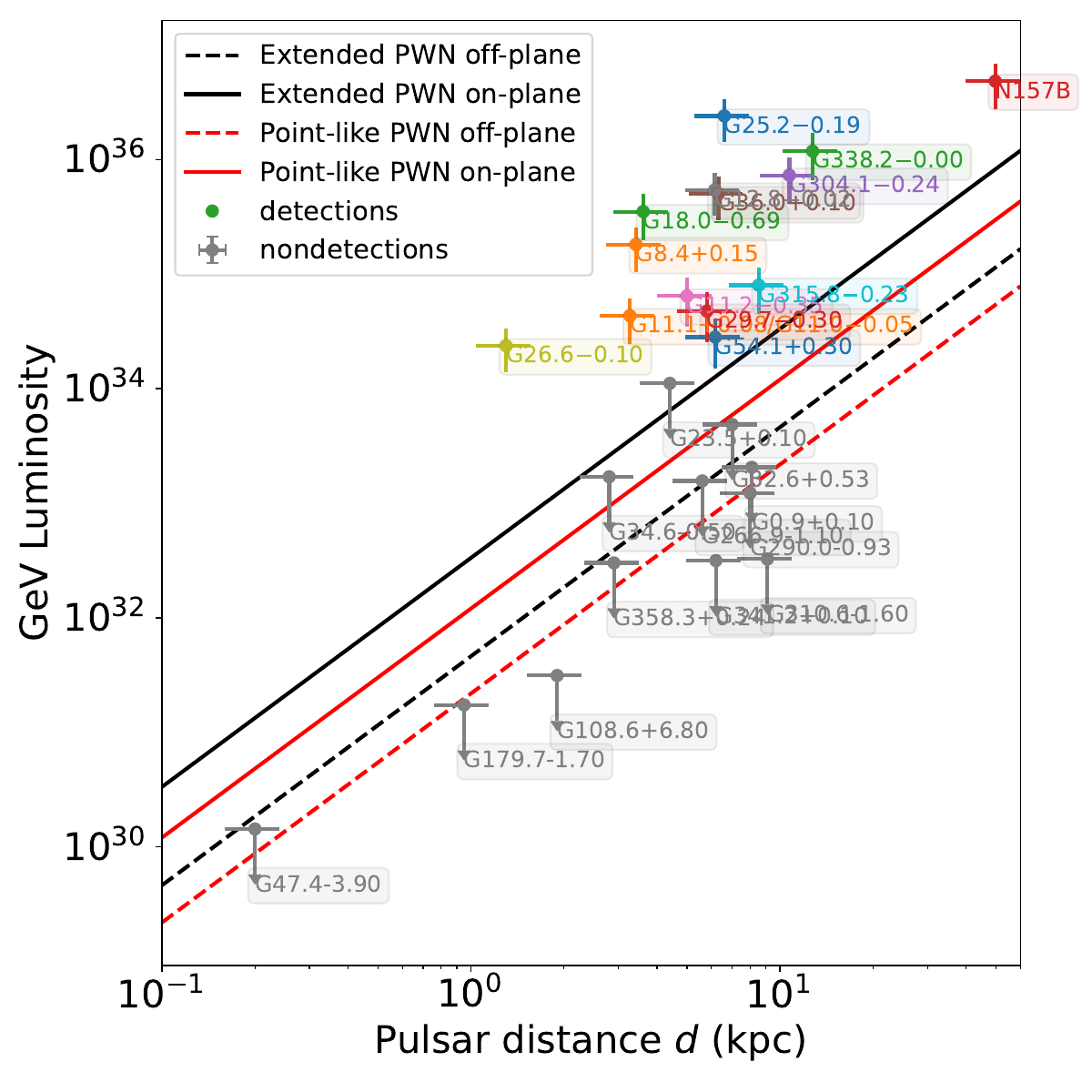}
\end{minipage}
\caption{{\it Left:} GeV index as a function of pulsar distance.  {\it Right:} 300\,MeV--2\,TeV luminosity as a function of pulsar distance. The sensitivity of the \fermi instrument varies based on the source sky location and source shape. We plot the sensitivity limits for four cases: point-like (extended) source located on (off) the Galactic plane. In all cases, a power-law spectrum with $\Gamma = 2$ is assumed. The extended source shape assumes the average size of the sample $\sim 0.4\,\degree$. For comparison we plot the nondetected sources that have associated pulsars in gray.}\label{fig:gev_index_lum_dist_plots}
\end{figure*} 

\begin{figure*}
\begin{minipage}[b]{0.5\linewidth}
\hspace{-0.15cm}
\includegraphics[width=0.8\linewidth]{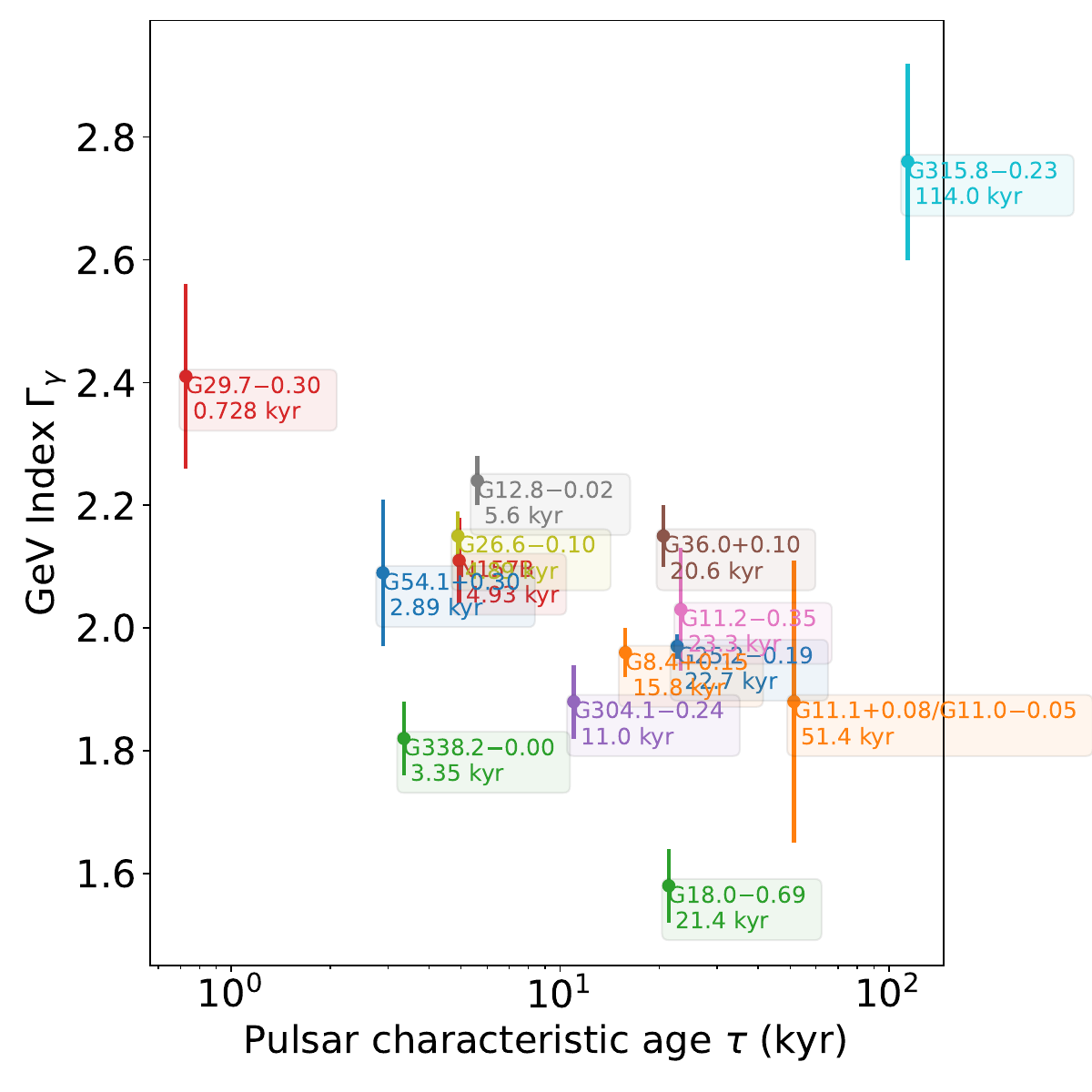}
\end{minipage}
\begin{minipage}[b]{0.5\linewidth}
\hspace{-1.5cm}
\includegraphics[width=1.2\linewidth]{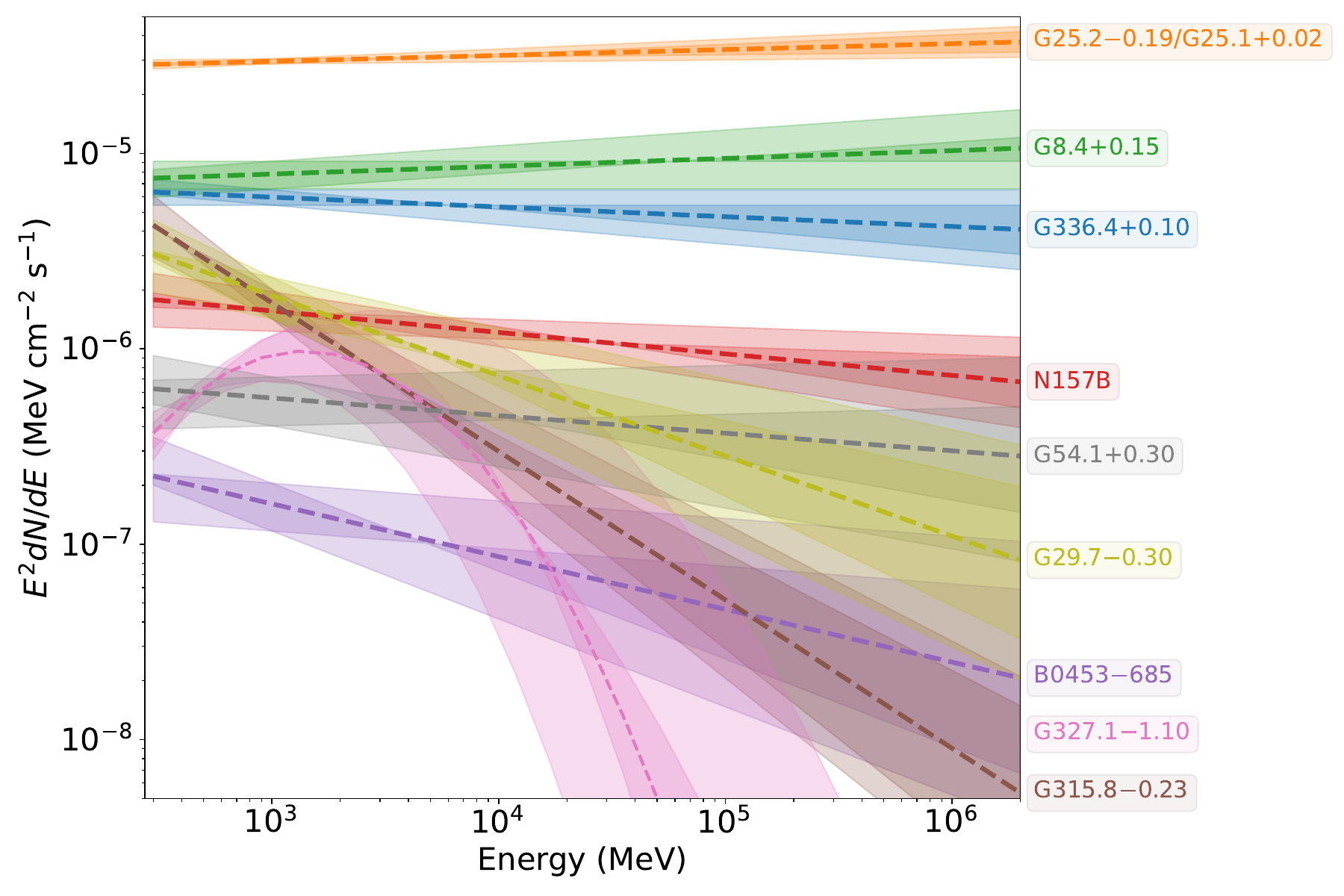}
\end{minipage}
\caption{{\it Left:} GeV index as a function of pulsar characteristic age.  {\it Right:} Best-fit spectral models for the 9 likely PWNe.}\label{fig:gev_index_distance_models}
\end{figure*} 
We explore the ways the distance may influence the high-energy spectral properties in both panels of Figure~\ref{fig:gev_index_lum_dist_plots}. Within uncertainty, there are no strong correlations present. However, the modest positive correlation, measured by the Pearson correlation coefficient $c_P$, that is derived for the 
GeV luminosity with pulsar distance ($ c_P = 0.9 \pm 0.5$) 
may be indicating a selection effect, where we are biased towards detecting only the closest PWNe. The TeV luminosity with pulsar distance ($c_P = 1.0 \pm 0.5$) and the TeV spectral index as a function of pulsar distance ($c_P = 0.5 \pm 0.2$) have similar implications. In Figure~\ref{fig:gev_index_distance_models}, left panel, the GeV index as a function of pulsar characteristic age is shown. There is a positive correlation $c_P = 0.5 \pm 0.05$, implying softer $\gamma$-ray spectra for older systems. This is an unexpected result since PWNe are expected to become $\gamma$-ray bright with age \citep[e.g.,][]{gelfand2009,slane2017}. It is worth noting that the correlation becomes negative ($c_P = -0.4 \pm 0.05$) if G315.8--0.23 is not considered. In any case, the result may be convoluted by possible pulsar contributions, see also the right panel of Figure~\ref{fig:gev_index_distance_models}, where the hardest spectra are observed from extended sources and the softest sources are all point-like. 

    

\section{Spectral flux measurements}\label{sec:seds}
We provide the spectral flux value $E^2 \frac{dN}{dE}$\footnote{Defined as the center energy $\sqrt{E_{1}E_{2}}$ squared multiplied by the differential flux at the center energy.} per bin for seven logarithmically-spaced energy bins for all 36 detected sources (Tables~\ref{tab:points} and \ref{tab:extent}) in Table~\ref{tab:spectral_fluxes}. We additionally provide the 95\% upper limit to the spectral flux $E^2 \frac{dN}{dE}$ per bin for nine logarithmically spaced energy bins for the 19 PWNe with no source detection (Table~\ref{tab:nondetect}) in Table~\ref{tab:spectral_ul_fluxes}, measured by placing a point source at each PWN position assuming a power-law spectrum. All flux values are in units MeV cm$^{-2}$ s$^{-1}$.

\tablewidth{0pt}
\tabletypesize{\footnotesize}
\setlength{\tabcolsep}{1pt}
\renewcommand*{\arraystretch}{1.0}
\setlength{\LTcapwidth}{\linewidth}
\begingroup
\begin{longrotatetable}
\begin{deluxetable*}{c|c|c|c|c|c|c|c}
\tablecaption{The spectral flux $E^2 \frac{dN}{dE}$ per bin for seven logarithmically spaced energy bins for all 36 detected sources (Tables~\ref{tab:points} and \ref{tab:extent}). All flux values are in units 10$^{-6}$ MeV cm$^{-2}$ s$^{-1}$.
}\label{tab:spectral_fluxes}
\tablecolumns{8}
\tablehead{
\colhead{PWN Name} &
\colhead{$E = 535\,$MeV} &
\colhead{$E = 1,705\,$MeV} &
\colhead{$E = 5,432\,$MeV} &
\colhead{$E = 17,303\,$MeV} &
\colhead{$E = 55,116\,$MeV} &
\colhead{$E = 175,560\,$MeV} &
\colhead{$E = 559,204\,$MeV}
}
\movetabledown=1.0cm
\startdata
			G8.40+0.15 & $8.87 \pm 1.04 \pm 3.12 \pm 7.02$ & $7.25 \pm 1.02 \pm 5.66$ & $8.04 \pm 1.27 \pm 3.91$ & $8.95 \pm 1.7 \pm 1.77$ & $11.7 \pm 2.75 \pm 1.37$ & $7.04 \pm 3.82 \pm 1.24$ & $13.3 \pm 8.33 \pm 1.07$ \\
			\multirow{2}{*}{G11.03$-$0.05$^\dag$} & \multirow{2}{*}{$7.84 \pm 1.05 \pm 3.14$} & \multirow{2}{*}{$7.48 \pm 1.06$} & \multirow{2}{*}{$2.59 \pm 1.23$} & \multirow{2}{*}{$1.55 \pm 1.52$} & \multirow{2}{*}{$3.48$} & \multirow{2}{*}{$4.63$} & \multirow{2}{*}{$15.3$} \\
            		G11.09+0.08 & & & & & & &  \\			
			G11.18$-$0.35 & $2.60$ & $2.32 \pm 0.555 \pm 0.732$ & $1.83 \pm 0.477 \pm 0.352$ & $1.29 \pm 0.517 \pm 0.114$ & $1.52 \pm 0.853 \pm 0.0485$ & $4.62 $ & $5.78 $ \\
			G12.82$-$0.02 & $26.0 \pm 1.15 \pm 3.44 \pm 4.92$ & $20.1 \pm 1.22 \pm 1.75$ & $10.4 \pm 1.45 \pm 0.612$ & $4.27 \pm 1.83 \pm 0.404$ & $4.84 \pm 2.57 \pm 0.936$ & $10.8 $ & $7.91 $ \\
			G15.40+0.10 & $14.9 \pm 0.908 \pm 2.72 \pm 6.52$ & $7.8 \pm 0.754 \pm 3.32$ & $2.56 \pm 0.725 \pm 1.32$ & $1.57 $ & $1.27 $ & $3.57 $ & $13.7 $ \\
			G16.73+0.08 & $7.37 \pm 0.851 \pm 2.55 \pm 4.26$ & $3.47 \pm 0.592 \pm 1.67$ & $1.67 \pm 0.487 \pm 0.297$ & $1.19 $ & $0.836 $ & $1.83 $ & $5.86 $ \\
			G18.00$-$0.69 & $6.19 \pm 1.38 \pm 4.15 \pm 2.21$ & $6.2 \pm 1.5 \pm 5.41$ & $8.36 \pm 1.93 \pm 2.79$ & $15.4 \pm 2.78 \pm 0.888$ & $20.1 \pm 4.4 \pm 3.57$ & $31.6 \pm 8.18 \pm 5.46$ & $14.7 $ \\
			G18.90$-$1.10 & $2.81 \pm 0.713 \pm 2.14 \pm 0.813$ & $3.8 \pm 0.508 \pm 0.244$ & $1.39 \pm 0.414 \pm 0.14$ & $0.517 $ & $1.17 $ & $1.79 $ & $5.84 $ \\			
			G20.20$-$0.20 & $7.19 \pm 0.866 \pm 2.6 \pm 1.22$ & $5.6 \pm 0.647 \pm 1.13$ & $1.09 \pm 0.443 \pm 0.262$ & $1.09 $ & $0.627 $ & $3.05 $ & $5.84 $ \\
			G24.70+0.60 & $6.19 \pm 0.878 \pm 2.63 \pm 1.87$ & $7.09 \pm 0.738 \pm 1.18$ & $7.98 \pm 0.876 \pm 0.453$ & $4.57 \pm 1.1 \pm 0.285$ & $4.01 \pm 1.71 \pm 0.439$ & $7.21 \pm 3.66 \pm 0.58$ & $19.3 $ \\
			\multirow{2}{*}{G25.24$-$0.19} & \multirow{2}{*}{$28.7 \pm 1.29 \pm 3.88 \pm 1.89$} & \multirow{2}{*}{$30.0 \pm 1.45 \pm 1.16$} & \multirow{2}{*}{$26.7 \pm 1.97 \pm 0.944$} & \multirow{2}{*}{$31.9 \pm 2.95 \pm 1.18$} & \multirow{2}{*}{$34.2 \pm 4.77 \pm 2.68$} & \multirow{2}{*}{$41.8 \pm 8.72 \pm 3.06$} & \multirow{2}{*}{$28.9 \pm 13.2 \pm 2.24$} \\
            		G25.10+0.02 & & & & & & &  \\			
			G26.60$-$0.10 & $13.9 \pm 1.1 \pm 3.29 \pm 1.81$ & $12.1 \pm 1.08 \pm 0.646$ & $8.38 \pm 1.26 \pm 0.352$ & $4.98 \pm 1.64 \pm 0.264$ & $11.7 \pm 2.99 \pm 1.65$ & $12.6 \pm 5.25 \pm 2.45$ & $6.45 $ \\
			G27.80+0.60 & $3.52 \pm 0.727 \pm 2.18 \pm 1.57$ & $3.38 \pm 0.502 \pm 0.726$ & $2.51 \pm 0.465 \pm 0.166$ & $0.582 \pm 0.397 \pm 0.0342$ & $0.547 $ & $1.85 $ & $5.87 $ \\
			G29.40+0.10 & $14.0 \pm 1.03 \pm 3.08 \pm 7.24$ & $8.4 \pm 0.894 \pm 4.45$ & $2.98 \pm 0.938 \pm 1.84$ & $1.35 \pm 1.17 \pm 0.694$ & $2.2 \pm 1.82 \pm 0.741$ & $4.93 $ & $5.88 $ \\
			G29.70$-$0.30 & $5.52 \pm 0.898 \pm 2.7 \pm 3.63$ & $2.64 \pm 0.616 \pm 2.07$ & $1.43 \pm 0.488 \pm 0.0184$ & $0.864 $ & $0.889 \pm 0.714 \pm 0.0572$ & $1.8 $ & $5.88 $ \\
			G36.01+0.10 & $12.0 \pm 1.1 \pm 3.31 \pm 2.53$ & $9.97 \pm 1.09 \pm 0.763$ & $7.06 \pm 1.28 \pm 1.21$ & $7.61 \pm 1.8 \pm 0.915$ & $8.19 \pm 2.56 \pm 0.704$ & $6.51 \pm 3.96 \pm 1.55$ & $21.3 $ \\
			G39.22$-$0.32 & $6.09 \pm 0.827 \pm 2.48 \pm 3.55$ & $3.04 \pm 0.578 \pm 1.51$ & $2.23 \pm 0.492 \pm 0.3$ & $0.659 \pm 0.437 \pm 0.057$ & $1.18 $ & $1.86 $ & $5.83 $ \\
                G49.2$-$0.30 & $15.8 \pm 0.884 \pm 2.65 \pm 3.63$ & $11.4 \pm 0.821 \pm 2.45$ & $3.47 \pm 0.754 \pm 0.747$ & $1.42 \pm 0.728 \pm 0.27$ & $1.48$ & $5.48$ & $5.66$ \\			
			G49.2$-$0.70 & $5.37 \pm 0.84 \pm 2.52 \pm 0.704$ & $3.75 \pm 0.754 \pm 0.981$ & $3.13 \pm 0.768 \pm 0.311$ & $1.34 \pm 0.708 \pm 0.127$ & $1.33 \pm 0.923 \pm 0.122$ & $1.75$ & $5.66$ \\
			G54.10+0.27 & $1.51$ & $0.65 \pm 0.232 \pm 0.226$ & $0.393 \pm 0.165 \pm 0.139$ & $0.608 \pm 0.201 \pm 0.0836$ & $0.664 $ & $0.824 \pm 0.645 \pm 0.0903$ & $1.65 $ \\
			G63.70+1.10 & $1.2 \pm 0.361 \pm 1.08 \pm 0.822$ & $1.62 \pm 0.276 \pm 0.358$ & $0.932 \pm 0.255 \pm 0.0842$ & $0.289 \pm 0.237 \pm 0.0245$ & $0.49 $ & $1.64 $ & $5.34 $ \\
			G65.73+1.18 & $2.51 \pm 0.408 \pm 1.22 \pm 0.427$ & $2.01 \pm 0.307 \pm 0.195$ & $0.612 \pm 0.243 \pm 0.0548$ & $0.724 $ & $0.495 $ & $2.11 \pm 1.69 \pm 0.0782$ & $5.33 $ \\
			G74.94+1.11 & $3.3 \pm 0.562 \pm 1.69 \pm 1.31$ & $2.15 \pm 0.449 \pm 0.511$ & $1.82 \pm 0.43 \pm 0.296$ & $1.26 \pm 0.461 \pm 0.142$ & $1.16 $ & $3.45 \pm 2.01 \pm 0.153$ & $5.17 $ \\
			G279.60$-$31.70 & $1.49 \pm 0.23 \pm 0.691 \pm 0.111$ & $1.76 \pm 0.224 \pm 0.154$ & $1.26 \pm 0.253 \pm 0.141$ & $0.636 \pm 0.298 \pm 0.0769$ & $0.861 \pm 0.599 \pm 0.143$ & $1.92 \pm 1.51 \pm 0.245$ & $12.9 $ \\
			G279.80$-$35.80 & $0.366$ & $0.17 \pm 0.0568 \pm 0.0461$ & $0.153 \pm 0.0577 \pm 0.0146$ & $0.103 $ & $0.11 $ & $0.348 $ & $1.23 $ \\			
			G304.10$-$0.24 & $4.31$ & $3.85 \pm 0.603 \pm 2.09$ & $3.35 \pm 0.712 \pm 0.763$ & $4.92 \pm 1.06 \pm 0.202$ & $7.6 \pm 1.85 \pm 0.287$ & $4.43 $ & $6.89 \pm 5.31 \pm 0.566$ \\
			G315.78$-$0.23 & $2.62 \pm 0.57 \pm 1.71 \pm 1.76$ & $1.32 \pm 0.371 \pm 0.588$ & $0.499 \pm 0.287 \pm 0.121$ & $0.247 $ & $1.07 $ & $2.03 $ & $4.93 $ \\
			G318.90+0.40 & $0.710$ & $1.15 \pm 0.235 \pm 0.214$ & $0.591 \pm 0.169 \pm 0.126$ & $0.128 \pm 0.117 \pm 1.12\times 10^{-6}$ & $0.41 $ & $0.924 $ & $1.26 $ \\
			G326.12$-$1.81 & $1.44$ & $1.57 \pm 0.319 \pm 0.491$ & $1.82 \pm 0.396 \pm 0.195$ & $1.96 \pm 0.587 \pm 0.152$ & $1.48 \pm 0.728 \pm 0.11$ & $3.45 $ & $4.41 $ \\			
			G327.15$-$1.04 & $2.26$ & $0.995 \pm 0.223 \pm 0.739$ & $0.31 \pm 0.151 \pm 0.115$ & $0.223 \pm 0.146 \pm 0.0322$ & $0.229 $ & $0.421 $ & $1.21 $ \\
			G328.40+0.20 & $20.9 \pm 1.03 \pm 3.1 \pm 4.53$ & $21.6 \pm 1.13 \pm 3.03$ & $17.0 \pm 1.45 \pm 1.8$ & $14.0 \pm 1.95 \pm 1.18$ & $12.4 \pm 2.8 \pm 1.07$ & $9.8 $ & $7.48 $ \\
			\multirow{2}{*}{G332.50$-$0.30} & \multirow{2}{*}{$12.8 \pm 0.989 \pm 2.97 \pm 2.2$} & \multirow{2}{*}{$12.5 \pm 0.958 \pm 1.32$} & \multirow{2}{*}{$12.3 \pm 1.19 \pm 0.978$} & \multirow{2}{*}{$13.1 \pm 1.74 \pm 0.578$} & \multirow{2}{*}{$15.1 \pm 2.83 \pm 1.02$} & \multirow{2}{*}{$13.1 \pm 4.8 \pm 2.09$} & \multirow{2}{*}{$16.8 \pm 8.52 \pm 1.46$} \\
            		G332.50$-$0.28 & & & & & & &  \\
			G336.40+0.10 & $5.86 \pm 0.89 \pm 2.67 \pm 1.79$ & $6.63 \pm 0.796 \pm 0.894$ & $4.96 \pm 0.888 \pm 0.621$ & $2.31 \pm 1.1 \pm 0.559$ & $7.51 \pm 2.19 \pm 0.551$ & $12.1 \pm 4.5 \pm 0.748$ & $13.3 $ \\
			G337.20+0.10 & $4.85$ & $2.29 \pm 0.68 \pm 0.16$ & $0.888 \pm 0.505 \pm 0.0871$ & $1.1 \pm 0.545 \pm 0.0508$ & $1.48$ & $1.61$ & $14.0$ \\
			G337.50$-$0.10 & $7.13 \pm 0.904 \pm 2.71 \pm 1.92$ & $3.27 \pm 0.658 \pm 0.88$ & $1.69 \pm 0.517 \pm 0.189$ & $0.981$ & $0.693$ & $3.06$ & $5.24$ \\			
			G338.20$-$0.00 & $4.07$ & $2.49 \pm 0.643 \pm 0.668$ & $3.19 \pm 0.576 \pm 0.215$ & $2.63 \pm 0.685 \pm 0.0969$ & $6.42 \pm 1.57 \pm 0.222$ & $8.19 \pm 3.22 \pm 0.415$ & $5.68 $ \\
\enddata
\tablecomments{The first quoted error is the $1 \sigma$ statistical error, the second error reflects the factor $\sim 3$ of the statistical error for the lowest energy bin, and the latter is the total systematic error. Flux values that lack quoted errors are instead the 95\% upper limit flux for that bin. \\
$^\dag$ corresponds to 4FGL~J1810.3--1925e, a candidate extended source in the 4FGL catalogs that is reported with three flags related to the source flux changing significantly when changing the diffuse model or analysis method. For this reason, we do not provide systematic errors as the method becomes unreliable for this source.}
\end{deluxetable*}
\end{longrotatetable}
\endgroup

\tablewidth{0pt}
\tabletypesize{\footnotesize}
\setlength{\tabcolsep}{2pt}
\renewcommand*{\arraystretch}{1.0}
\setlength{\LTcapwidth}{\linewidth}
\begingroup
\begin{longrotatetable}
\begin{deluxetable*}{l|c|c|c|c|c|c|c|c|c}
\tablecaption{The 95\% upper limit to the spectral flux $E^2 \frac{dN}{dE}$ per bin for nine logarithmically spaced energy bins for the 19 PWNe with no source detection (Table~\ref{tab:nondetect}). \\ All flux values are in units MeV cm$^{-2}$ s$^{-1}$. \label{tab:spectral_ul_fluxes}}
\tablecolumns{8}
\tablehead{
\colhead{PWN Name} &
\colhead{$E = 477\,$MeV} &
\colhead{$E = 1,205\,$MeV} &
\colhead{$E = 3,044\,$MeV} &
\colhead{$E = 7,690\,$MeV} &
\colhead{$E = 19,429\,$MeV} &
\colhead{$E = 49,086\,$MeV} &
\colhead{$E = 124,017\,$MeV} &
\colhead{$E = 313,325\,$MeV} &
\colhead{$E = 791,608\,$MeV}
}
\movetabledown=1cm
\startdata
G0.87+0.08 & 2.84 $\times 10^{-7}$ &  7.12 $\times 10^{-7}$ &  1.59 $\times 10^{-6}$ &  5.79 $\times 10^{-7}$ &  2.44 $\times 10^{-7}$ &  8.08 $\times 10^{-7}$ &  1.65 $\times 10^{-6}$ &  3.32 $\times 10^{-6}$ &  4.33 $\times 10^{-6}$  \\
G23.50+0.10 & $3.89 \times 10^{-6}$ & $3.07 \times 10^{-6}$ & $1.63 \times 10^{-6}$ & $1.46 \times 10^{-6}$ & $3.93 \times 10^{-7}$ & $9.18\times 10^{-7}$ & $1.10\times 10^{-6}$ & $1.71\times 10^{-6}$ & $2.71\times 10^{-6}$ \\
G32.64+0.53 & 8.87 $\times 10^{-7}$ &  1.18 $\times 10^{-6}$ &  7.53 $\times 10^{-7}$ &  5.30 $\times 10^{-7}$ &  1.16 $\times 10^{-6}$ &  1.27 $\times 10^{-6}$ &  4.51 $\times 10^{-7}$ &  9.62 $\times 10^{-7}$ &  3.79 $\times 10^{-6}$  \\
G34.56--0.50 & 4.31 $\times 10^{-6}$ & 2.81 $\times 10^{-6}$ & 2.55 $\times 10^{-6}$ & 3.03 $\times 10^{-6}$ & 7.56 $\times 10^{-7}$ & 2.06 $\times 10^{-6}$ & 1.59 $\times 10^{-6}$ & 4.19 $\times 10^{-6}$ & 1.09 $\times 10^{-5}$ \\
G47.38--3.88 & 4.60 $\times 10^{-7}$ &  3.94 $\times 10^{-7}$ &  4.12 $\times 10^{-7}$ &  1.44 $\times 10^{-7}$ &  3.95 $\times 10^{-7}$ &  1.65 $\times 10^{-7}$ &  4.61 $\times 10^{-7}$ &  9.0 $\times 10^{-7}$ &  2.45 $\times 10^{-6}$ \\
G74.00--8.50 & 1.13 $\times 10^{-7}$ &  2.06 $\times 10^{-7}$ &  2.16 $\times 10^{-7}$ &  4.00 $\times 10^{-7}$ &  2.43 $\times 10^{-7}$ &  5.82 $\times 10^{-7}$ &  1.48 $\times 10^{-6}$ &  3.943 $\times 10^{-6}$ &  1.02 $\times 10^{-5}$  \\
G93.3+6.90 & 2.85 $\times 10^{-7}$ &  4.05 $\times 10^{-7}$ &  2.41 $\times 10^{-7}$ &  3.62 $\times 10^{-7}$ &  2.57 $\times 10^{-7}$ &  1.54 $\times 10^{-7}$ &  8.51 $\times 10^{-7}$ &  7.49 $\times 10^{-7}$ &  2.03 $\times 10^{-6}$  \\
G108.60+6.80 & 5.80 $\times 10^{-7}$ &  2.77 $\times 10^{-7}$ &  8.44 $\times 10^{-8}$ &  7.65 $\times 10^{-8}$ &  1.29 $\times 10^{-7}$ &  3.93 $\times 10^{-7}$ &  2.67 $\times 10^{-7}$ &  7.35 $\times 10^{-7}$ &  1.82 $\times 10^{-6}$ \\
G141.2+5.00 &  4.40 $\times 10^{-7}$ &  3.03 $\times 10^{-7}$ &  5.70 $\times 10^{-8}$ &  8.61 $\times 10^{-8}$ &  1.55 $\times 10^{-7}$ &  1.32 $\times 10^{-7}$ &  3.13 $\times 10^{-7}$ &  7.0 $\times 10^{-7}$ &  1.84 $\times 10^{-6}$ \\
G179.72--1.69 & 1.27 $\times 10^{-6}$ &  3.43 $\times 10^{-7}$ &  1.05 $\times 10^{-7}$ &  7.50 $\times 10^{-8}$ &  2.35 $\times 10^{-7}$ &  6.76 $\times 10^{-7}$ &  4.03 $\times 10^{-7}$ &  1.00 $\times 10^{-6}$ &  2.53 $\times 10^{-6}$ \\
G189.10+3.00 & 4.68 $\times 10^{-7}$ &  3.82 $\times 10^{-6}$ &  1.20 $\times 10^{-6}$ &  1.60 $\times 10^{-6}$ &  1.95 $\times 10^{-6}$ &  2.31 $\times 10^{-6}$ &  1.43 $\times 10^{-6}$ &  4.29 $\times 10^{-6}$ &  2.70 $\times 10^{-6}$ \\
G266.97--1.00 & 1.18 $\times 10^{-6}$ & 1.12 $\times 10^{-6}$ & 2.93 $\times 10^{-7}$ & 2.32 $\times 10^{-7}$ & 2.60 $\times 10^{-7}$ & 2.01 $\times 10^{-7}$ & 9.11 $\times 10^{-7}$ & 2.98 $\times 10^{-6}$ & 8.93 $\times 10^{-6}$ \\
G290.00--0.93 & 6.49 $\times 10^{-7}$ &  4.83 $\times 10^{-7}$ &  4.93 $\times 10^{-7}$ &  1.11 $\times 10^{-7}$ &  2.44 $\times 10^{-7}$ &  8.61 $\times 10^{-7}$ &  3.22 $\times 10^{-7}$ &  9.64 $\times 10^{-7}$ &  2.24 $\times 10^{-6}$ \\
G310.60--1.60 &  1.28 $\times 10^{-7}$ &  4.75 $\times 10^{-8}$ &  5.32 $\times 10^{-8}$ &  9.87 $\times 10^{-8}$ &  3.30 $\times 10^{-7}$ &  3.38 $\times 10^{-7}$ &  1.11 $\times 10^{-6}$ &  9.03 $\times 10^{-7}$ &  2.13 $\times 10^{-6}$ \\
G322.50--0.10 & 1.95 $\times 10^{-6}$ &  9.32 $\times 10^{-7}$ &  5.91 $\times 10^{-7}$ &  5.45 $\times 10^{-7}$ &  1.53 $\times 10^{-7}$ &  2.70 $\times 10^{-7}$ &  3.46 $\times 10^{-7}$ &  1.16 $\times 10^{-6}$ &  2.20 $\times 10^{-6}$ \\
G341.20+0.90 & 4.44 $\times 10^{-7}$ &  2.85 $\times 10^{-7}$ &  1.61 $\times 10^{-7}$ &  1.73 $\times 10^{-7}$ &  1.62 $\times 10^{-7}$ &  2.70 $\times 10^{-7}$ &  8.44 $\times 10^{-7}$ &  1.03 $\times 10^{-6}$ &  2.43 $\times 10^{-6}$ \\
G350.20--0.80 & 1.90 $\times 10^{-6}$ &  1.13 $\times 10^{-6}$ &  8.65 $\times 10^{-7}$ &  7.39 $\times 10^{-7}$ &  5.73 $\times 10^{-7}$ &  5.49 $\times 10^{-7}$ &  6.94 $\times 10^{-7}$ &  1.01 $\times 10^{-6}$ &  2.40 $\times 10^{-6}$ \\
G358.29+0.24 &  1.01 $\times 10^{-6}$ &  8.49 $\times 10^{-7}$ &  6.52 $\times 10^{-7}$ &  6.77 $\times 10^{-7}$ &  3.69 $\times 10^{-7}$ &  4.37 $\times 10^{-7}$ &  8.43 $\times 10^{-7}$ &  9.51 $\times 10^{-7}$ &  2.58 $\times 10^{-6}$ \\
G358.60--17.20 &  1.02 $\times 10^{-7}$ &  7.12 $\times 10^{-8}$ &  5.21 $\times 10^{-8}$ &  7.15 $\times 10^{-8}$ &  2.11 $\times 10^{-7}$ &  2.06 $\times 10^{-7}$ &  3.58 $\times 10^{-7}$ &  9.57 $\times 10^{-7}$ &  2.48 $\times 10^{-6}$ \\
\enddata
\end{deluxetable*}
\end{longrotatetable}
\endgroup


Finally, we provide the 300\,MeV--1\,TeV SEDs 
for the 9 detected $\gamma$-ray sources classified as likely PWNe. The complete figure set of all 36 SEDs is provided in the online journal. The best-fit fluxes with the 1$\,\sigma$ statistical (orange) and total systematic (black) uncertainties are plotted. The lowest energy flux error in blue represents the factor $\sim 3$ of the statistical error (see Table~\ref{tab:spectral_fluxes}). The dashed white line and the blue uncertainty band (1\,$\sigma$) represent the best-fit spectral model (Table~\ref{tab:tev_table}).

\begin{figure}[!h]
\centering
\includegraphics[width=1.0\linewidth]{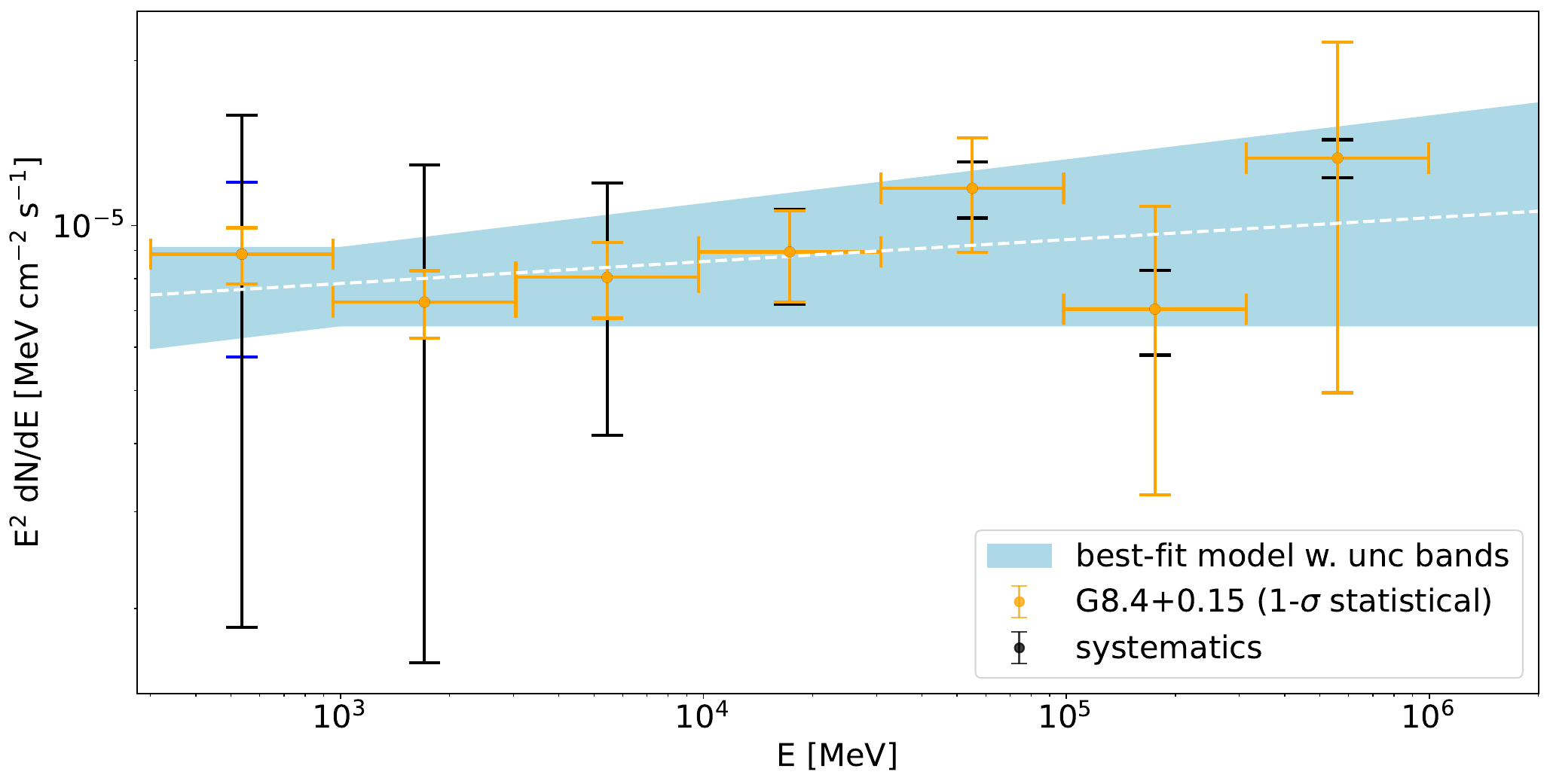}
\caption{300\,MeV--1\,TeV SED for PWN~G8.40+0.15, corresponding to 4FGL J1804.7--2144e. The complete figure set of all 36 SEDs is provided in the online journal.}
\end{figure}

\begin{figure}[!h]
\centering
\includegraphics[width=1.0\linewidth]{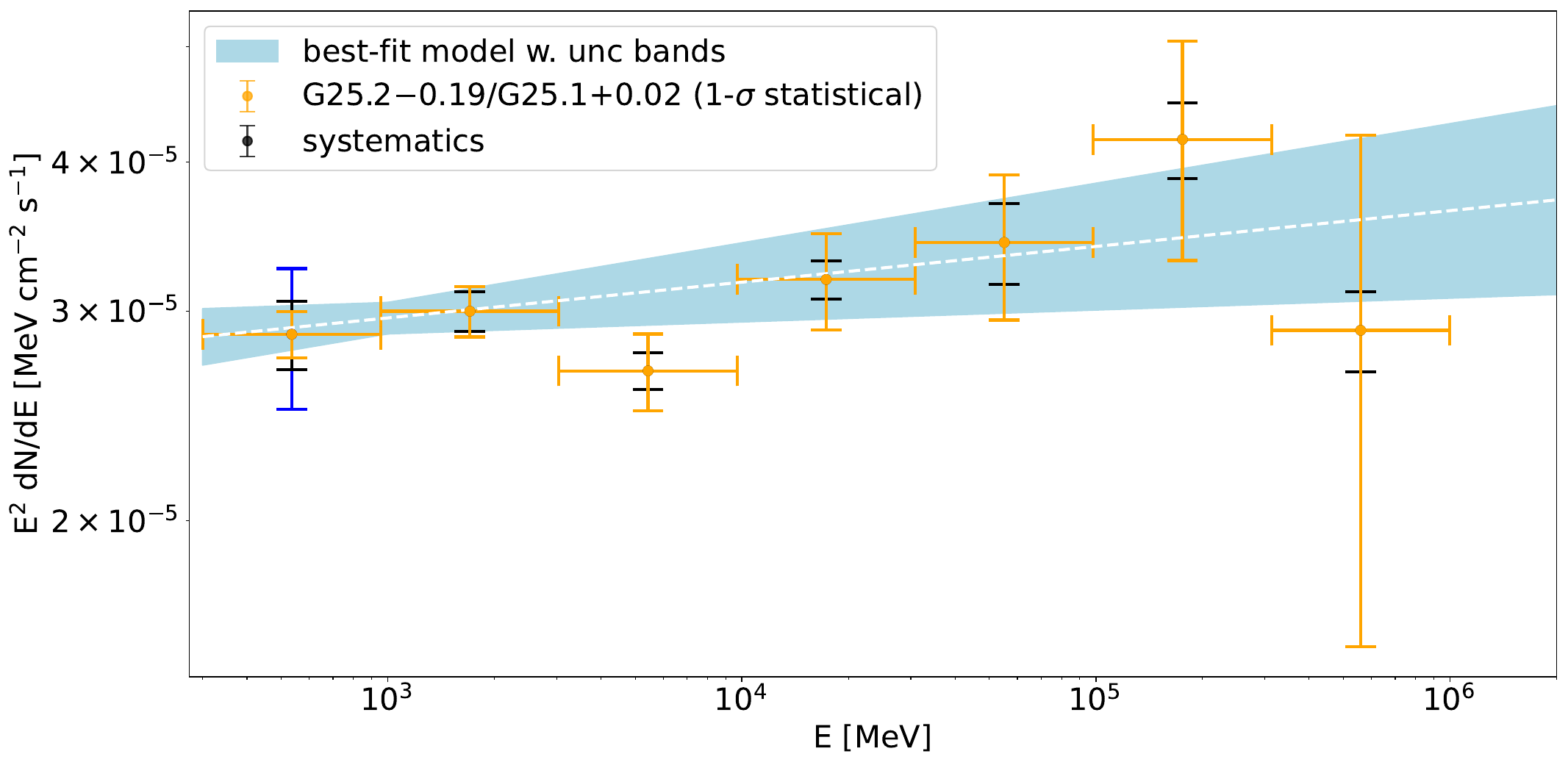}
\caption{300\,MeV--1\,TeV SED for TeV PWN HESS~J1837--069, corresponding to 4FGL J1836.5--0651e. Two lower energy PWN counterparts G25.24--0.19 and G25.10+0.02 coincide with the extended Fermi source. The complete figure set of all 36 SEDs is provided in the online journal.}
\end{figure}

\begin{figure}[!h]
\centering
\includegraphics[width=1.0\linewidth]{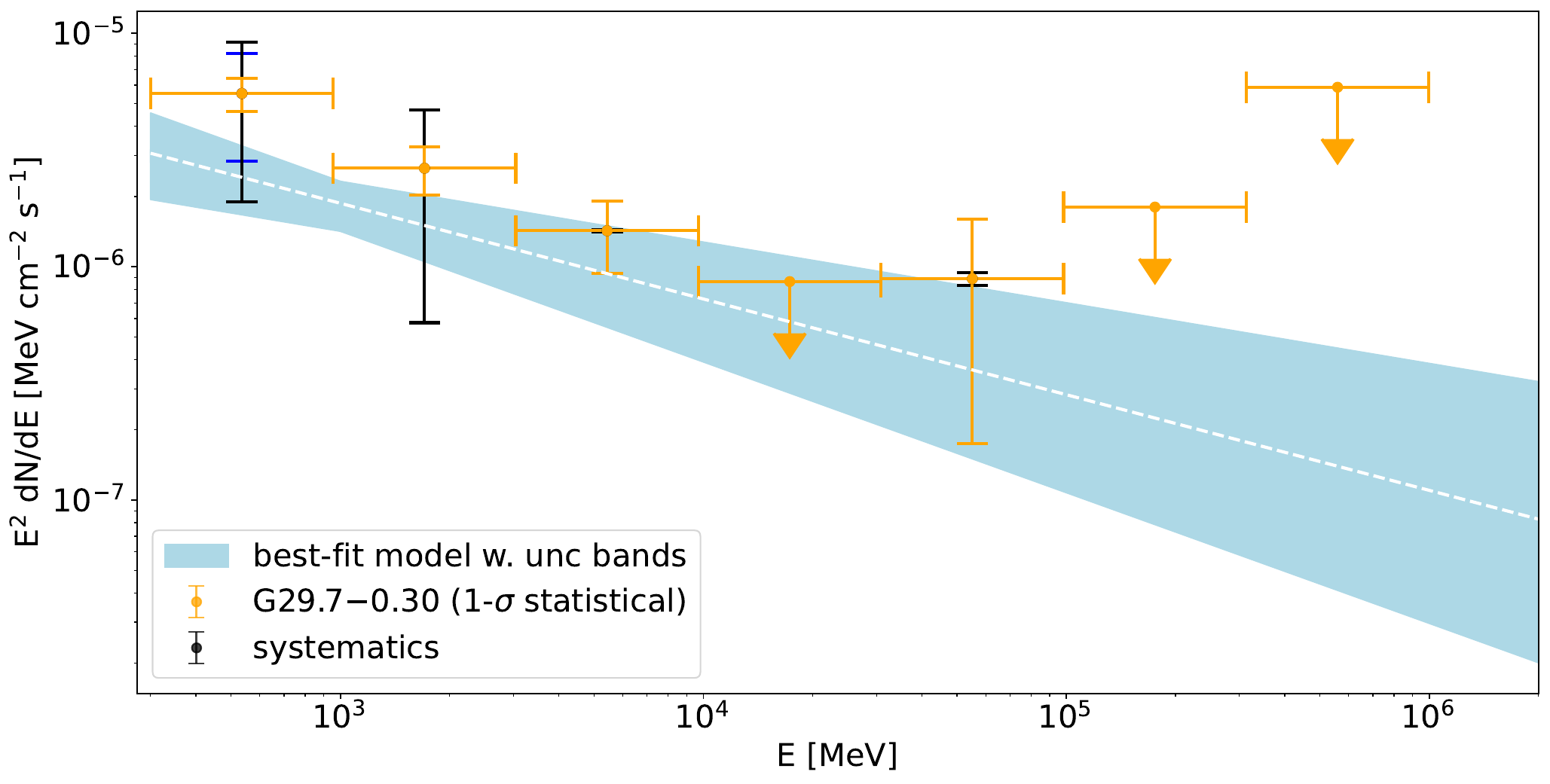}
\caption{300\,MeV--1\,TeV SED for PWN~Kes~75 (G29.70--0.30). The complete figure set of all 36 SEDs is provided in the online journal.
}
\end{figure}

\begin{figure}[!h]
\centering
\includegraphics[width=1.0\linewidth]{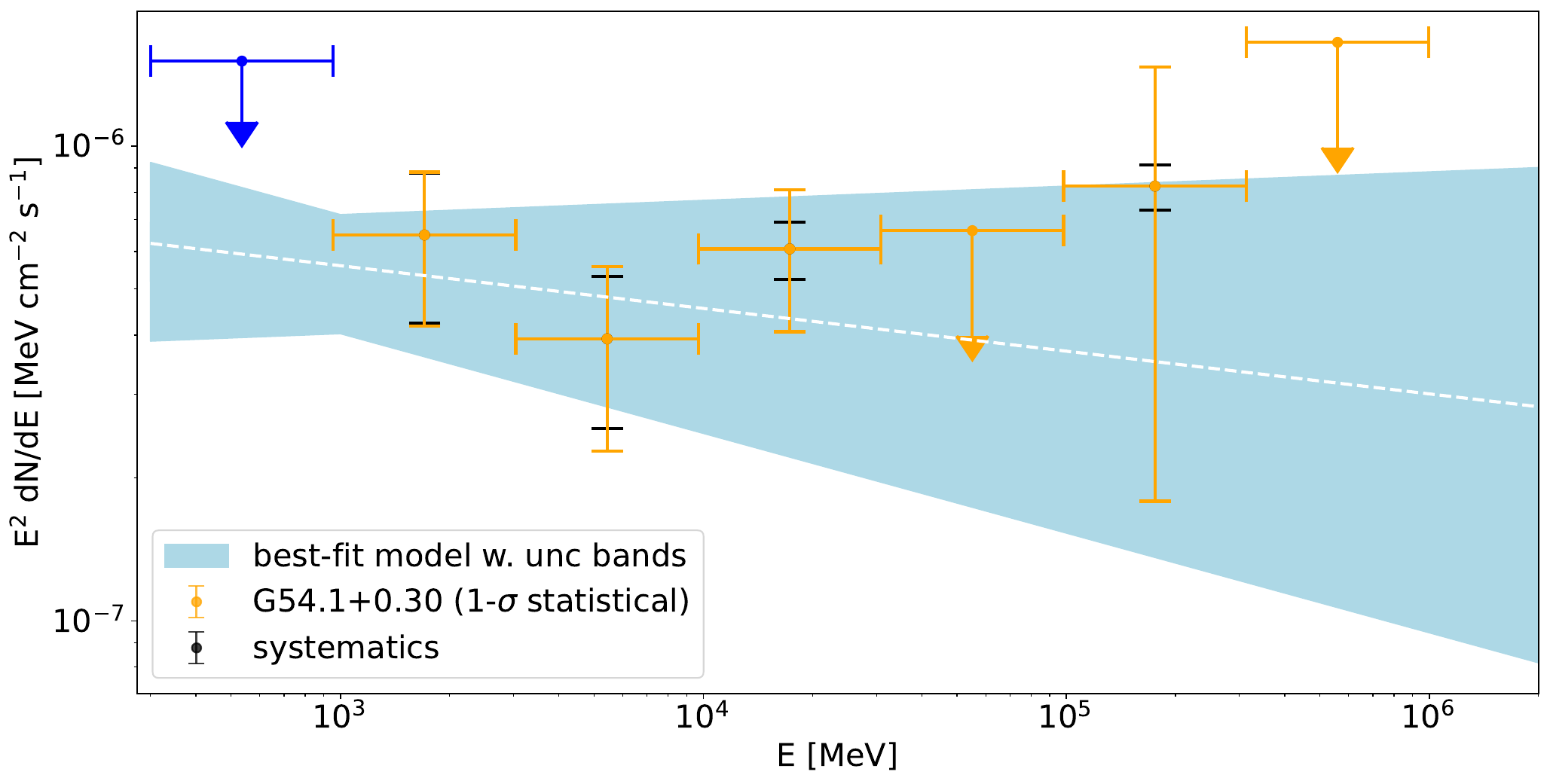}
\caption{300\,MeV--1\,TeV SED for PWN~G54.10+0.27, corresponding to 4FGL J1930.5+1853 in 4FGL--DR3. The complete figure set of all 36 SEDs is provided in the online journal.}
\end{figure}

\begin{figure}[!h]
\centering
\includegraphics[width=1.0\linewidth]{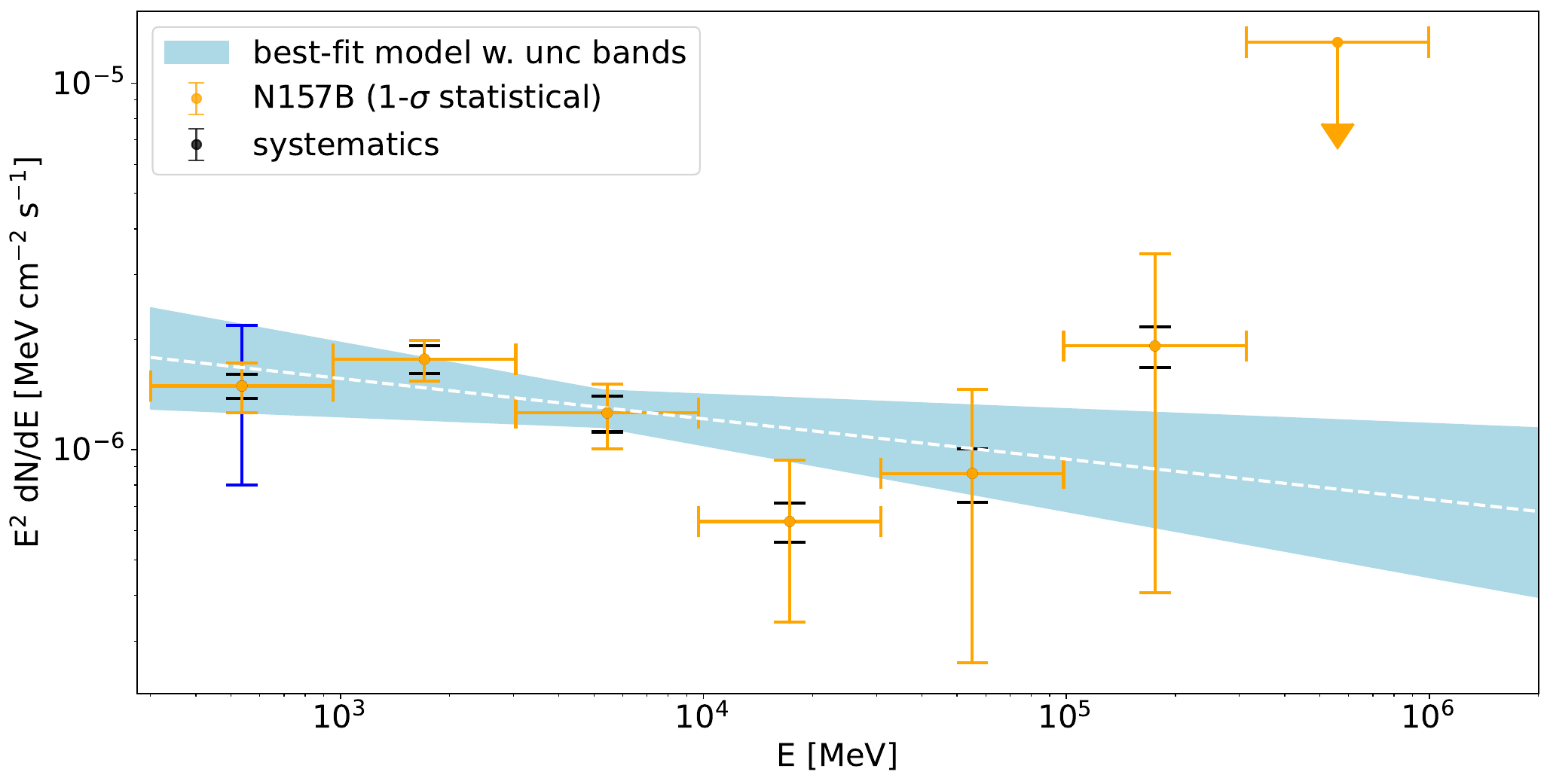}
\caption{300\,MeV--1\,TeV SED for PWN N~157B, corresponding to 4FGL J0537.8--6909. The complete figure set of all 36 SEDs is provided in the online journal.}
\end{figure}

\begin{figure}[!h]
\centering
\includegraphics[width=1.0\linewidth]{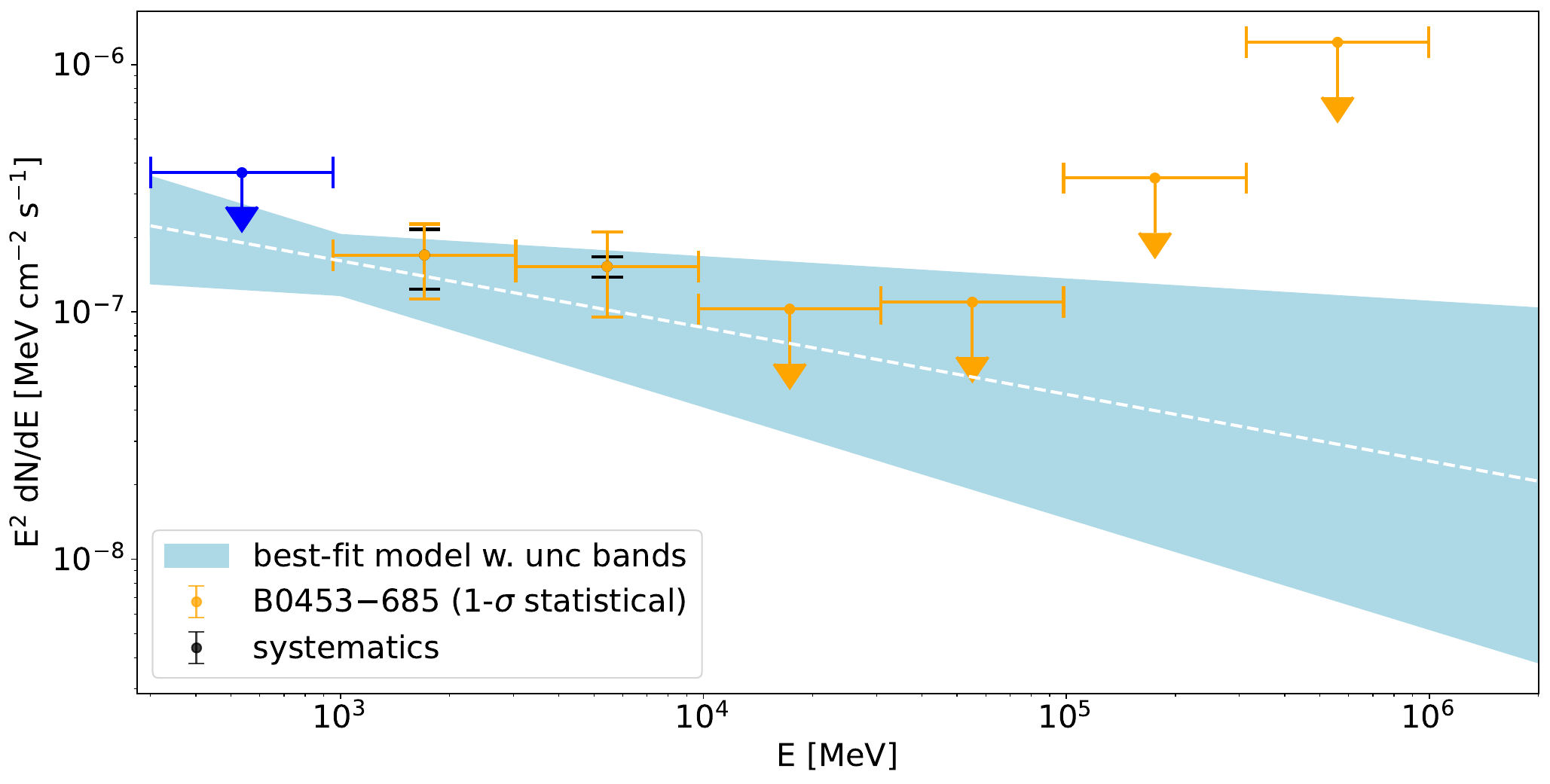}
\caption{300\,MeV--1\,TeV SED for PWN~B0453--685. The complete figure set of all 36 SEDs is provided in the online journal.}
\end{figure}

\begin{figure}[!h]
\centering
\includegraphics[width=1.0\linewidth]{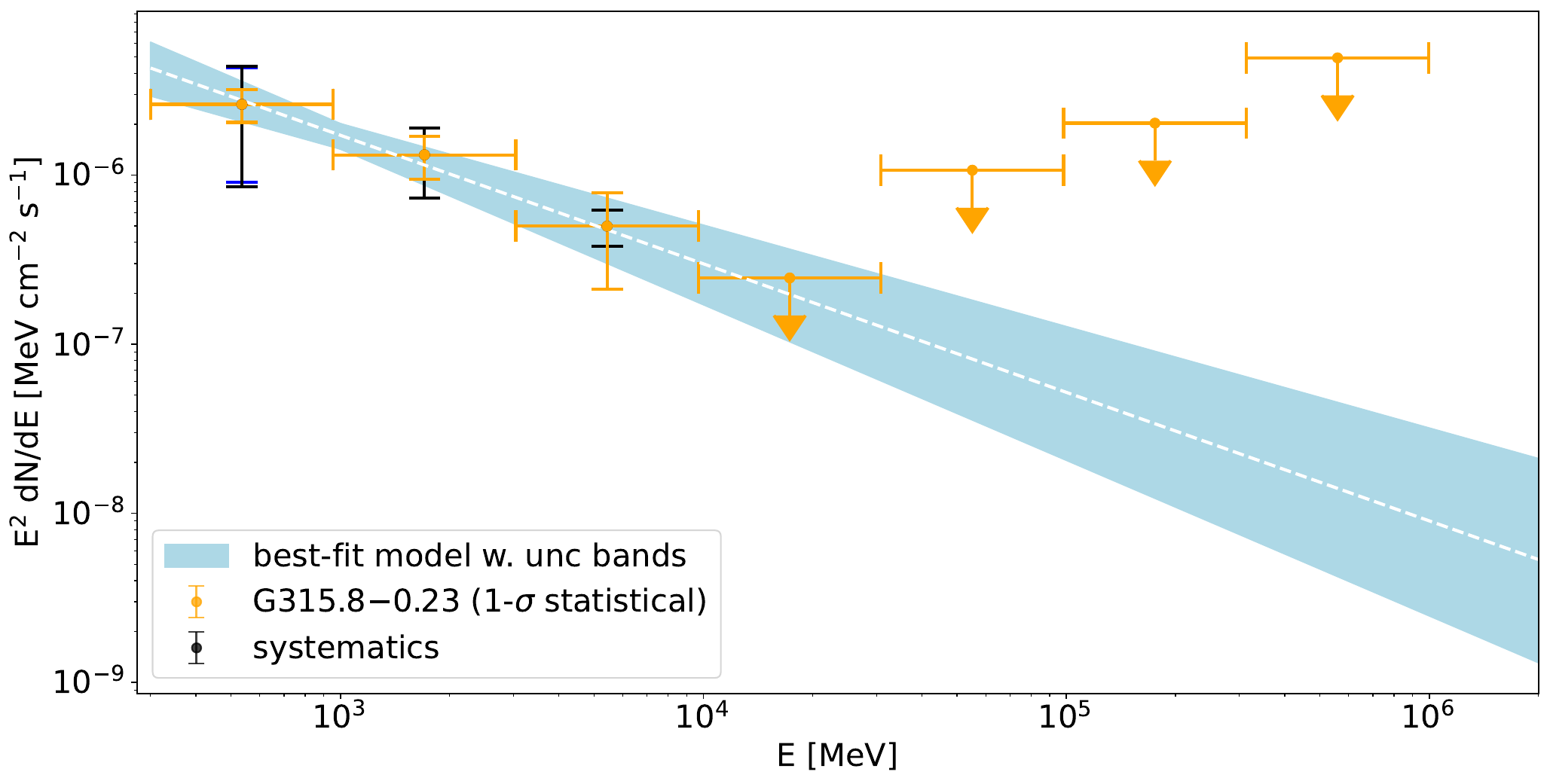}
\caption{300\,MeV--1\,TeV SED for PWN~G315.78--0.23 (the Frying Pan), corresponding to 4FGL J1435.8--6018. The complete figure set of all 36 SEDs is provided in the online journal.}
\end{figure}

\begin{figure}[!h]
\centering
\includegraphics[width=1.0\linewidth]{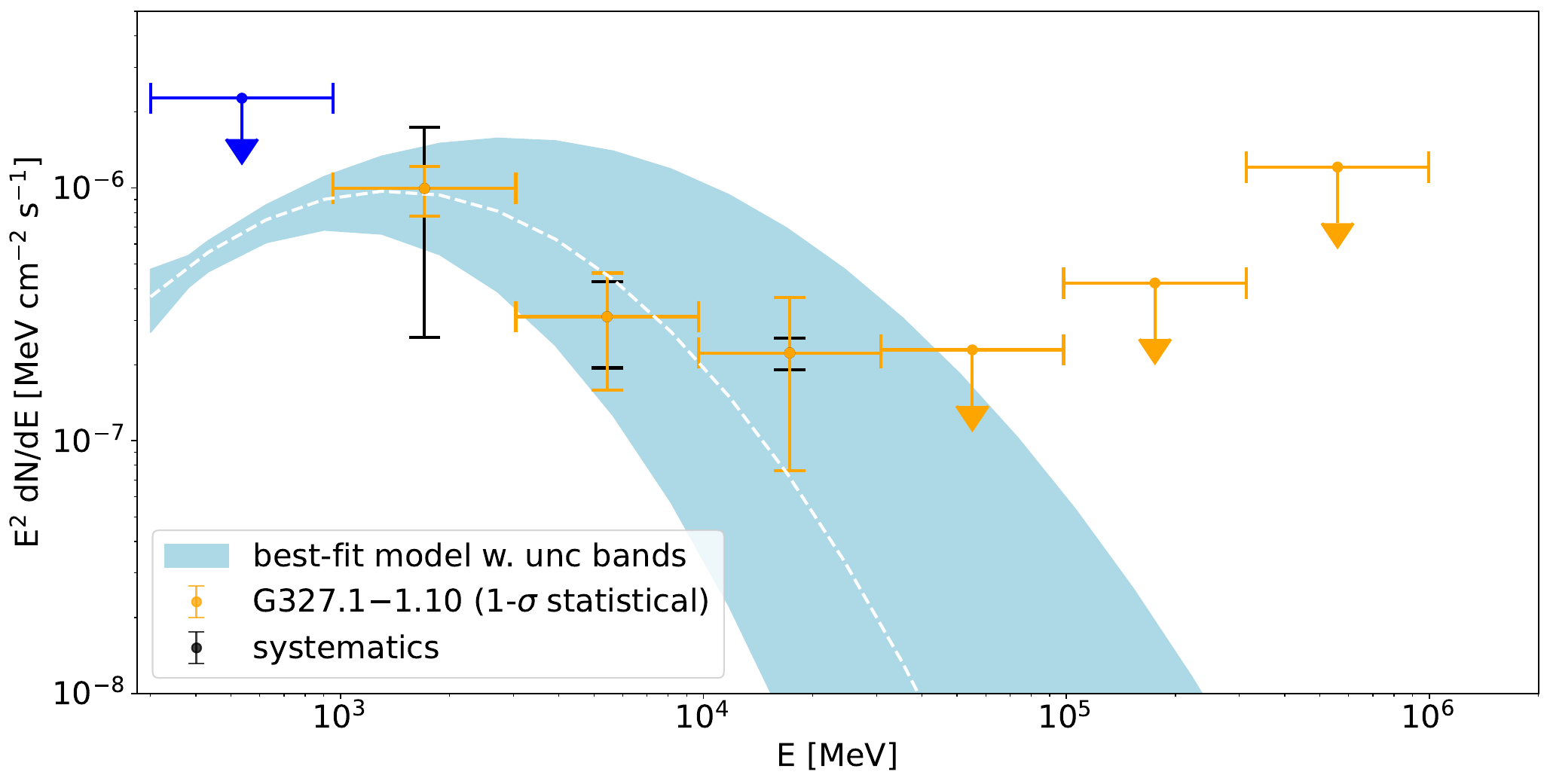}
\caption{300\,MeV--1\,TeV SED for PWN~G327.15--1.04, corresponding to 4FGL J1554.4--5506 in 4FGL--DR3. The complete figure set of all 36 SEDs is provided in the online journal.}
\end{figure}

\begin{figure}[!h]
\centering
\includegraphics[width=1.0\linewidth]{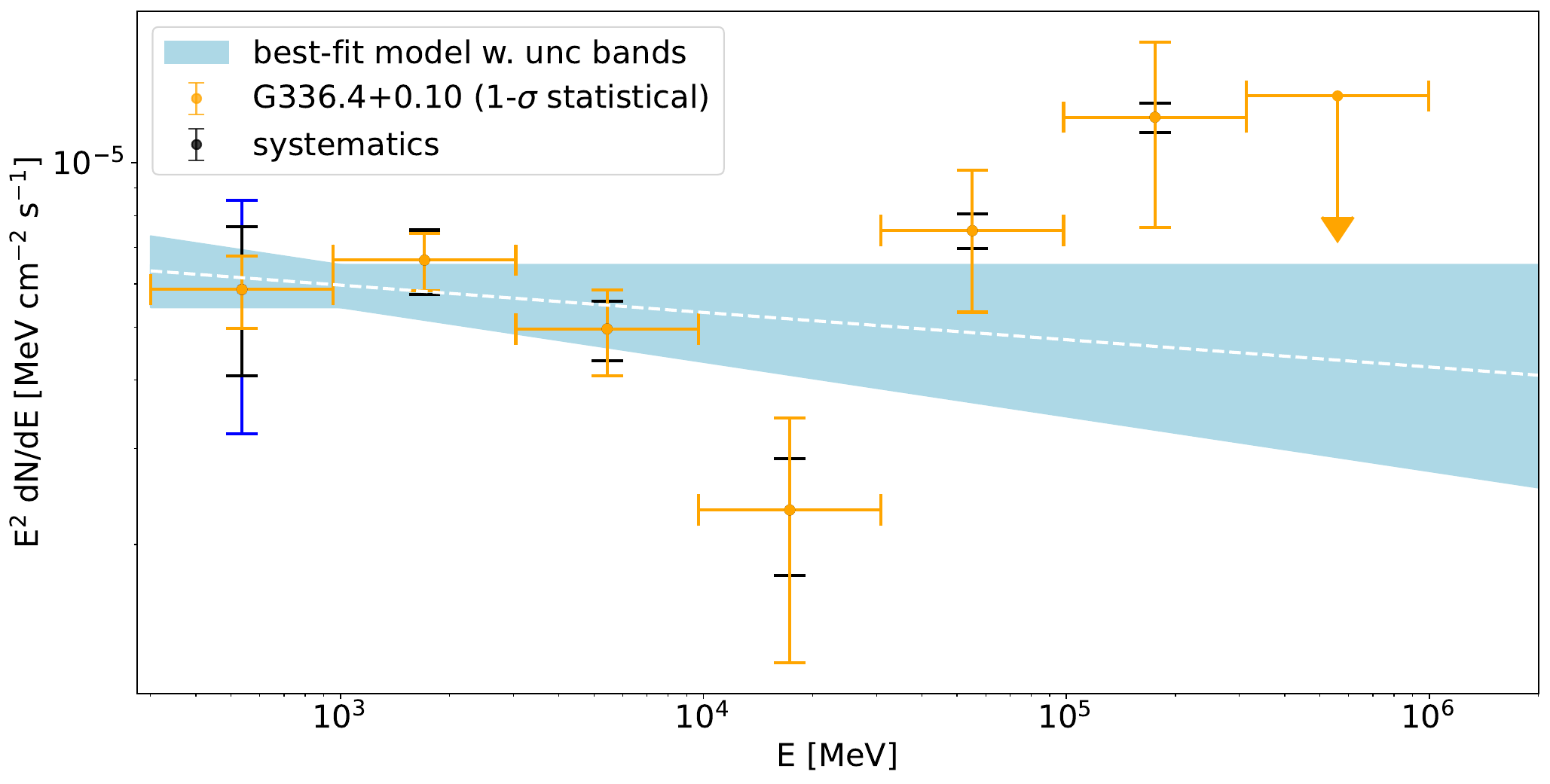}
\caption{300\,MeV--1\,TeV SED for PWN~G336.40+0.10, corresponding to 4FGL J1631.6--4756e. The complete figure set of all 36 SEDs is provided in the online journal.}
\end{figure}

\bibliographystyle{aasjournal}
\bibliography{ref.bib}

\end{document}